\documentclass[english,aps,manuscript,nofootinbib,aps,prc,tightenlines,floats,floatfix]{revtex4}
\usepackage[T1]{fontenc}
\usepackage[latin9]{inputenc}
\usepackage{float}
\usepackage{amsmath}
\usepackage{graphicx}
\usepackage{amssymb}
\usepackage{color}

\makeatletter

\providecommand{\tabularnewline}{\\}

\newcommand{\nn}{\nonumber}

\@ifundefined{textcolor}{}
{%
 \definecolor{BLACK}{gray}{0}
 \definecolor{WHITE}{gray}{1}
 \definecolor{RED}{rgb}{1,0,0}
 \definecolor{GREEN}{rgb}{0,1,0}
 \definecolor{BLUE}{rgb}{0,0,1}
 \definecolor{CYAN}{cmyk}{1,0,0,0}
 \definecolor{MAGENTA}{cmyk}{0,1,0,0}
 \definecolor{YELLOW}{cmyk}{0,0,1,0}
 }

\usepackage{amsbsy}
\usepackage{amsfonts}
\usepackage{float}
\usepackage{rotating}

\usepackage{bm}
\makeatletter
\usepackage{graphics}
\usepackage{epsf}
\usepackage{bm}

\def\s#1{\setbox0=\hbox{$#1$}  
   \dimen0=\wd0     
   \setbox1=\hbox{/} \dimen1=\wd1  
   \ifdim\dimen0>\dimen1   
      \rlap{\hbox to \dimen0{\hfil/\hfil}} 
      #1     
   \else     
      \rlap{\hbox to \dimen1{\hfil$#1$\hfil}} 
      /      
   \fi}      %

\makeatother

\usepackage{babel}

\def\beq{\begin{equation}}
\def\eeq{\end{equation}}
\def\bea{\begin{eqnarray}}
\def\eea{\end{eqnarray}}
\def\beqa{\begin{equation}\begin{array}{l}}
\def\eeqa{\end{array}\end{equation}}

\def\Figref#1{Fig.~\ref{fig:#1}}

\def\Tabref#1{Table \ref{tab:#1}}

\def\slap{p \hspace{-2mm} \slash}

\def\half{\mbox{\small{$\frac{1}{2}$}}}

\def\ga{\gamma} 
\def\de{\delta} \def\De{\Delta}\def\vDe{\varDelta}

\def\w{\omega}

\def\nn{\nonumber}

\def\im{\mbox{Im}}

\begin{document}
\preprint{MKPH-T-11-15}

%
%
%
%

\title{The nucleon and $\Delta$(1232) form factors at low momentum-transfer
  and small pion masses }

\author{T. Ledwig$^{1}$\footnote{E-mail: ledwig@kph.uni-mainz.de}}
\author{J. Martin-Camalich$^{2}$}
\author{V. Pascalutsa$^{1}$}
\author{M. Vanderhaeghen$^{1}$}

\affiliation{$^{1}$Institut für Kernphysik, Johannes Gutenberg-Universität, D-55099
Mainz, Germany\\
 $^{2}$Departamento de Fisica Teorica and IFIC, Universidad de Valencia-CSIC, Spain.\\
    Department of Physics and Astronomy, University of Sussex, BN1 9Qh, Brighton, UK.}

%
%
%
%

\begin{abstract}
An expansion of the electromagnetic form factors of the nucleon and $\Delta(1232)$
in small momentum transfer and pion mass is performed 
in a manifestly-covariant EFT framework consistent with chiral symmetry and analyticity.
We present the expressions for 
the nucleon and $\Delta(1232)$  electromagnetic form factors, charge radii, and
electromagnetic moments 
in the framework of SU(2) baryon chiral perturbation
theory, with nucleon and $\Delta$-isobar degrees of freedom, to
next-to-leading order. Motivated by the results for the proton electric
radius obtained from the muonic-hydrogen atom and electron-scattering process, we extract
values for the second derivative of the electric form factor which is a
genuine prediction of the $p^3$ B$\chi$PT. The chiral behavior of 
radii and moments is studied and compared to that obtained in the heavy-baryon
framework and lattice QCD. The chiral behavior of $\Delta(1232)$-isobar properties
exhibits cusps and singularities at the threshold of $\Delta\to\pi N$
decay, and their physical significance is discussed. 
\end{abstract}
\maketitle

\date{\today}

%
%
%
%

\tableofcontents

\section{Introduction}

The physics of nucleon form factors is about 60 years old \cite{Hofstadter:1955ae,Hofstadter:1956qs} and yet surprises
in this venue are not unusual till now.  Just last year the most precise atomic
measurement of the proton charge radius yielded \cite{Pohl:2010zza}:
  $r_{Ep}=\sqrt{\left<r^2\right>}=0.84184(67)$ fm, 
in unexpected disagreement with the best electron-scattering result \cite{Bernauer:2010wm}: 
$r_{Ep} = 0.879(8)$ fm.
Much effort since then have been focused on finding a ``missing" correction in the
muonic-hydrogen result, e.g. 
\cite{Jentschura:2010ha,Distler:2010zq,Carlson:2011zd,Miller:2011yw}. We, on the other hand, will attempt to provide here some 
grounds for an improvement of the electron-scattering analysis.
The electron-scattering measurement of the proton charge radius is done
by determining the slope of the proton form factors at zero momentum
transfer: $Q^2\equiv -q^2 = 0$. In reality the measurements are done at small but finite 
momentum transfer, $Q^2\geq 0.01$ GeV$^2$,
and an interpolation to zero is required. The simplest one is based
on Taylor expansion in $Q^2$, 
\beq
G_{Ep}(Q^2) = 1 + \frac{\left<r^2\right>}{6} Q^2 +   \frac{\left<r^4\right>}{120} Q^4 + \ldots,
\eeq
where $\left<r^n\right>$ is the $n$th moments of the proton charge distribution which
values are fitted to data. 
However, the validity of such an expansion, its radius of convergence, is limited
by the nearest singularity in the complex $Q^2$ plane, which, if we neglect
the radiative corrections, is located at $Q^2=-4m_\pi^2$,
the the two-pion production threshold. This simply means that a polynomial 
fit is limited to $|Q^2| \ll 4m_\pi^2 \approx 0.08$ GeV$^2$, where the database
is scarce.  One can extend the interpolation range only by including the effect of the pion-production channels explicitly. This can in principle be done using dispersion theory,
see e.g.~\cite{Mergell:1995bf,Hammer:2006mw,Pacetti:2007zz}. 
For that, however, one needs the information in the timelike
region, which is also not accurate enough, and is usually complemented
in a model-dependent fashion. Nonetheless,
some of the state-of-the-art dispersion analyses 
\cite{Hammer:2003ai,Belushkin:2005ds} had 
obtained the smaller value of $r_{Ep}$ (well before the muonic hydrogen result appeared!),
which reinforces the motivation to include the pion-production effects in the interpolation
of low-$Q^2$ data. 
Here we approach this issue in the framework 
of chiral perturbation theory ($\chi$PT) \cite{Weinberg:1978kz, Gasser:1983yg}. 
The $\chi$PT itself does
not have a prediction for the proton charge radius, its leading-order value is given by 
a combination of low-energy constants (LECs), which are
free parameter of the theory to be matched to QCD. 
However, the leading order pion-loop contributions are fixed in terms of
well-known parameters and provide a prediction of the analytic structure of the
form factors at small $Q^2$.  In this work we shall only
present the relevant $\chi$PT calculations; their impact on the charge radius extraction will
be studied elsewhere.

Another set of issues concerning the electromagnetic form factors comes from the side of lattice QCD,
which presently is the only method to do {\em ab initio} calculations of the low-$Q^2$
hadron structure.  The latest lattice QCD calculations of the nucleon  
\cite{Yamazaki:2009zq,Syritsyn:2009mx,Alexandrou:2011db,Collins:2011,Bratt(2010):LHPCLatticeNucleon}
and $\Delta(1232)$ \cite{Aubin:2008qp, Alexandrou:2008bn,Alexandrou:2009hs,Alexandrou:2009nj,Alexandrou:2010uk}
electromagnetic (e.m.) properties call for 
a better analysis of the pion-mass and volume dependencies. The most troublesome
are the results for the nucleon charge radii, which show little dependence on the pion mass
and a large discrepancy with experiment upon a naive extrapolation to the physical
pion mass. $\chi$PT predicts charge radii to diverge in chiral limit ($m_\pi \to 0$)
and therefore from its point of view it is plausible that the correct chiral
extrapolation and finite-volume corrections will reconcile the lattice results
with experiment. 

Presently, both $m_\pi$ and
finite-volume dependencies are usually computed using the heavy-baryon 
$\chi$PT (HB$\chi$PT) \cite{Jenkins:1990jv}, 
where the chiral expansion is accompanied with an expansion in the inverse baryon
mass. The latter expansion can be poorly convergent (see, e.g. \cite{Becher:1999he,Pascalutsa:2011fp}) and the 
so-called manifestly Lorentz-invariant schemes \cite{Becher:1999he, Gegelia:1999qt}, which avoid the heavy-baryon expansion,
gain popularity in practice. In this work we adopt the 
extended on-mass shell scheme (EOMS) \cite{Fuchs:2003qc}, which has the advantage of 
preserving analyticity. As a result, our expressions for the form factors will satisfy
the usual dispersion relations written in $Q^2$, as well as  the dispersion
relation of Ref.~\cite{Ledwig:2010nm} written in $m_\pi^2$:
 \begin{subequations}
  \bea
 G(Q^2, m_\pi^2) &= & \frac{1}{\pi}\int\limits_{0}^\infty \!d q^2\, \frac{\im\, G(-q^2,m_\pi^2)}{q^2+Q^2}\left(\frac{-Q^2}{q^2} \right)^n
 \\
 &=&-\frac{1}{\pi}\int\limits_{-\infty}^0 \! d \tilde m_\pi^2\, \frac{\im \,
 G(Q^2,\tilde m_\pi^2)}{\tilde m_\pi^2-m_\pi^2} \left(\frac{m_\pi^2}{\tilde m_\pi^2} \right)^n,
 \eea
 \end{subequations}
where $0$ in the integration limits is indicative of the threshold position, $n$
is the number of subtractions; $Q^2$ and $m_\pi^2$ are positive.
The earlier $\chi$PT analyses of nucleon and $\De$-isobar form factors 
were based on either the heavy-baryon approach \cite{Bernard:1998gv,Jiang:2009jn},
or the infrared-regularization scheme \cite{Kubis:2000zd}, where the above dispersion relations
can only be satisfied approximately, unless a special care is taken as, e.g., in \cite{Kaiser:2003qp}.  
Ref.~\cite{Fuchs:2003ir} contains thusfar the only $SU(2)$ calculation of
nucleon form factors in the EOMS whereas calculations of the octet- and
decuplet-baryon em moments has been reported in the context of $SU(3)$
B$\chi$PT in \cite{Geng:2008mf,Geng:2009hh,Geng:2009ys}. Here we have recalculated the contributions 
found in~\cite{Fuchs:2003ir}, included the leading-order corrections due
to $\De$-isobar, and computed all the  $\De$(1232)-isobar form factors
to next-to-leading order. 

 In Sect.~II, we summarize the ideas of chiral expansion in the single-baryon sector
and specify the contributions calculated in this work. In Sect.\ III and IV we consider
the pion-mass dependence of, respectively, the nucleon and $\Delta$ electromagnetic 
radii and moments, and compare it with the HB$\chi$PT results and lattice-QCD results where possible. Some conclusions are
presented in Sect.\ V. Appendix A contains the notation and definitions, while 
Appendices  B and C contain analytical expressions of the contributions to, respectively, 
the nucleon and the $\De$ form factors obtained in this work.

%
%
%
%

\section{Form factors in baryon $\chi$PT\label{sec:Chiral-Lagrangians}}
The chiral effective-field theory to which we refer as to $\chi$PT 
 is an effective-field theory of low-energy QCD, as it contains the
most general form of interaction among the lightest hadrons --- most notably, pions ---
in a way consistent with symmetries of QCD Green's functions  \cite{Weinberg:1978kz, Gasser:1983yg}. A special role
is enjoyed by chiral symmetry which insures that pions couple through a derivative
couplings while the symmetry breaking terms are accompanied with powers of $m_\pi^2$.
When the scale of spontaneous chiral symmetry breaking,  $4\pi f_\pi \approx $ 1 GeV,  is much larger
than the scale of the explicit symmetry breaking, $m_\pi$, as is observed in nature, one 
may set up a systematic expansion of any observable quantity
in powers of $E/(4\pi f_\pi)$ and $m_\pi/(4\pi f_\pi)$,
where $E$ is the characteristic relative-energy of external legs
 in a given process. These ratios of light to heavy scales
are commonly denoted as $p$. To a given order in $p$, a finite number of terms, 
accompanied by a finite number of 
low-energy constants (LECs), contribute. Simple power-counting
rules exist to select the necessary contributions to any given order in $p$.

\subsection{Power counting in the single-baryon sector}
The inclusion of the nucleon fields was initially done by Gasser, Sainio and Svarc~\cite{Gasser:1987rb}, who note that the nucleon mass $M_N$ invalidates the usual power-counting arguments.
For instance, the one-loop nucleon self-energy graph, with the leading $\pi NN$ couplings,
counts as order $p^3$, but in the actual calculation the positive powers of $M_N$ appear
and make this contribution of order $p^2$. It was later on realized that such "power-counting
violating" terms have no physical effect since their contribution is always compensated by LECs
present at that order in the expansion of physical quantity \cite{Gegelia:1999qt}. 
One can set up a scheme where the troublesome terms are absorbed by a renormalization
of available LECs, e.g.\ the EOMS \cite{Fuchs:2003qc}.

A neat way to get rid
of positive powers of $M_N$ from the outset is provided by the  
HB$\chi$PT \cite{Jenkins:1990jv}.  In HB expansion, which is in a way
similar to semi-relativistic treatments,  in addition to the positive power of $M_N$
one drops a number of contributions with negative power of $M_N$. These contributions are
typically of the form 
\beq
\left(\frac{m_\pi}{M_N}\right)^n \Big[ a + b \ln \frac{m_\pi}{M_N} \Big]  , 
\eeq
with $n$ higher than the order of $p$ to which the expansion is  made. 
As long as the constants $a$ and $b$ are of order of unity (natural size) relative
to the coefficients of the given-order term,  these terms are indeed 
of the size of higher-order corrections. There are examples, however, where $a, \, b$
are unnaturally large and the expansion fails as the result (see, e.g.
 \cite{Becher:1999he,Pascalutsa:2011fp}). In these cases,
the expansion in $p$ might only converge if one refrains from the HB expansion.

A popular manifest-Lorentz-invariant 
scheme where the power-counting-violating terms do not arise is the infrared
regularization (IR) of Becher and Leutwyler~\cite{Becher:1999he}, which 
has been applied to nucleon form factors by Kubis and Meissner \cite{Kubis:2000zd}. 
The IR procedure can be described as follows.

\begin{itemize}
\item[o] {\em An equivalent formulation of the IR:}\\
 The negative-pole contribution of nucleon propagator in a give loop graph
is deleted by hand. As the result, the graphs with nucleon propagators only
vanish, since the contour can always be closed in the half-plane which does not
have a pole. In the graphs where both the nucleon propagators enter with a pion propagator, e.g.,
\beq
S_\pi(k) S_N(p) \equiv \frac{1}{k^2 - m_\pi^2} \frac{\slap+M_N}{p^2 - M_N^2} 
\eeq
the nucleon propagator is replaced as follows:
\beq
S_N(p) \to \frac{\slap+M_N}{p^2 - M_N^2 - S_\pi^{-1}(k)} = S_N(p) 
\left[1+\frac{1}{S_\pi(k) \, (p^2-M_N^2)} \right].
\eeq
In any one-loop graph containing $N_\pi$ pion propagators,
\beq
S_\pi(k_1) \cdots S_\pi (k_{N_\pi})\,,
\eeq
each nucleon propagator
changes as follows:
\beq
S_N(p) \to   S_N(p) 
 \left[1- (-1)^{N_\pi} \prod_{n=1}^{N_\pi} 
\frac{1}{S_\pi(k_n)\, (p^2-M_N^2)}\right]\,.
\eeq
\end{itemize}
This formulation is more convenient to  check
Ward-Takahashi identities since the normal propagator preserve gauge invariance and the additional part vanishes upon
closing the loop integration contour in the half-plane which is free of
poles. It is not difficult to see that the "modified" IR procedure
\cite{Schindler:2003xv}, introduced to define IR beyond one loop, 
does not satisfy the e.m. gauge symmetry exactly, but only to a given order in the chiral expansion. The violating terms are of higher order from the viewpoint of heavy-baryon expansion, but not in a covariant framework.

One apparent drawback of IR is that it changes the analytic structure
of the loop integrals  such that unphysical cuts appear. The unphysical cuts lie
far outside the region of $\chi$PT interest, but they still have an effect on that region
as can be seen for example through a dispersive representation. 
Namely, if the quantity in question obeys a dispersion relation, let say in energy $s$,
\beq
G(s) = \frac{1}{\pi} \int\limits_{s_0}^\infty\! ds'\,  \frac{\im\, G(s')}{s'-s},
\eeq
then in the IR scheme it would take the form:
\beq
G^{(\mathrm{IR})}(s) = \frac{1}{\pi} \int\limits_{s_0}^\infty\! ds'\,  \frac{\im\, G(s')}{s'-s}
+  \frac{1}{\pi} \int\limits_{-\infty}^{f_0} \! ds'\, \frac{\im\, G^{(\mathrm{IR})}(s')}{s'-s},
\eeq
such that, even if $f_0$ is far away from the region of interest (i.e.,  $f_0\ll s$ and $s\approx s_0 $), an
unphysical contribution is generated and its smallness is hard to assess {\it a priory}.
The imaginary part over the physical cut is the same in IR, EOMS, or any other relativistic
scheme. In fact, the whole difference between the IR and EOMS is the unphysical cut contribution.

A common problem of Lorentz-covariant schemes is that the superficial index
of divergence $\w$ may exceed the chiral power-counting index $n$, and thus
an UV-divergence may appear $\w-n$ orders lower than the LEC which renormalizes
it. This problem is often viewed as an inconsistency of the covariant approach,
but in fact it only means one needs to specify the renormalization scheme for all LECs from
the outset. In HB$\chi$PT, $\w=n$, because the time-derivatives of the heavy (nucleon) field are eliminated. On the other hand, the HB$\chi$PT results can readily be reproduced from
covariant ones by expanding the latter in the inverse baryon masses.

Since the nucleon is easily excited into the $\De(1232)$-resonance, the excitation energy
$\vDe=M_\De -M_N \ll 4\pi f_\pi$,  the $\chi$PT with nucleons is not complete without the $\De$-isobar degrees of freedom. The power-counting for the $\De$-isobar contributions depends
on how the two light scales $m_\pi $ and  $\vDe$ compare to each other. $m_\pi \sim \vDe$
leads to the "small-scale-expansion" (SSE) \cite{Hemmert:1996xg}, while $m_\pi \ll \vDe$ leads to
the "$\de$-expansion" \cite{Pascalutsa:2002pi,Long:2009wq}. 
In the absence of one-particle-reducible
graphs, as is in the case of form factors, the two power-countings yield very similar results.
In the $\de$-expansion, where a one-particle-irreducible graph
with $L$ loops, $N_\pi$ pion propagators, $N_N$ nucleon propagators, $N_\De$ propagators,
and $V_k$ vertices containing $k$ powers of pion momentum (and electric charge), counts as:
\beq
p^{4L - 2N_\pi - N_N  - N_\De - \sum_k k V_k}\, \left( \frac{p}{\vDe}\right)^{N_\De}\label{PC1}.
\eeq
In the SSE the power-counting index of such graphs would be
\beq
p^{4L - 2N_\pi - N_N  -  N_\De - \sum_k k V_k }\,\label{PC2},
\eeq
and hence for $p\sim \vDe$ the two countings coincide.  The pion mass insertions which
are relevant for the pion-mass-dependence calculation will still render 
the countings to be different. In this case, however, the $\de$-expansion is not appropriate
as the pion-mass dependence needs usually to be assessed in the range of $m_\pi \sim \vDe$.
For this purpose we adopt the SSE counting.

In this work we have calculated the $p^3$ graphs shown in \Figref{diagramsNucleon}
and  \Figref{diagramsDelta}. The resulting expressions are collected in Appendix B and C,
respectively. Below we list  the terms of the effective chiral Lagrangian that were
used in the calculation of these loops.

\begin{figure}[bht]
\begin{center}
\includegraphics[scale=0.35]{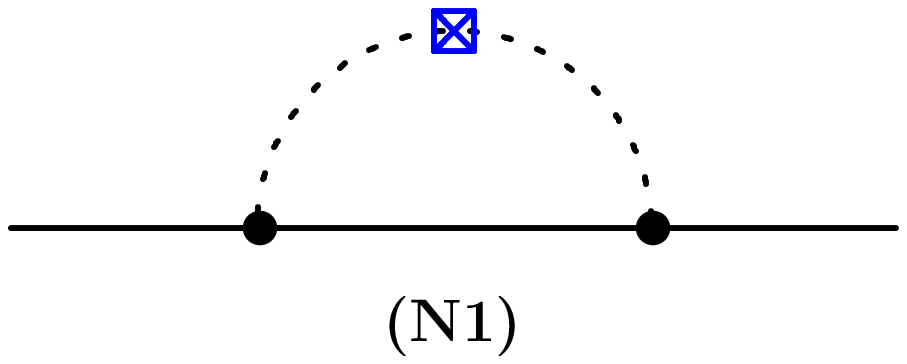}~~~~~\includegraphics[scale=0.33]{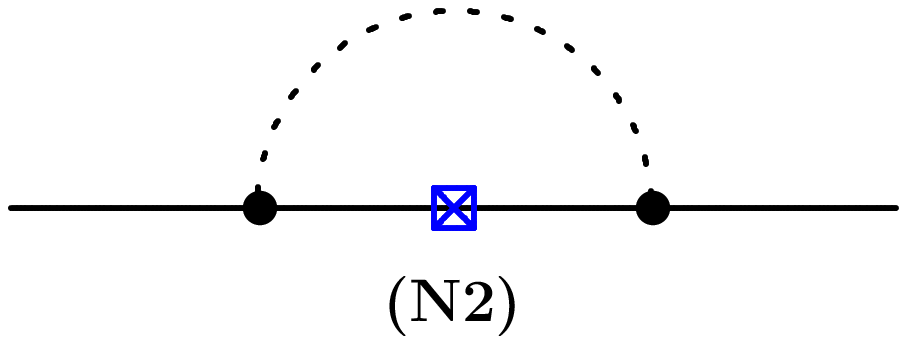}~~~~~\includegraphics[scale=0.33]{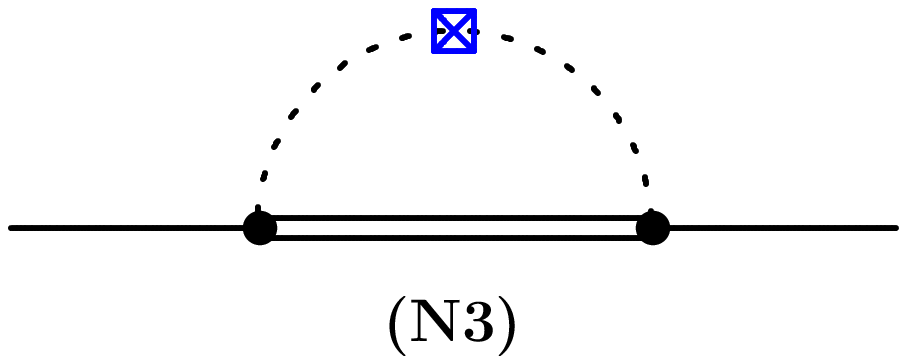}~~~~~\includegraphics[scale=0.35]{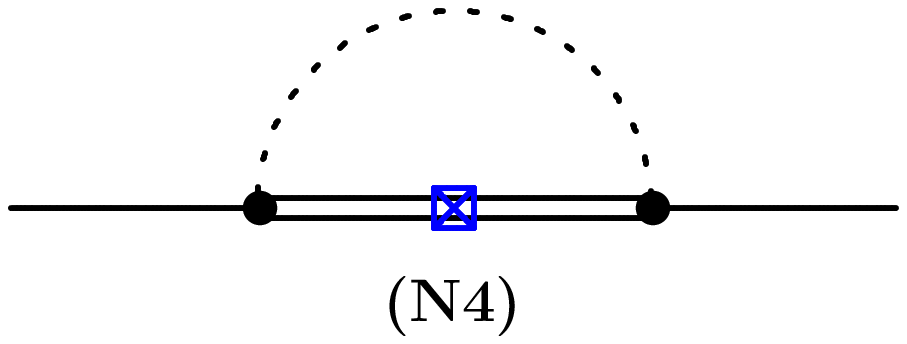}\\
~\\
\includegraphics[scale=0.35]{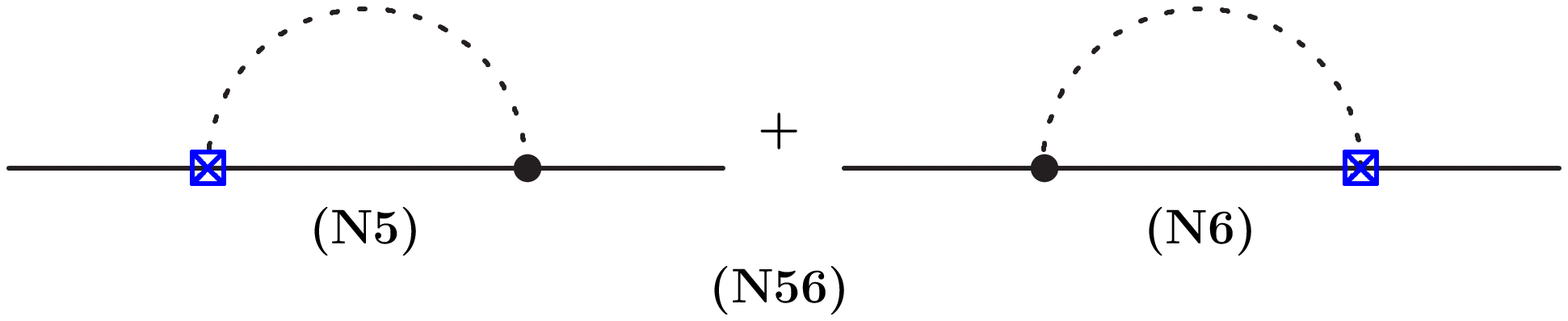}~~~~~~~~~~~\includegraphics[scale=0.33]{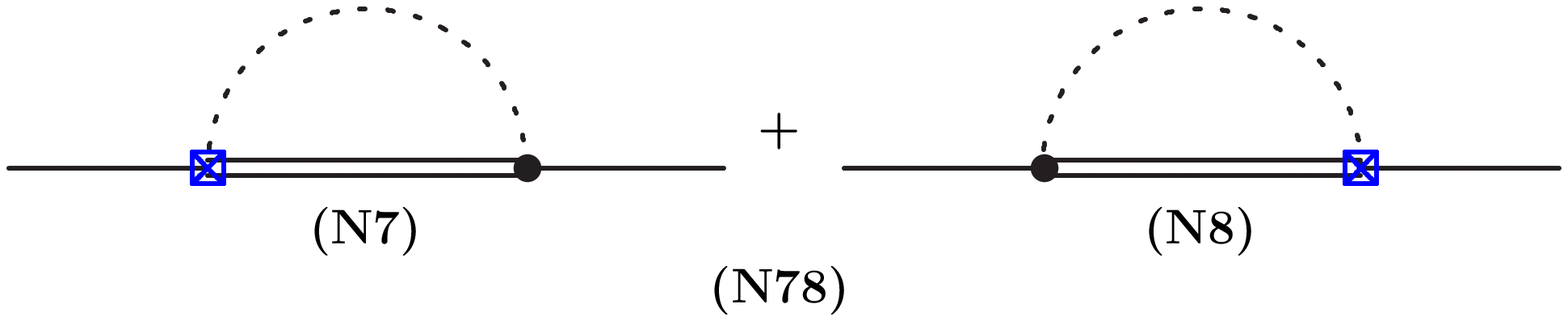}\\
\hspace{1cm}\\
\includegraphics[scale=0.35]{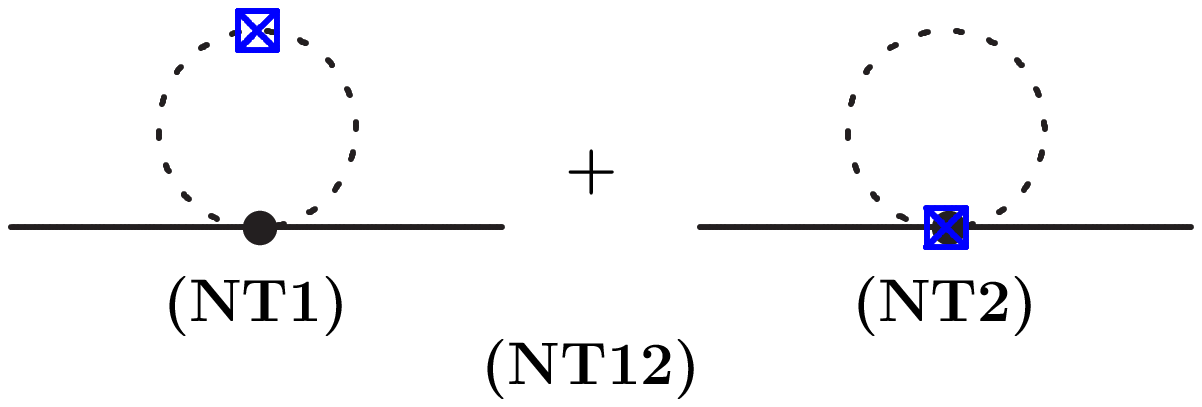}
\end{center}
\caption{\label{fig:diagramsNucleon} Order-$p^3$ corrections to 
the nucleon form factors. Single-lines denote the nucleon, double-lines
the $\Delta$-isobar, and dashed lines the pion propagators. The photon
coupling is denoted by the blue square and the $N\pi$ or $\Delta\pi$
vertices by dots.}
\end{figure}

\begin{figure}[htb]
\begin{center}
\includegraphics[scale=0.33]{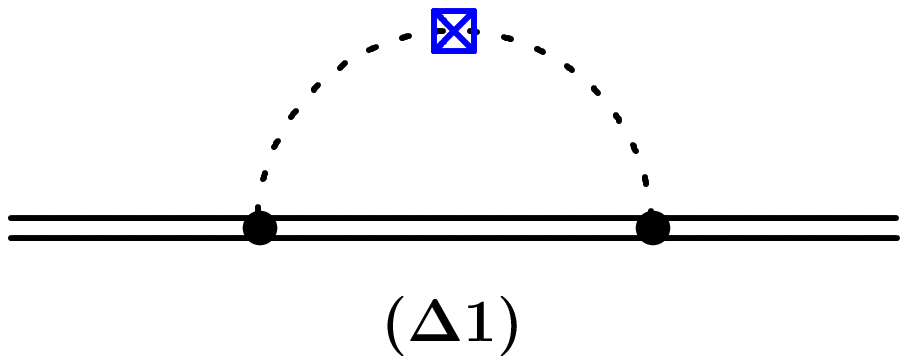}~~~~~\includegraphics[scale=0.33]{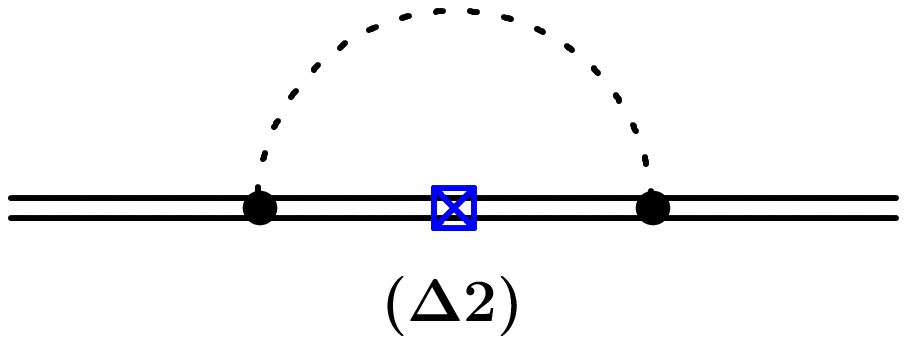}~~~~~\includegraphics[scale=0.33]{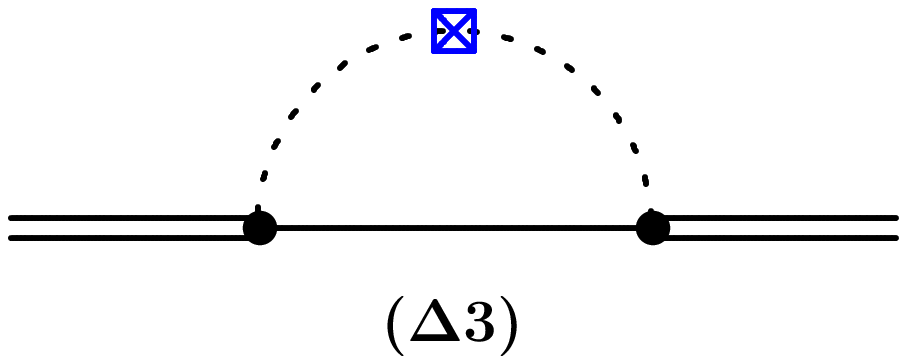}~~~~~\includegraphics[scale=0.33]{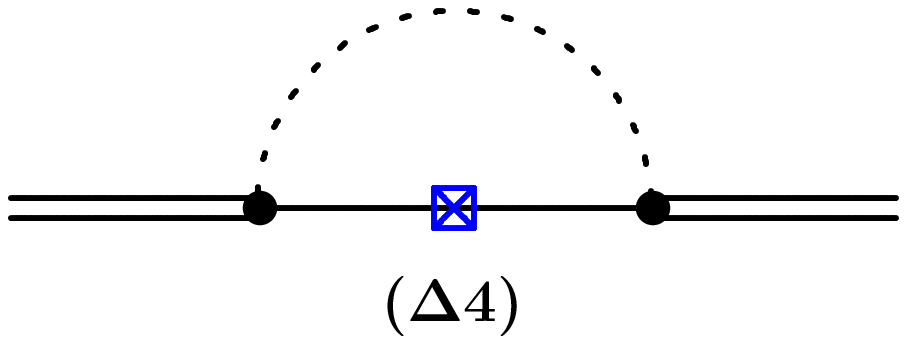}\\
~\\
\includegraphics[scale=0.33]{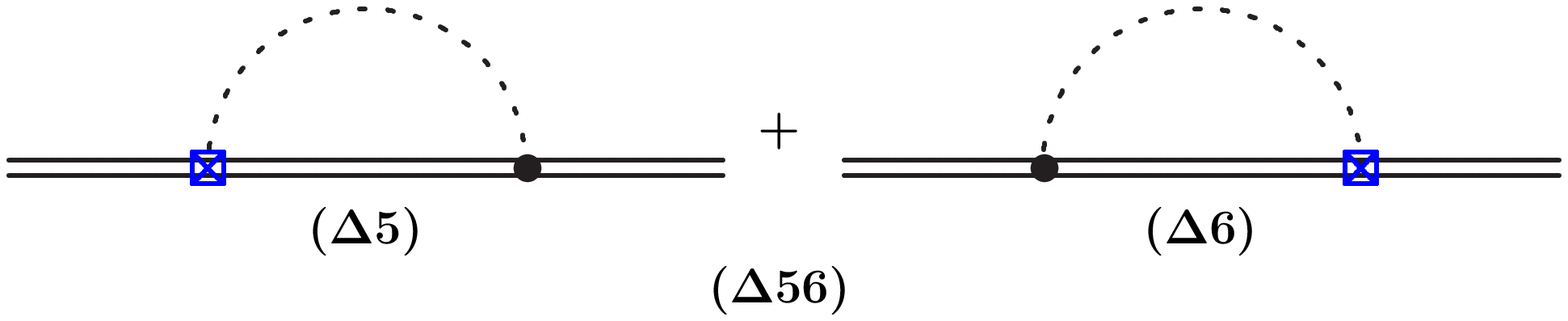}~~~~~~~~~~~\includegraphics[scale=0.33]{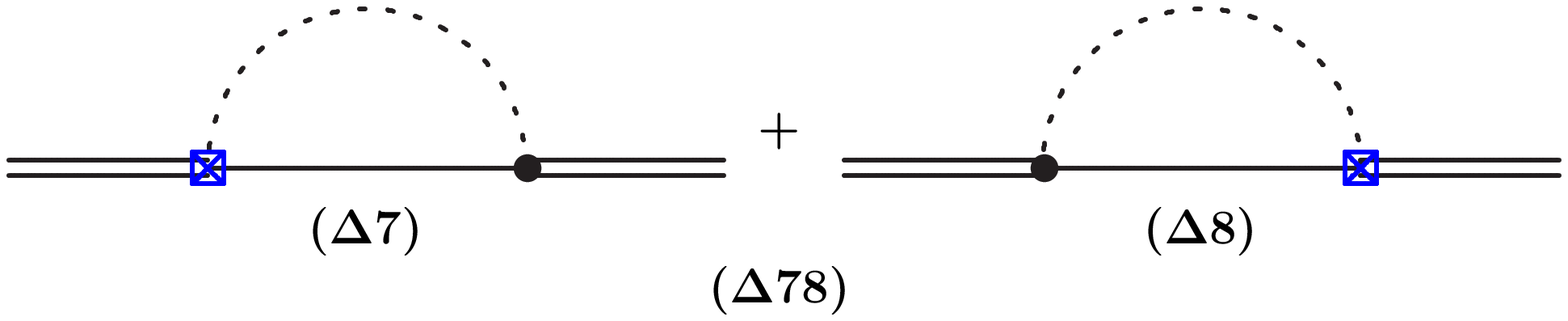}\\
\hspace{1cm}\\
\includegraphics[scale=0.33]{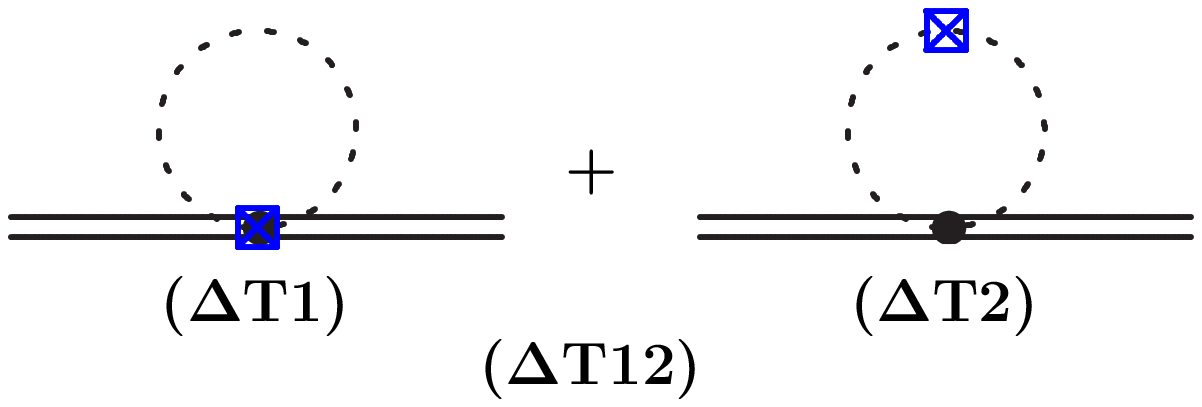}
\end{center}
\caption{\label{fig:diagramsDelta} 
Order-$p^3$ corrections to 
the $\Delta$-isobar form factors.  }
\end{figure}

\subsection{Details of the effective Lagrangian and loop results}
The effective Lagrangian is written in terms of pion, nucleon, $\Delta$-isobar and photon fields,
$\pi^{a}$, $N$, $\Delta_{\mu}$, $A_{\mu}$, and --- expanded to the appropriate power in 
the number of these fields, pion derivatives and mass -- reads as follows:
\begin{eqnarray}
\label{eq:totLagrangian}
\mathcal{L} & = & \mathcal{L}_{N}^{(1)}+\mathcal{L}_{\Delta}^{(1)}+\mathcal{L}_{\Delta N\pi}^{(1)}+\mathcal{L}_{\pi}^{(2)}\,, \nn\\
\mathcal{L}_{N\pi}^{(1)} & = & \overline{N}(i\s D-M_{N})N-\frac{g_{A}}{2f_{\pi}}\overline{N}\tau^{a}\left(\s D^{ab}\pi^{b}\right)\gamma_{5}N\,,\nn\\
\mathcal{L}_{\Delta\pi}^{(1)} & = & \overline{\Delta}_{\mu}(i\gamma^{\mu\nu\alpha}D_{\alpha}-M_{\Delta}\gamma^{\mu\nu})\Delta_{\nu}+\frac{H_{A}}{2f_{\pi}M_{\Delta}}\varepsilon^{\mu\nu\alpha\lambda}\overline{\Delta}_{\mu}\mathcal{T}^{a}(D_{\alpha}\Delta_{\nu})D_{\lambda}^{ab}\pi^{b}\,\\
\mathcal{L}_{\pi\Delta N}^{(1)} & = & i\frac{h_{A}}{2f_{\pi}M_{\Delta}}\overline{N}T^{a}\gamma^{\mu\nu\lambda}\left(D_{\mu}\Delta_{\nu}\right)\left(D_{\lambda}^{ab}\pi^{b}\right)+\mbox{h.c.}\,,\nn\\
\mathcal{L}_{\pi}^{(2)} & = &
\half (D_{\mu}^{ab}\pi^{b})(D_{ac}^{\mu}\pi^{c})-\half m_{\pi}^{2}\pi_{a}\pi^{a}\,\nn,
\end{eqnarray}
where the definitions of the iso-spin and Dirac matrices
are given in Appendix \ref{App:Pre} and 
the covariant derivatives  are:
\begin{eqnarray}
D_{\mu}^{ab}\pi^{b} & = & \delta^{ab}\partial_{\mu}\pi^{b}+ieQ_{\pi}^{ab}A_{\mu}\pi^{b}\,,\nn\\
D_{\mu}N & = & \partial_{\mu}N+ieQ_{N}A_{\mu}N+\frac{i}{4f_{\pi}^{2}}\epsilon^{abc}\tau^{a}\pi^{b}\left(\mbox{\ensuremath{\partial}}_{\mu}\pi^{c}\right)\,,\\
D_{\mu}\Delta_{\nu} & = &
\partial_{\mu}\Delta_{\nu}+ieQ_{\Delta}A_{\mu}\Delta_{\nu}+\frac{i}{2f_{\pi}^{2}}\epsilon^{abc}\mathcal{T}^{a}\pi^{b}\left(\mbox{\ensuremath{\partial}}_{\mu}\pi^{c}\right)\,\nn,
\end{eqnarray}
with $e>0$. 
Further details can be found in~Sect.\ 4 of Ref.~\cite{Pascalutsa:2006up}.

\begin{table}[H]
\begin{center}
\begin{tabular}{|c|c|c|c|c|c|c|}
\hline 
$g_{A}$ & $H_{A}$ & $h_{A}$ & $f_{\pi}$ {[}MeV{]} & $m_{\pi}$ {[}MeV{]} & $M_{N}$ {[}MeV{]} & $M_{\Delta}$ {[}MeV{]}\tabularnewline
\hline
\hline 
$1.27$ & $2.28$ & $2.85$ & $92.4$ & $139.6$ & $939$ & $1232$\tabularnewline
\hline
\end{tabular}
\end{center}
\caption{\label{tab:Parameters} List of parameters appearing in the loops
and their numerical values.}
\end{table}

The parameters of the Lagrangian are considered to be known and their physical
values are listed in  \Tabref{Parameters}.
The value of $H_{A}=(9/5)g_{A}\simeq2.28$ is taken from the large-$N_{c}$
limit and the value of $h_{A}=2.85$ is fixed by the experimental
$\Delta(1232)$-isobar decay width $\Gamma_\De=0.115$GeV, cf.~\cite{Pascalutsa:2005nd}.

The couplings to the $\Delta$-isobar are chosen to be
consistent with the covariant construct of the free Rarita-Schwinger theory 
and hence do not invoke the unphysical degrees of 
freedom of vector-spinor field \cite{Pascalutsa:1998pw,Pascalutsa:1999zz,Pascalutsa:2000kd}. However, the minimal coupling of the photon, here the $\ga\De\De$ coupling, is
 the well-known exception. We attempt to correct this problem by
adding non-minimal $\gamma\Delta\Delta$
coupling \cite{Deser:2000dz,Alexandrou:2009hs}:
\begin{equation}
\mathcal{L}_{\gamma\Delta\Delta}^{{\rm
    nm}}=\frac{e}{M_{\Delta}}\overline{\Delta}_{\mu}(i\kappa_{1}F^{\mu\nu}-\kappa_{2}\gamma_{5}\tilde{F}^{\mu\nu})\Delta_{\nu}\,\,\,,\label{Eq:LagNM}
\end{equation}
where $F^{\mu\nu}=\partial^{\mu}A^{\nu}-\partial^{\nu}A^{\mu}$ and
$\tilde{F}^{\mu\nu}=\varepsilon^{\mu\nu\rho\lambda}\partial_{\rho}A_{\lambda}$
are the electromagnetic field strength tensor and its dual with $\varepsilon_{0123}=+1$.
This non-minimal couplings, for $\kappa_{1}=\kappa_{2}=1$ is the one found in
$N=2$ supergravity (SUGRA), which is known to overcome the above-mentioned consistency
problem.

As the result, the $\gamma\Delta\Delta$ vertex becomes:
\begin{eqnarray}
\Gamma_{\gamma\Delta\Delta}^{\mu\nu\alpha}(p',p) & = & e Q_{\Delta}\Big[
-i\gamma^{\mu\nu\alpha}+i\frac{\kappa_{1}}{M_\De}\left(g^{\alpha\nu}q^{\mu}-g^{\alpha\mu}q^{\nu}\right)+\frac{\kappa_{2}}{M_\De}\varepsilon^{\mu\nu\rho\alpha}q_{\rho}\gamma_{5}\Big]\,,
\end{eqnarray}
with $q=p^{\prime}-p$, and denote   $\kappa_{nm}\equiv \kappa_{1}=\kappa_{2}$
in the resulting expressions of the Appendix. In this way by putting $\kappa_{nm}=0$
or 1 we recover either the result of the minimal coupling or of the `truncated SUGRA'.
We want to note, that only  the SUGRA choice ensures that the e.m.\ moments of the $\De$(1232)
take natural values at the tree level, see~\cite{Alexandrou:2009hs,Lorce:2009bs}
for more details. 

With the above and the notation in Appendix A, our results from the graphs in \Figref{diagramsNucleon} for 
the iso-vector (V) nucleon anomalous magnetic moment $\kappa_V$
and for the Dirac $\langle r_{1}^{2}\rangle_V$ and Pauli $\langle r_{2}^{2}\rangle_V$
radii are then:

\begin{eqnarray}
\kappa_V & = & \frac{e}{2M_{N}}\left[\overset{\circ}{\kappa}_V+F_{2}^V(0)\right]\,\,\,,\label{eq:kappa}\\
\langle r_{1}^{2}\rangle_V & = & \overset{\circ}{r}_V+6\left.\frac{d}{dq^{2}}\right|_{q^{2}=0}F_{1}^V(q^{2})\,\,\,,\label{eq:r1}\\
\langle r_{2}^{2}\rangle_V & = &
\frac{6}{\kappa_V}\left.\frac{d}{dq^{2}}\right|_{q^{2}=0} F_{2}^V(q^{2})\,\,\,,\label{eq:FINALr2}
\end{eqnarray}
with

\begin{eqnarray}
F_{i}^V(q^{2}) & = & \left[
  F_{i}^{N1}(q^{2})+F_{i}^{N2}(q^{2})+F_{i}^{N3}(q^{2})+F_{i}^{N4}(q^{2}) \right.\nn\\
 &  & \left.+F_{i}^{N56}(q^{2})+F_{i}^{N78}(q^{2})+F_{i}^{NT12}(q^{2})\right]^{p-n}\,\,\,.
\end{eqnarray}
We list all expressions for the $F_{i}^{Nj}(q^{2})$ in Appendix
\ref{APP:NucleonFF}, and  the $\overset{\circ}{\kappa}_V$ and the $\overset{\circ}{r}_V$ are the low energy constants (LECs) for the nucleon anomalous
magnetic moment and the Dirac radius \cite{Bernard(1998):HBChPT}. 
We fix these by constraining Eq. (\ref{eq:kappa}) and
Eq. (\ref{eq:r1}) to their phenomenological
values at the physical pion mass: $\kappa_V=3.7$ \cite{PDG(2008)}
and $\langle r_{1}^2\rangle_V=\left(0.765\mbox{ fm}\right)^{2}$
\cite{Mergell:1995bf,Belushkin(2007):NuclRadii}. A LEC for the Pauli radius enters at a
$p^4$ B$\chi$PT calculation. From an EFT viewpoint, the iso-vector and
iso-scalar nucleon combinations have a very different behavior. In the case of
the iso-scalar, unlike to the iso-vector part, sizeable two-loop corrections
are known to appear \cite{Bernard(1998):IsoScalar}. We will not discuss the
iso-scalar quantities in our one-loop calculation. 

Accordingly, our results from the graphs in 
\Figref{diagramsDelta} for the $\Delta^+(1232)$-isobar magnetic moment
$\mu_{\Delta}$, electric quadrupole moment $\mathcal{Q}_{\Delta}$, magnetic
octupole moment $\mathcal{O}_{\Delta}$ and the charge radius $\langle r_{E0}^{2}\rangle$  are:

\begin{eqnarray}
\mu_{\Delta} & = &
\frac{e}{2M_{N}}r\left[\overset{\circ}{\mu}_\Delta+F_{2}^\Delta(0)\right]\,\,\,\label{EFTMDMDelta},\\
\mathcal{Q}_{\Delta} & = & \frac{e}{M_{\Delta}^{2}}\left[\overset{\circ}{\mathcal{Q}}_\Delta-\frac{1}{2}F_{3}^\Delta(0)\right]\,\,\,,\label{eq:EQMwithDiagrams}\\
\mathcal{O}_{\Delta} & = & \frac{e}{2M_{\Delta}^{3}}\left[\overset{\circ}{\mathcal{O}}_\Delta+F_{2}^{\Delta}(0)-\frac{1}{2}\left(F_{3}^{\Delta}(0)+F_{4}^{\Delta}(0)\right)\right]\,\,\,\label{EFTMOMDelta},\\
\langle r_{E0}^{2}\rangle & = & \overset{\circ}{r_{E0}}+6\left[\left.\frac{d}{dq^{2}}\right|_{q^{2}=0}F_{1}^{\Delta}(q^{2})-\frac{1}{4M_{\Delta}^{2}}F_{2}^{\Delta}(0)-\frac{1}{12M_{\Delta}^{2}}F_{3}^{\Delta}(0)\right]\,\,\,\label{EFTCRDelta},
\end{eqnarray}
with

\begin{eqnarray}
F_{i}^{\Delta}(q^{2}) & = &
\left[ F_{i}^{\Delta 1}(q^{2})+F_{i}^{\Delta 2}(q^{2})+F_{i}^{\Delta
    3}(q^{2})+F_{i}^{\Delta 4}(q^{2})
\right.\nn\\
 &  &
+\left.F_{i}^{\Delta 56}(q^{2})+F_{i}^{\Delta 78}(q^{2})+F_{i}^{\Delta T12}(q^{2})\right]^{\Delta^+}\,\,\,.
\end{eqnarray}
All expressions for the $F_{i}^{\Delta j}(q^{2})$ are given in Appendix
\ref{APP:DeltaFF}. The quantities $\overset{\circ}{\mu}_\Delta$, $\overset{\circ}{\mathcal{Q}}_\Delta$
, $\overset{\circ}{\mathcal{O}}_\Delta$ and $\overset{\circ}{r_{E0}}$
are the LECs for the $\Delta^{+}(1232)$-isobar moments and its charge radius.

We estimate the error coming from terms higher order in
$m^2_\pi$ by adding $\pm n \cdot  m_\pi^2$ to our results where $n$ is taken to
be of {\it natural} size, i.e. $n=1$.

In the upcoming sections we will use the following parameters for better reading:

\begin{equation}
\mu=\frac{m_\pi}{M_{sc}} \;\;\;,\;\;\; R=\frac{M_\Delta}{M_N}\;\;\;, \;\;\;
r=\frac{M_N}{M_\Delta}\;\;\;, \;\;\;\tilde{q}^2=\frac{q^2}{M_N^2}\;\;\;,
\;\;\;\Delta=M_\Delta - M_N\;\;\;,
\end{equation}
\begin{equation}
\delta=\frac{\Delta}{M_\Delta}\;\;\;, \;\;\;C_{NN} = \frac{g_A M_{sc}}{8f_\pi\pi}\;\;\;, \;\;\; C_{N\Delta} = \frac{h_A M_{sc}}{8f_\pi\pi}\;\;\;,
\;\;\; C_{\Delta \Delta} = \frac{H_A M_{sc}}{8f_\pi\pi}\;\;\;,
\end{equation}
where $M_{sc}$ is the relevant mass scale for the observables in question,
i.e. $M_{sc}=M_N$ for the nucleon quantities and $M_{sc}=M_\Delta$ for the
$\Delta(1232)$ ones. We work in $d=4-2\varepsilon$ dimensions.

\subsection{Chiral structure and renormalization}

As discussed in detail in Subsection A, we employ the EOMS scheme
\cite{Fuchs:2003qc} to renormalize the loops in Figs.(\ref{fig:diagramsNucleon}, \ref{fig:diagramsDelta}).
We cancel the ultra-violet divergences so that the renormalized LECs are equal
to their "physical" values in the chiral limit. Within this renormalization prescription the divergences proportional to $L=-\frac{1}{\epsilon}+\gamma_{E}+\ln\frac{M_{sc}^{2}}{4\pi\Lambda^{2}}$
($\tilde{MS}$ scheme), as
well as the finite $m_{\pi}$ constant terms,  are absorbed into the corresponding LECs.

In App. \ref{APP:NucleonFF} and \ref{APP:DeltaFF} we give all nucleon and
$\Delta(1232)$-isobar quantities renormalized with $\tilde{MS}$, the renormalization of the
power-counting breaking terms is done in this section. We will see that all
renormalized LECs will not change much by including various contributions.

For the nucleon iso-vector quantities to the order $p^3$ there are LECs for
the Dirac radius and the anomalous magnetic moment  while one for the Pauli
radius enters at the order $p^4$, cf. \cite{Bernard(1998):HBChPT}. Schematically the chiral structures are:


\begin{eqnarray}
\langle r_1^2 \rangle_V &=&\overset{\circ}{ r_V } + c_{1} + \alpha_{1} \ln \mu
+  \beta_{1} \mu  + \mathcal{O} (\mu^2 ) \,\,\,,\\
\kappa_V &=&\overset{\circ}{\kappa_V} + c_\kappa +  \beta_{\kappa} \mu
 + \mathcal{O}( \mu^2 )\,\,\,,\\
\langle r_2^2 \rangle_V &=&  \gamma_2 \frac{1}{\mu} + c_{2} + \alpha_{2} \ln
\mu +  \beta_{2} \mu + \mathcal{O}(\mu^2) \,\,\,,
\end{eqnarray}
with $c_i$, $\alpha_i$, $\beta_i$ and $\gamma_i$ as some definite constants given in the
next section. In the chiral limit both radii diverge with $\ln \mu$ and $1/\mu$,
respectively. The constants $c_1$ and $c_\kappa$ have to be renormalized and
are listed in Appendix B. In Tab. \ref{tab:LECs} we see how the values of the LECs $\overset{\circ}{r_V}$ and $\overset{\circ}{\kappa_V}$ change by renormalizing the
above constants when taking into account: only virtual nucleons, virtual nucleons and $\Delta(1232)$ with minimal photon
coupling, and virtual nucleons and $\Delta(1232)$ with truncated SUGRA. The renormalized values of the LECs do not change much by including
the different contributions. Further, our results for the nucleon case are compatible to the $p^4$ calculation in \cite{Fuchs:2003ir}. 

In the case of the $\Delta(1232)$ electromagnetic quantities there are LECs
for all multipole moments and the charge radius:
$\overset{\circ}{\mu_\Delta}$, $\overset{\circ}{\mathcal{Q}_\Delta}$,
$\overset{\circ}{\mathcal{O}_\Delta}$ and $\overset{\circ}{r_{E0}}$. Schematically the chiral
structures are:
\begin{eqnarray}
\mu_{\Delta} &=& \overset{\circ}{\mu_\Delta} + c_\mu + \beta_\mu \mu  + \mathcal{O}(\mu^2)  \,\,\,,\\
\mathcal{Q}_{\Delta} &=& \overset{\circ}{\mathcal{Q}_\Delta} + c_\mathcal{Q} +
\alpha_\mathcal{Q} \ln\mu + \beta_\mathcal{Q} \mu + \mathcal{O}(\mu^2)\,\,\,, \\
\mathcal{O}_{\Delta} &=& \overset{\circ}{\mathcal{O}_\Delta} + c_\mathcal{O} + \beta_\mathcal{O} \mu + \mathcal{O}(\mu^2)  \,\,\,,\\
\langle r^2_{E0} \rangle &=& \overset{\circ}{r_{E0}} + c_{r} + \alpha_{r}
\ln \mu + \beta_{r} \mu  + \mathcal{O}(\mu^2)\,\,\,.
\end{eqnarray} 
In the chiral limit only the $\Delta(1232)$ EQM and its charge radius diverge with $\ln \mu$
whereas the MOM is finite. In the MOM the logarithm coming from the
$F_3^\Delta(0)$ is exactly canceled by the same term appearing in
$F_4^\Delta(0)$. 

Again, the $c_i$ have to be renormalized and are explicitly listed in
the Appendix C. We use the following values for the physical point to see the
changes of the LECs with respect to including the various
contributions: $\mu_{\Delta}=2.7\,\mu_N$,
$\mathcal{Q}_{\Delta}=-1.87\,\frac{e}{M_{\Delta}^{2}}$. The $\Delta^+(1232)$
magnetic dipole moment is taken from 
\cite{DeltaEXPERIMENT:MDM} and the value for the quadrupole moment is a large
$N_c$ estimate, see Section \ref{ResultsDelta}. For the octupole
moment and the charge radius no information is available and we use: $\mathcal{O}_{\Delta}=0$ and $\langle r^2_{E0}
\rangle = 0$. The numbers for these quantities correspond
to the bare change of the LECs.

In total we see that the renormalized LECs change within a reasonable degree.

\begin{table}
\caption{\label{tab:LECs} Values of the LECs by considering various
contributions to the observables. We use the following values for the nucleon iso-vector and
$\Delta^+(1232)$ quantities at the physical point: $\langle
r_1^2 \rangle_V=0.585$ fm$^2$, $\kappa_V=3.7$, $\mu_{\Delta}=2.7\,\mu_N$, 
$\mathcal{Q}_{\Delta}=-1.87\,\frac{e}{M_{\Delta}^{2}}$, $\mathcal{O}_{\Delta}=0$ and $\langle r^2_{E0}\rangle = 0$. In the second column the $N/\Delta$ means to
take only virtual nucleons ($\Delta(1232)$) contributions for the nucleon
($\Delta(1232)$) quantities, the third column to include both virtual baryons
with minimal $\gamma\Delta\Delta$ coupling and the fourth to take the
truncated SUGRA.}
\begin{tabular}{|c|c|c|c|}
\hline 
 & $N/\Delta$ & $N+\Delta$ minimal & $N+\Delta$ non-minimal\tabularnewline
\hline
\hline 
$\overset{\circ}{ r_{V}}/\mbox{fm}^{\,2}$ & $-0.69$ & $-0.76$ & $-0.74$\tabularnewline
\hline 
$\overset{\circ}{\kappa_{V}}$ & $5.03$ & $5.05$ & $5.13$\tabularnewline
\hline 
$\overset{\circ}{\mu_\Delta}/\mu_N$ & $2.78$ & $2.73$ & $2.87$\tabularnewline
\hline 
$\overset{\circ}{\mathcal{Q}_\Delta}/\frac{e}{M_{\Delta}^{2}}$ & $-3.50$ & $-3.49$ & $-3.72$\tabularnewline
\hline 
$\overset{\circ}{\mathcal{O}_\Delta}/\frac{e}{2M_{\Delta}^{3}}$ & $-0.40$ & $-0.33$ & $-0.12$\tabularnewline
\hline 
$\overset{\circ}{ r_{E0}}/\mbox{fm}^{\,2}$ & $-0.091$ & $-0.086$ & $-0.0831$\tabularnewline
\hline
\end{tabular}
\end{table}

\section{Recovering the HB$\chi$PT results}

The nucleon electromagnetic quantities were studied
within the $SU(2)$ heavy baryon $\chi$PT approach in
Ref. \cite{Bernard(1998):HBChPT} while the $\Delta(1232)$-isobar ones in the HB$\chi$PT SU(3)
calculation Ref. \cite{Jiang:2009jn}.  The HB$\chi$PT approach is an expansion
in powers of $1/M_N$ where only the leading term is kept. 
We compare our covariant B$\chi$PT results with these studies and see that in
this limit our formulas reduce to the HB$\chi$PT expressions.
In App. \ref{AppNucleon} we give our full results and discuss in this section
only the terms up to the second order in $m_\pi$.

\subsection{Nucleon electromagnetic form factors}

To compare our results to the HB$\chi$PT study \cite{Bernard(1998):HBChPT} , we expand the HB$\chi$PT expression in $m_\pi$ and
$\Delta$ and absorb all constant terms into the LECs:

\begin{eqnarray}
 \langle r^2_1\rangle^{(HB)}_V \frac{1}{6}&=& \overset{\circ}{r_V}\frac{1}{6} -\frac{ (1 + 5 g_A^2)}{48 f_\pi^2 \pi^2}\ln
\frac{m_\pi}{M_N}-\frac{m_\pi^2}{\Delta^2}\frac{5h_A^2}{864f_\pi^2\pi^2}(1+2\ln\frac{m_\pi}{2\Delta})\,\,\,,\label{dF1HB}\\
\kappa_V^{(HB)} &=& \overset{\circ}{\kappa}_V  -\frac{g_A^2 M_N m_\pi }{4 f_\pi^2 \pi}-\frac{h_A^2M_N}{72f_\pi^2 \pi^2}   \frac{m_\pi^2}{\Delta}  (1-2\ln \frac{m_\pi}{2\Delta})\,\,\,,\label{F2HB}\\
 \langle r^2_2\rangle^{(HB)}_V\frac{\kappa_V^{(HB)}}{6} &=& \frac{g_A^2 M_N}{48 f_\pi^2 \pi m_\pi}-\frac{h_A^2M_N}{216f_\pi^2\pi^2\Delta}
\ln\frac{m_\pi}{2\Delta} - \frac{h_A^2M_Nm_\pi^2}{864f_\pi^2\pi^2\Delta^3}
(1+2\ln\frac{m_\pi}{2\Delta})\,\,\,.\label{dF2HB}
\end{eqnarray}
We want to note that the $\Delta$ expansion is done only for the comparison purpose.
The corresponding parts of our work with only nucleon contributions are:

\begin{eqnarray}
\langle r^2_1\rangle^{(N)}_V \frac{1}{6} &=&    \frac{\overset{\circ}{r_V}}{6} - \left(\frac{1 + 5 g_A^2}{48 f_\pi^2 \pi^2} 
                                     - \frac{ 11 g_A^2 m_\pi^2}{24 f_\pi^2 \pi^2 M_N^2 }\right) \ln \frac{m_\pi}{M_N}
                             + \frac{35 g_A^2 m_\pi}{ 192 f_\pi^2 \pi M_N} 
                             - \frac{5 g_A^2 m_\pi^2}{ 192 f_\pi^2 \pi^2 M_N^2 }
                             \,,\\
\kappa^{(N)}_V &=&                  \overset{\circ}{\kappa_V} 
                             - \frac{g_A^2 M_N m_\pi}{ 4 f_\pi^2 \pi} 
                             - \frac{g_A^2 m_\pi^2}{8 f_\pi^2 \pi^2} 
                              - \frac{ 7 g_A^2 m_\pi^2}{8 f_\pi^2 \pi^2} \ln \frac{m_\pi}{M_N}  \,\,\,,\\
\langle r^2_2\rangle^{(N)}_V\frac{\kappa_V}{6} &=&    \frac{g_A^2M_N}{48 f_\pi^2 \pi m_\pi} 
                             +  \frac{29 g_A^2}{96 f_\pi^2 \pi^2}
                             -  \frac{ 35 g_A^2 m_\pi}{128 f_\pi^2 \pi M_N} 
                             +  \frac{23 g_A^2 m_\pi^2}{ 288 f_\pi^2 \pi^2 M_N^2} \nn \\&&
                             + \left( \frac{g_A^2}{4 f_\pi^2 \pi^2} -  \frac{5 g_A^2 m_\pi^2}{ 8 f_\pi^2 \pi^2 M_N^2}\right) \ln \frac{m_\pi}{M_N}\label{r2}\,\,\,.
\end{eqnarray}
The HB$\chi$PT results are reproduced and all additional terms are of
higher order in $1/M_N$, however, some of these are numerically as important as the HB$\chi$PT
expressions.

Expanding our expressions with $\Delta(1232)$ contributions and minimal
$\gamma\Delta\Delta$ coupling in $m_\pi$ gives: 

\begin{eqnarray}
\langle r^2_1 \rangle_V^{(\Delta)}\frac{1}{6} &=& \frac{m_\pi^2}{\Delta^2}
\frac{C_{N\Delta}^2}{162R^4M_N^2}\Big(-30 + 50 R - 170 R^2 + 243 R^3 - 
   496 R^4 + 991 R^5 \nn\\&& - 574 R^6 - 254 R^7 + 
   180 R^8 + (96 R^2 - 216 R^3) \ln \frac{m_\pi}{M_N}\nn\\&& + (-96 R^2 + 192 R^3 - 96 R^4 + 240 R^5 + 400 R^6 - 
      2236 R^7 + 1328 R^8\nn\\&& + 508 R^9 - 360 R^{10}) \ln R+ (40 - 25 R + 14 R^2 - 107 R^3 + 108 R^4\nn\\&& - 60 R^5 - 
      100 R^6 + 559 R^7 - 332 R^8 - 127 R^9 + 
      90 R^{10}) \ln \frac{\Delta^2(R+1)^2}{M_N^2}\Big),
\end{eqnarray}

\begin{eqnarray}
\kappa_V^{(\Delta)} &=&\frac{m_\pi^2}{\Delta M_N} \frac{C_{N\Delta}^2}{81R^4}\Big(-140 + 40 R - 328 R^2 + 4 R^3 + 12 R^4 + 500 R^5 + 56 R^6\nn\\&& - 
   216 R^7 + 16 R^5 (45 + 20 R - 76 R^2 - 7 R^3  + 27 R^4) \ln R\nn\\&& + 
   4 (5 + 5 R + 26 R^2 - 18 R^3 - 27 R^4 - 45 R^5\nn\\&&  - 20 R^6 + 76 R^7 + 
      7 R^8 - 27 R^9) \ln \frac{\Delta^2(R+1)^2}{M_N^2} + 144 R^2 \ln \frac{m_\pi}{M_N}\Big)\,\,\,,
\end{eqnarray}

\begin{eqnarray}
\langle r^2_2\rangle^{(\Delta)}_V\frac{\kappa^{(\Delta)}_V}{6} &=&   
\frac{M_N}{\Delta}\frac{C_{N\Delta}^2}{M_N^2486R^4}\Big(90 - 370 R + 451 R^2 + 935 R^3 - 1214 R^4 - 78 R^5 + 
   861 R^6\nn\\&& - 807 R^7 - 30 R^8 + 162 R^9 - 
   12 R^3 (20 + 10 R\nn\\&& + 210 R^2 - 300 R^3 + 52 R^4 + 146 R^5 - 
      148 R^6 - 5 R^7 + 27 R^8) \ln R \nn\\&&+ 
   3 R (50 - 71 R - 179 R^2 + 242 R^3 + 210 R^4 - 300 R^5 + 52 R^6\nn\\&&+ 
      146 R^7 - 148 R^8 - 5 R^9 + 27 R^{10})
      \ln\frac{\Delta^2(R+1)^2}{M_N^2}-144 \ln  \frac{m_\pi}{M_N} \Big)
\nn\\&& + \frac{m_\pi^2 M_N}{\Delta^3 (1 + R)}\frac{C_{N\Delta}^2}{M_N^2486R^4} \Big(40 R - 
    152 R^2 + 582 R^3 - 1912 R^4 + 2540 R^5\nn\\&& + 612 R^6 - 4614 R^7 + 
    2820 R^8 + 660 R^9 - 648 R^{10} + 
    24 R^3 (-10 - 10 R\nn\\&& + 50 R^2 + 90 R^3 - 337 R^4 + 62 R^5 + 
       412 R^6 - 262 R^7 - 55 R^8 + 54 R^9) \ln R \nn\\&&- 
    6 R (25 - 68 R + 32 R^2 - 8 R^3 + 43 R^4 + 90 R^5 - 337 R^6\nn\\&& + 
       62 R^7 + 412 R^8 - 262 R^9 - 55 R^{10} + 
       54 R^{11}) \ln \frac{\Delta^2(R+1)^2}{M_N^2}\nn\\&& + (-432 R^2 + 1008 R^3 - 1152 R^4 + 
       432 R^5) \ln \frac{m_\pi}{M_N}\Big)\,\,\,.
\end{eqnarray}
Expanding these expressions also in $\Delta$ yields the $\Delta(1232)$
contributions of Eqs.(\ref{dF1HB}-\ref{dF2HB}).
Compared to the minimal $\gamma\Delta\Delta$ coupling, the non-minimal contributions give terms that are of higher
$1/M_N$ order than those already present and do therefore not
appear in a HB$\chi$PT calculation.

In total, our results reduce in the limit $M_N\to\infty$ to the corresponding
HB$\chi$PT expressions of \cite{Bernard(1998):HBChPT}. We also see explicitly
that in the HB$\chi$PT numerical sizeable contributions, in that approach subleading in
$1/M_N$, are dropped.

\subsection{$\Delta(1232)$ electromagnetic form factors}

In the case of the $\Delta(1232)$ em quantities, there exists the HB$\chi$PT SU(3)
calculation Ref. \cite{Jiang:2009jn}. We will compare our covariant
formulae with this non-relativistic study. 

To do that we expand our results of App. \ref{APP:DeltaFF} to the
second order in $m_\pi$ below the $\Delta(1232) \to \pi N$ threshold:

\begin{eqnarray}
\mu_{\Delta} \frac{1}{r}&=&\overset{\circ}{\mu_{\Delta}}+ \frac{C^2_{\Delta \Delta}}{972} \Big( - 288
\mu \pi  +  \mu^2 (-5440 + 7514 \kappa_{nm} + (-2976 + 6864 \kappa_{nm}) \ln \mu) \Big)\nn\\&& +
\frac{C^2_{N\Delta}}{36(r-1)}\Big( i \pi (6  + 12  r - 18  r^2 - 48  r^3 + 24  r^4 + 
    78  r^5 - 18  r^6 - 60  r^7 + 6  r^8 + 18  r^9)\nn\\&&  + 
 \mu^2 (6 - 18 r + 72 r^3 - 
    72 r^5 + i \pi (-12  - 48  r^3 + 108  r^5 - 72  r^7) + 
    24 \ln \mu\nn\\&& + (96 r^3 - 216 r^5 + 144 r^7) \ln r + (-6 - 24 r^3 + 54 r^5 -
    36 r^7) \ln (r^2-1)^2) \Big)\,\,\,,
\end{eqnarray}

\begin{eqnarray}
\mathcal{Q}_{\Delta}&=& \overset{\circ}{\mathcal{Q}_{\Delta}} -
\frac{C^2_{\Delta \Delta}}{486} \Big(- 48 \mu \pi + 192 \ln \mu + 
 \mu^2 (-684 - 1794 \kappa_{nm}  + (3240 + 1872 \kappa_{nm}) \ln \mu ) \Big)\nn\\&&
-\frac{C^2_{N\Delta}}{18(r-1)^2}\Big( i\pi (-6   + 6   r + 14   r^2 - 18   r^3 - 8   r^4 + 
      24   r^5 - 8   r^6 - 18   r^7 + 14   r^8\nn\\&& + 6   r^9 - 
      6   r^{10})  + 
  \mu^2 (6 - 7 r - 4 r^2 + 25 r^3 - 12 r^4 - 30 r^5 + 
      24 r^6\nn\\&& + i\pi (4   - 6   r + 8   r^2 - 20   r^3 + 40   r^5 - 
         24   r^6 - 30   r^7 + 24   r^8) + (-8 + 12 r) \ln \mu\nn\\&& + (-16 r^2 + 40 r^3 - 80 r^5 + 48 r^6 + 60 r^7 - 
         48 r^8) \ln r + (2 - 3 r + 4 r^2 - 10 r^3 + 20 r^5\nn\\&& - 12 r^6 - 15 r^7 + 
         12 r^8) \ln(r^2-1)^2)\Big)\,\,\,,
\end{eqnarray}

\begin{eqnarray}
\mathcal{O}_{\Delta}&=& \overset{\circ}{\mathcal{O}_{\Delta}} +
\frac{C^2_{\Delta \Delta}}{2916} \Big( - 288 \mu \pi + 
 \mu^2 (10168 - 2470 \kappa_{nm} + (-27024 + 14352 \kappa_{nm}) \ln \mu)\Big)
\nn\\&&+\frac{C^2_{N\Delta}}{36(r^2-1)} \Big( i \pi (-6   - 10   r + 6   r^2 + 18   r^3 + 8   r^4 + 
    6   r^5 - 4   r^6 - 26   r^7 - 10   r^8 + 12   r^9 + 
    6   r^{10})\nn\\&&  + 
 \mu^2 (-2 + 26 r + 30 r^2 + 10 r^3 - 36 r^5 - 
    24 r^6 + i \pi (4   - 16   r - 16   r^2 + 24   r^3 + 32   r^4\nn\\&& + 
       28   r^5 + 12   r^6 - 36   r^7 - 24   r^8) - 
    8 \ln \mu + (32 r + 32 r^2 - 48 r^3 - 64 r^4 - 56 r^5 - 24 r^6 + 
       72 r^7\nn\\&& + 48 r^8) \ln r + (2 - 8 r - 8 r^2 + 12 r^3 + 16 r^4\nn\\&& + 14 r^5 + 6 r^6 - 
       18 r^7 - 12 r^8) \ln(r^2-1)^2)\Big)\,\,\,,
\end{eqnarray}

\begin{eqnarray}
\langle r^2_{E0}\rangle &=& \overset{\circ}{ r_{E0}} 
+6\frac{C^2_{\Delta \Delta}}{81M_\Delta^2} (-100 - \frac{144}{H_A^2}) \ln \mu
+6\frac{C^2_{\Delta \Delta}}{1944M_D^2}  \Big(2760 \mu \pi \nn\\&& +\mu^2 (16476 + 13680 \ln \mu) + \kappa_{nm} 
    \mu^2 (13026 + 9360 \ln \mu)\Big)\nn\\&&
+6\frac{C^2_{N\Delta}}{72M_\Delta^2(r-1)^2}  \Big(  i \pi (-48   + 54   r + 116   r^2 - 162   r^3 - 80   r^4 + 
    222   r^5 - 56   r^6 - 174   r^7\nn\\&& + 128   r^8 + 60   r^9 - 
    60   r^{10})  + 
 \mu^2 (78 - 106 r - 22 r^2 + 214 r^3 - 84 r^4 - 300 r^5 + 
    240 r^6 \nn\\&&+ i\pi (4   - 24   r + 80   r^2 - 200   r^3 + 
       364   r^5 - 204   r^6 - 300   r^7 + 240   r^8) + (-8 + 
       48 r) \ln \mu\nn\\&& + (-160 r^2 + 400 r^3 - 728 r^5 + 408 r^6 + 600 r^7 - 
       480 r^8) \ln r + (2 - 12 r + 40 r^2\nn\\&& - 100 r^3 + 182 r^5 - 102 r^6 - 
       150 r^7 + 120 r^8) \ln(r^2-1)^2)\Big)\,\,\,,
\end{eqnarray}
where the factor $r$ in $\mu_{\Delta}$ comes from defining the quantity in
$\mu_N$ and we included the factor $1/2$ for the $\mathcal{Q}_{\Delta}$ and the
$6$ for the $\langle r^2_{E0} \rangle$. We will now compare
certain ratios of coefficients within our formulas against the same ratios
extracted from \cite{Jiang:2009jn}. 

Starting with $\langle r^2_{E0}\rangle$, the ratio between the
$\Delta T1$ and $\Delta 1$ contribution of \Figref{diagramsDelta} in \cite{Jiang:2009jn} is $25 g_{\Delta\Delta}^2/81$ where it is in our
work for the $\Delta^{+}(1232)$: $100H_A^2/144$. Together with the ratio of the $\Delta(1232)$ kinetic to interacting
term of $2/3$ in our Lagrangian compared to the one used in
\cite{Jiang:2009jn}, we obtain the same $\Delta T1/\Delta 1$ ratio.
Further, our ratio of the
$\mu$ term in $\mu_{\Delta}$ to the $\ln \mu$ term in $\langle
r^2_{E0}\rangle $ is $\frac{288}{972}/\frac{100}{81}$
which equals that of \cite{Jiang:2009jn}, i.e. 
$\frac{8}{27}/\frac{100}{81}$ where the factor $6$ of the radius
definition is not included in \cite{Jiang:2009jn}.
Comparing in this manner the formulas, we obtain an agreement for all
coefficients of the various $\mu$ and $\ln \mu$ terms in the various moments
and charge radius.

In total, we conclude that our formulas reduce in the limit of $M_N\to \infty$ to the
HB$\chi$PT non-relativistic ones. Comparing numerically several terms of our formulas
against the leading $1/M_N$ parts show that
sometimes sizeable contributions are dropped in HB$\chi$PT.

\section{Covariant baryon $\chi$PT results\label{sec:Results}}
In this section we present our main results. We study the nucleon
form factors at the physical point for small momentum transfer with respect to
the extraction of the proton electric radius from experimental data. Further,
we also study the chiral behavior of the nucleon and $\Delta(1232)$ form factors for
$Q^2=0$ with $m_\pi^2< 0.3$ GeV$^2$ and compare them to available lattice QCD results.
%
%
%
%

\subsection{Nucleon electromagnetic form factors\label{sec:Results:Nucleon}}

As discussed in the introduction the inclusion of pion-production effects in
the interpolation of low-$Q^2$ data for the proton electric form factor can be
addressed within the framework of the baryon $\chi$PT. It is the second
derivative of $G^p_{E}(q^2)$ which is a genuine prediction of this theory and
constraints the analytic structure of the form factor at small
space-like momentum transfers, $-q^2 \leq 0.01$ GeV$^2$. 
Including such constraints in the proton charge radius extraction from
electron-scattering data could have a quantitative impact.

The proton electric form factor expanded to second order in $q^2$ is:
\begin{eqnarray}
G^p_E(q^2) &=& F_1^p(q^2) + \frac{q^2}{4M_N^2} F_2^p(q^2)= 1 + q^2\frac{\langle r^2_E
  \rangle_p}{6} +\frac{1}{2} q^4 \frac{d^2}{[dq^2]^2}G^p_E(0) + \mathcal{O}(q^6)\,\,\,, \label{ddGE}
\end{eqnarray}
where we obtain for $\frac{d^2}{[dq^2]^2}G^p_E(0)$ from
App. (\ref{AppNucleon}) the following result at the physical point:

\begin{eqnarray}
\frac{d^2}{[dq^2]^2}G_E(0) \cdot \textrm{GeV}^{\, 4}  &=& 2.089 I_{N1} - 0.001 I_{N2} + 0.640I_{NT1} \nn \\
&&   + 0.580 I_{N3} - 0.021  I_{N4} -\kappa_{nm}1.797 I_{N4}\,\,\,,
\end{eqnarray}
with $I_i$ as the iso-spin factors. Since the important contributions, apart
from the truncated SUGRA, come from the diagrams N1, NT1 and N3, we can make a direct comparison to the HB$\chi$PT results of Ref. \cite{Bernard(1998):HBChPT}:

\begin{eqnarray}
\frac{d^2}{[dq^2]^2}G_E^{p(HB)}(0) &=&  \frac{1 + 7 g_A^2}{960 f_\pi^2  m_\pi^2 \pi^2} 
            \nn \\&&- g_{PND}^2\frac{7\left(m_\pi^2 - \Delta^2 + \Delta \sqrt{-m_\pi^2 + \Delta^2} 
            \ln \frac{\Delta + \sqrt{-m_\pi^2 + \Delta^2}}{m_\pi}\right) }{1080 f_\pi^2 \pi^2 (m_\pi^2 - \Delta^2)^2}
              \,\,\,.
\end{eqnarray}
The first term comes from the diagrams N1, NT1 and is also the $1/M_N$ leading term in our results.
Expanding our results in $\Delta$ coincides with the leading $1/M_N$ part of
the same expansion in the HB$\chi$PT. 

In Tab. \ref{tab:ddGE} we compare our covariant numbers with those of the
HB$\chi$PT. These numbers are to be entered for $\frac{d^2}{[dq^2]^2}G^p_E(0)$ in Eq. (\ref{ddGE}). In the case of only virtual nucleons, second column, the difference of the
results are the additional terms of higher $1/M_N$ orders in the diagram N1. 
In HB$\chi$PT the second derivative is dominated by $\frac{d^2}{[dq^2]^2}F_1^p(0)$ where the
contributions coming from $\frac{d}{dq^2}F_2^p(0)$ are of subleading order. However, these terms are in case of the $\Delta(1232)$ contributions, third
column, the main cause of the difference. The non-minimal $\gamma\Delta\Delta$
coupling contributions are not present in the HB$\chi$PT. The large number
comes from the $m_\pi$-constant terms in $\frac{d^2}{[dq^2]^2}F_1^p(0)$ and
$\frac{d}{dq^2}F_2^p(0)$. These terms are of the same size as the
corresponding constant appearing in the Pauli radius by considering only
nucleons, Eq. (\ref{r2}), and are also discussed with respect to the chiral
behavior later. Using the numbers of Tab. \ref{tab:ddGE} to constrain
the extrapolation of
experimental data in the region of $Q^2 \leq 0.01$ GeV$^2$ could have a
quantitative impact on the extracted number for $\langle r^2_E \rangle_p$.

\begin{table}
\caption{\label{tab:ddGE}Contributions to the second derivative of the proton
  electric charge radius, $\frac{d^2}{[dq^2]^2}G^p_E(0)$, from the covariant baryon $\chi$PT in units of $\mbox{GeV}^{\,-4}$. The
  columns correspond to: second one to taking only virtual nucleons, third one
  to taking only virtual $\Delta(1232)$ with minimal $\gamma\Delta\Delta$
  coupling, fourth to the truncated SUGRA and last one to the sum of all contributions. }
\begin{tabular}{|c|c|c|c|c|}
\hline 
Diagrams & N1+N2+NT1 & N3+N4 & N4 nm & Sum\tabularnewline
\hline
\hline 
Covariant B$\chi$PT & $4.80$ & $-0.45$ & $4.79$ & $9.14$\tabularnewline
\hline 
HB$\chi$PT & $7.83$ & $-1.30$ & $--$ & $6.53$\tabularnewline
\hline
\end{tabular}
\end{table}

\begin{figure}
\caption{\label{fig:IV-nucl}Nucleon iso-vector anomalous magnetic moment
  $\kappa_V$ and the Dirac $\langle r^2_1 \rangle_V$ and Pauli $\langle r^2_2
  \rangle_V$ radii. The results correspond to: solid (red) curve to nucleon
and $\Delta(1232)$ contributions with truncated SUGRA; long-dashed (blue)
curve to our result with virtual nucleons and $\Delta(1232)$ with minimal coupling;
short-dashed (green) curve to only virtual nucleon. The
lattice results are taken from: blue down-triangles
\cite{Syritsyn:2009mx}, black right-triangles \cite{Alexandrou:2011db}, brown
up-triangles \cite{Collins:2011}, green left-triangles \cite{Bratt(2010):LHPCLatticeNucleon}. The blue circles denote the
phenomenolgical values.}

\begin{center}
\includegraphics[scale=0.4]{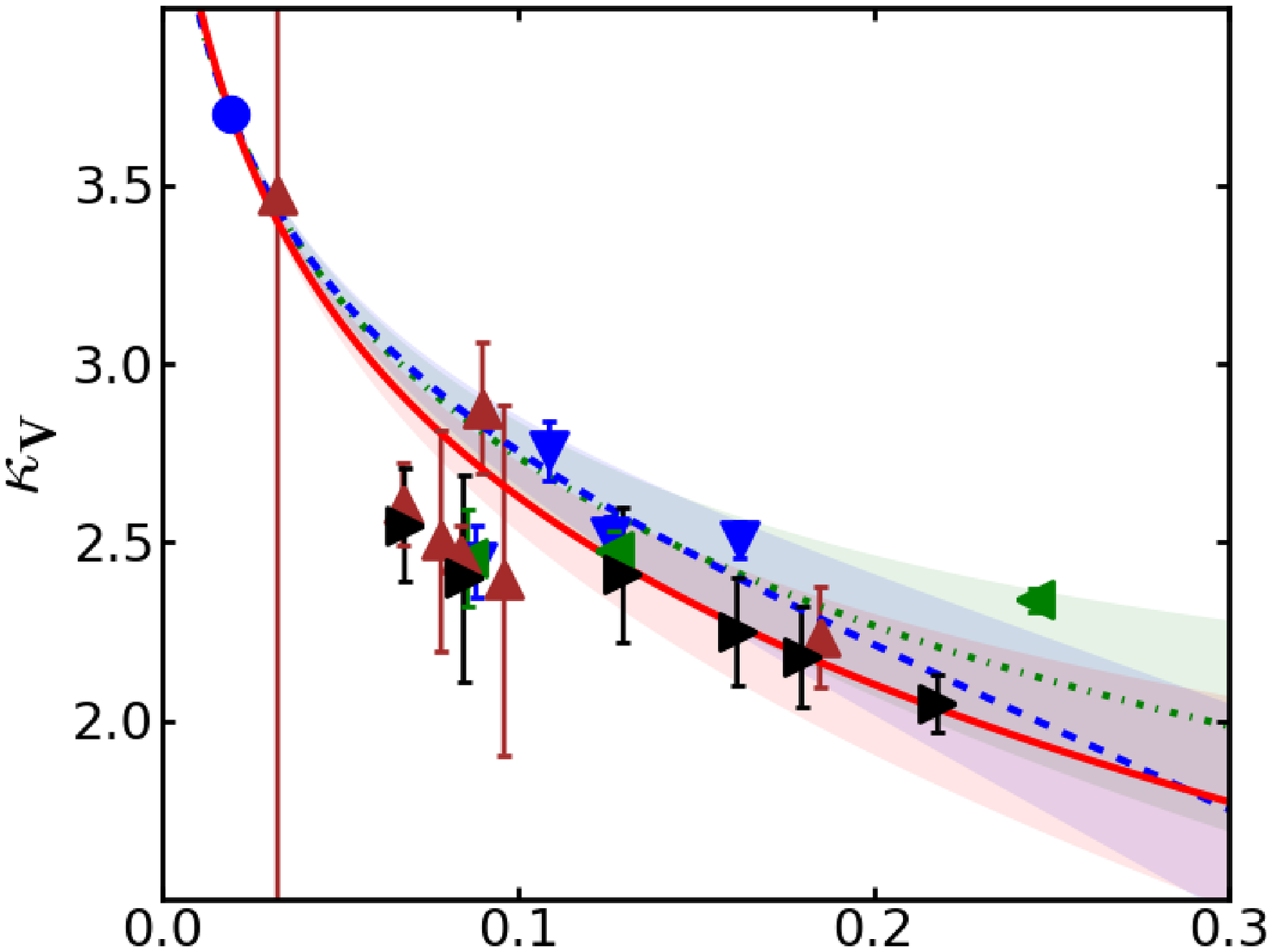}\includegraphics[scale=0.4]{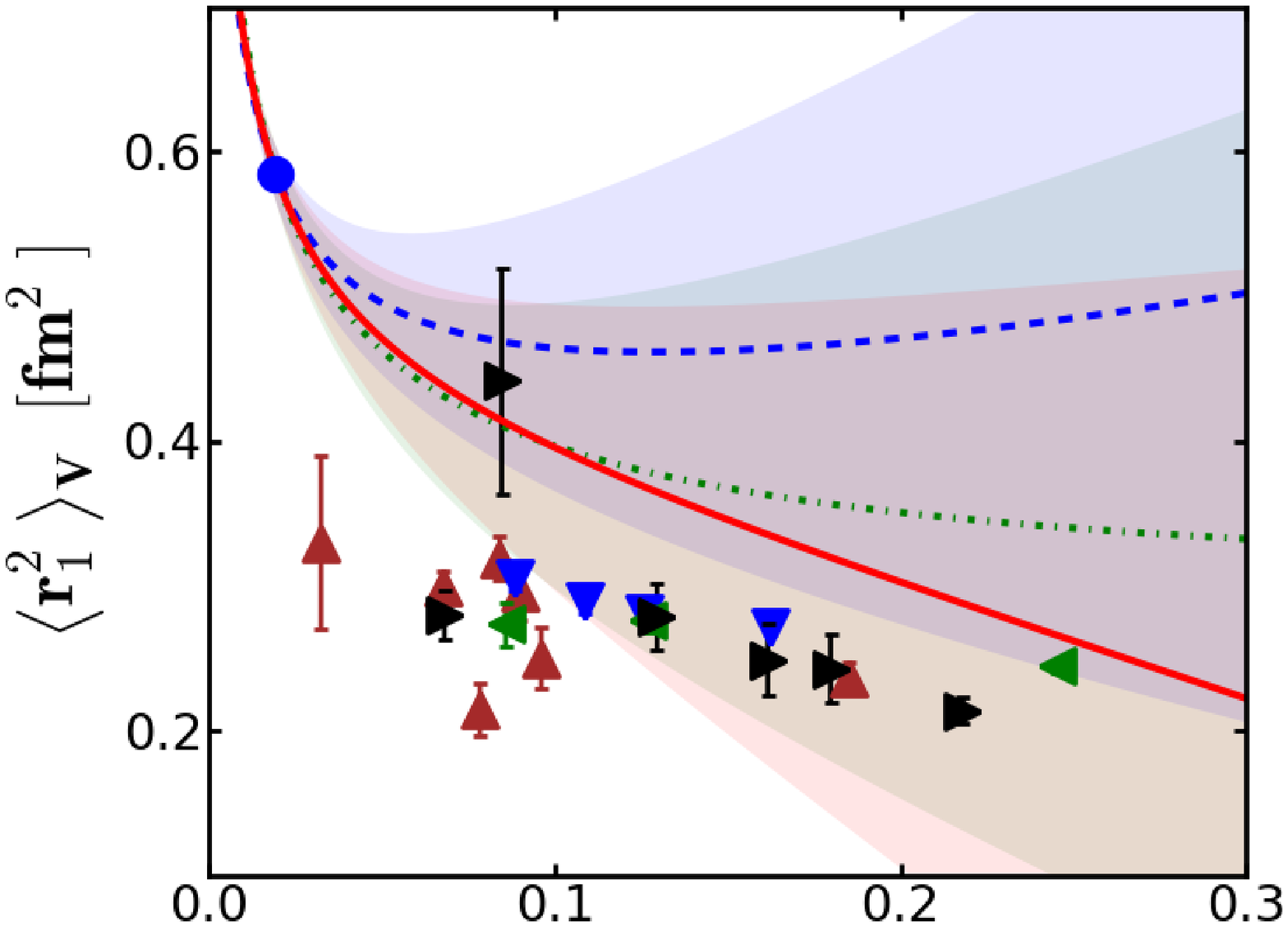}\\
\includegraphics[scale=0.4]{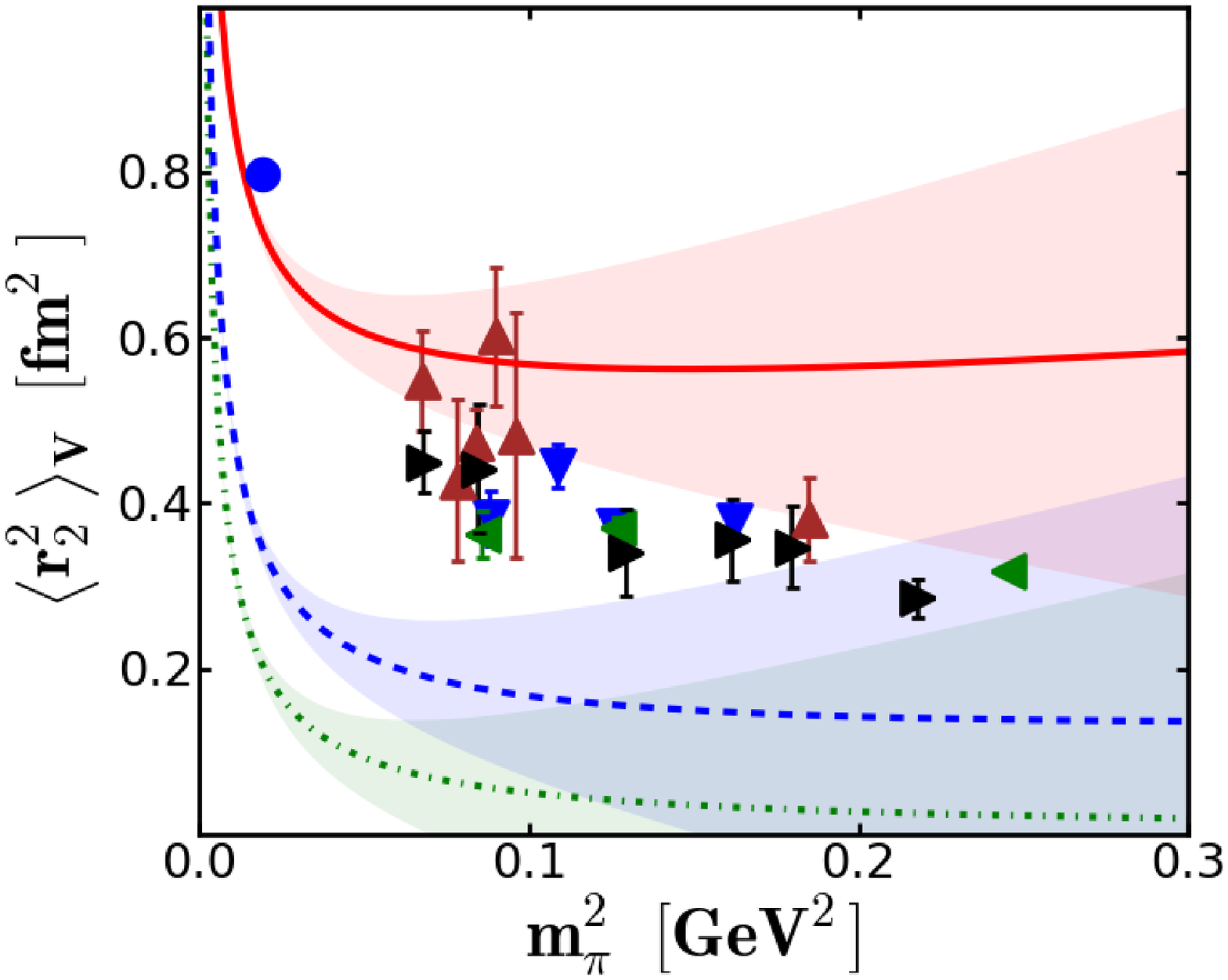}\includegraphics[scale=0.4]{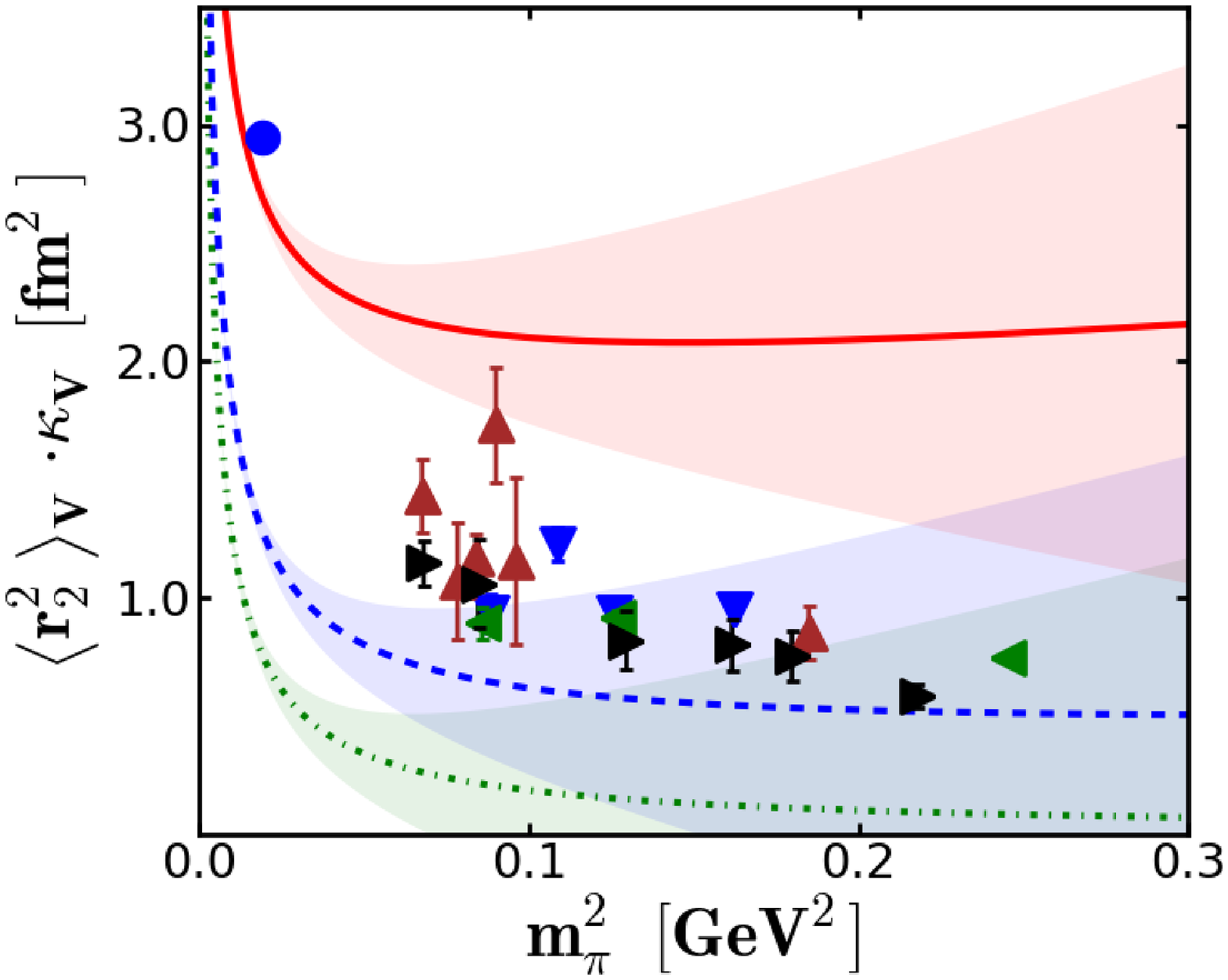}
\end{center}
\end{figure}

Another application of our B$\chi$PT form factor results is to study their
chiral behavior in comparison to lattice QCD calculations.
In Fig. \ref{fig:IV-nucl} we show the nucleon iso-vector
quantities $\kappa_{V}$, $\langle r^2_{1}\rangle_{V}$ and $\langle r^2_{2}\rangle_{V}$.
The red solid curve corresponds to taking all contributions, truncated SUGRA, while the blue long-dashed
curve to taking virtual baryons with
strictly minimal photon couplings. The green short-dashed curve corresponds
to the calculation with only virutal nucleons. The lQCD results are those of
the LHPC collaboration \cite{Bratt(2010):LHPCLatticeNucleon,Syritsyn:2009mx},
of the work \cite{Alexandrou:2011db} and of 
the QCDSF/UKQCD collaboration \cite{Collins:2011}. In our $p^3$ B$\chi$PT calculation appear
LECs for the quantities $\kappa_{V}$ and $\langle
r_{1}^{2}\rangle_{V}$ and we constrain our
results to the experimental values: $\kappa_V=3.7$ \cite{PDG(2008)} and $\langle
r_1^2 \rangle_V=0.585$ fm$^2$ \cite{Mergell:1995bf,Belushkin(2007):NuclRadii}.

For the iso-vector anomalous magnetic moment $\kappa_V$ and the Dirac radius
$\langle r_1^2\rangle_V$, we see that the lQCD and
our B$\chi$PT results agree within the $\chi$PT error for pion masses above
$m_{\pi}^{2}=0.1\,\mbox{GeV}^{2}$. However, the data points for the smaller
pion masses are not reached. A similar behavior is found in
\cite{Syritsyn:2009mx,Collins:2011,Alexandrou:2011db}.
In these works the LHPC results were tried to fit by a HB$\chi$PT small scale expansion
calculation with inclusion of explicit $\Delta(1232)$-isobar and
a covariant NNLO B$\chi$PT without explicit $\Delta(1232)$-isobar.
Conclusions in \cite{Syritsyn:2009mx,Collins:2011} are that the lattice LHPC 
data could not be fitted simultaneously.

In the case of the Pauli radius $\langle r_{2}^2\rangle_{V}$ to the order $p^3$ there appears no
LEC, hence the results are predictions. A LEC enters at the order $p^4$.
We see that our truncated SUGRA results nearly hits the experimental value. The reason
is the $m_\pi$-constant contribution from the non-minimal $\gamma\Delta\Delta$
coupling. This term is of the same
size as the constant
$\frac{29 g_A^2}{96 f_\pi^2 \pi^2}$ coming from the usual virtual nucleon
contribution, Eq. (\ref{r2}). However there, the two negative $m_\pi$ terms 
are the cause of the small B$\chi$PT result with only nucleons.  
The whole $m_\pi$ dependance of $\langle r_{2}\rangle_{V}^{2}$ coming from all
considered diagrams is dominated by the nucleon diagrams $N1$ and $N2$. The
$\Delta(1232)$ contributions add merely
a small constant term in the case of minimal photon coupling and a large term
for the non-minimal coupling. In a $p^4$ calculation these parts would be
renormalized.

For the Pauli-radius, our B$\chi$PT study and the lQCD results have the fact in common that the $m_{\pi}^{2}$
dependance is nearly linear for pion masses $m_{\pi}^{2}\gtrsim0.1$
$\mbox{GeV}^{2}$. However, the absolute values of $\langle r_{2}^{2}\rangle_{V}$
are rather different indicating that some unknown components in the
B$\chi$PT or lQCD calculations are missing.

In this and other works finite volume lQCD results are compared to infinite volume B$\chi$PT
ones. Finite volume effects on the B$\chi$PT side are known to be
missing and could be one part of the explanation for the above discrepancies,
especially for the small pion mass region.
Further studies are presently done in that direction.
%
%
%
%

\subsection{$\Delta(1232)$ electromagnetic form factors \label{ResultsDelta}}

\begin{figure}[H]
\caption{\label{fig:DELTAresutls}The magnetic dipole MDM, electric quadrupole
EQM, magnetic octupole MOM moments of the $\Delta^{+}(1232)$ and
its charge radius CR. Both the results for the real part are shown with (red solid curves) and without (blue dashed curves) inclusion
of the $\gamma\Delta\Delta$ non-minimal coupling, respectively. The
green short-dashed curves depict the imaginary parts of these quantities.
The blue circle correspond to the experimental value $\mu_{\Delta^{+}}=\left(2.7\pm1.5\right)\,\mu_{N}$,
a large-$N_{c}$ estimation $\mathcal{Q}_{\Delta^{+}}=-1.87\,\frac{e}{M_{\Delta}^{2}}$
and $\mathcal{O}_{\Delta^{+}}=0$. The lQCD data of \cite{Aubin:2008qp}
are denoted by green triangles while those of \cite{Alexandrou:2008bn}
are depicted by orange rectangles. The grey bands are described in
the text.}
\begin{center}
\includegraphics[scale=0.4]{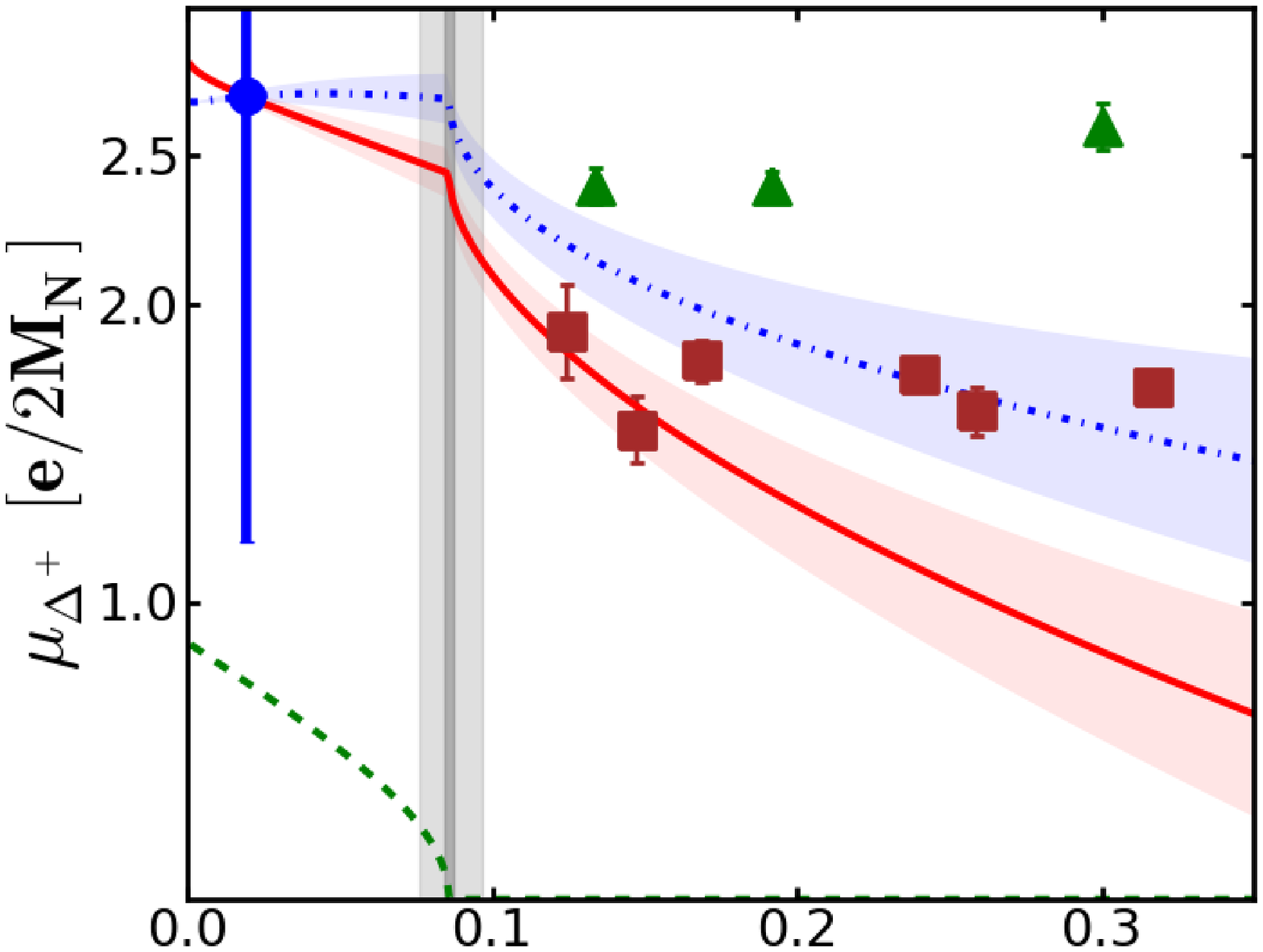}\includegraphics[scale=0.4]{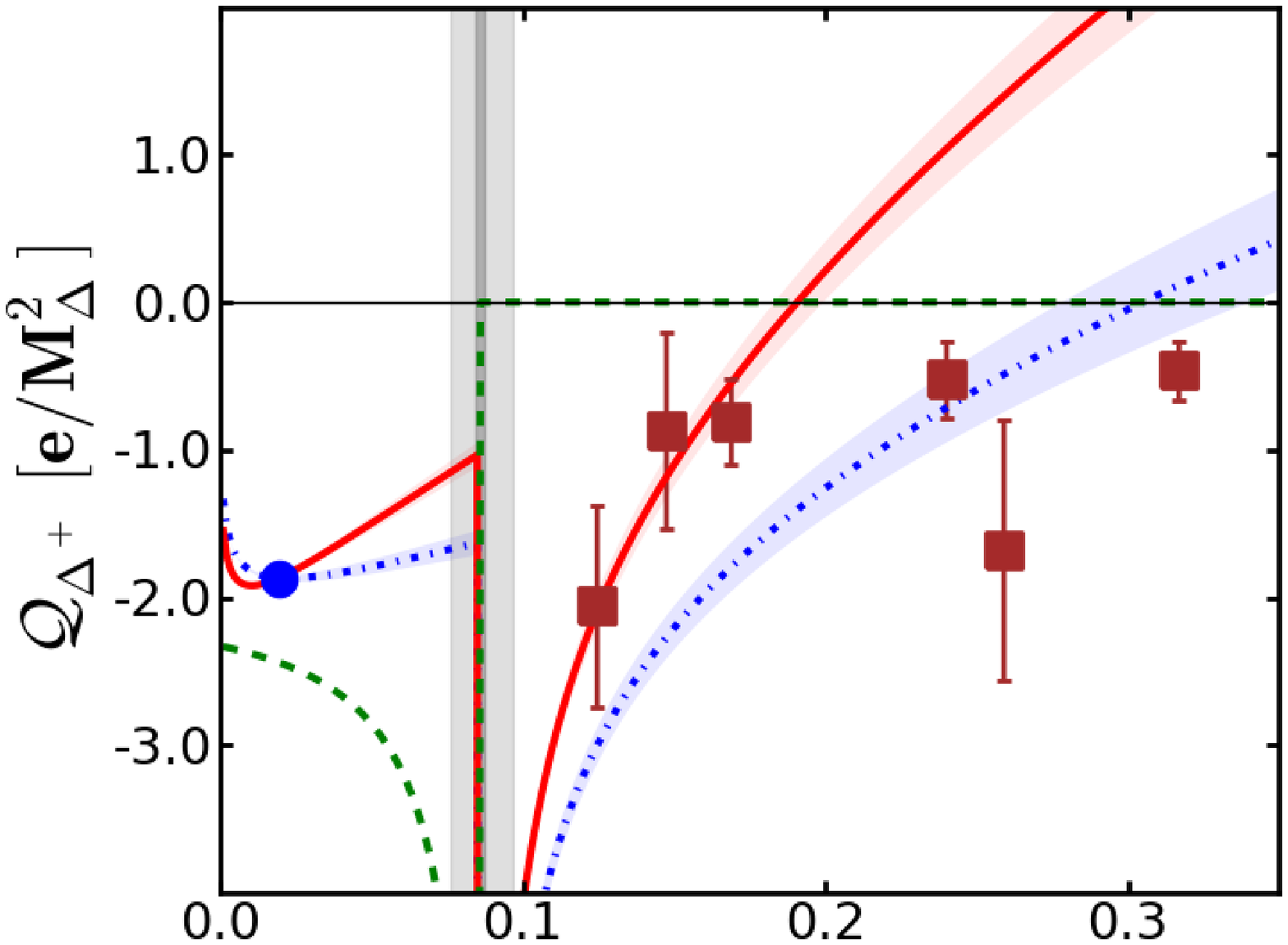}\\
\includegraphics[scale=0.4]{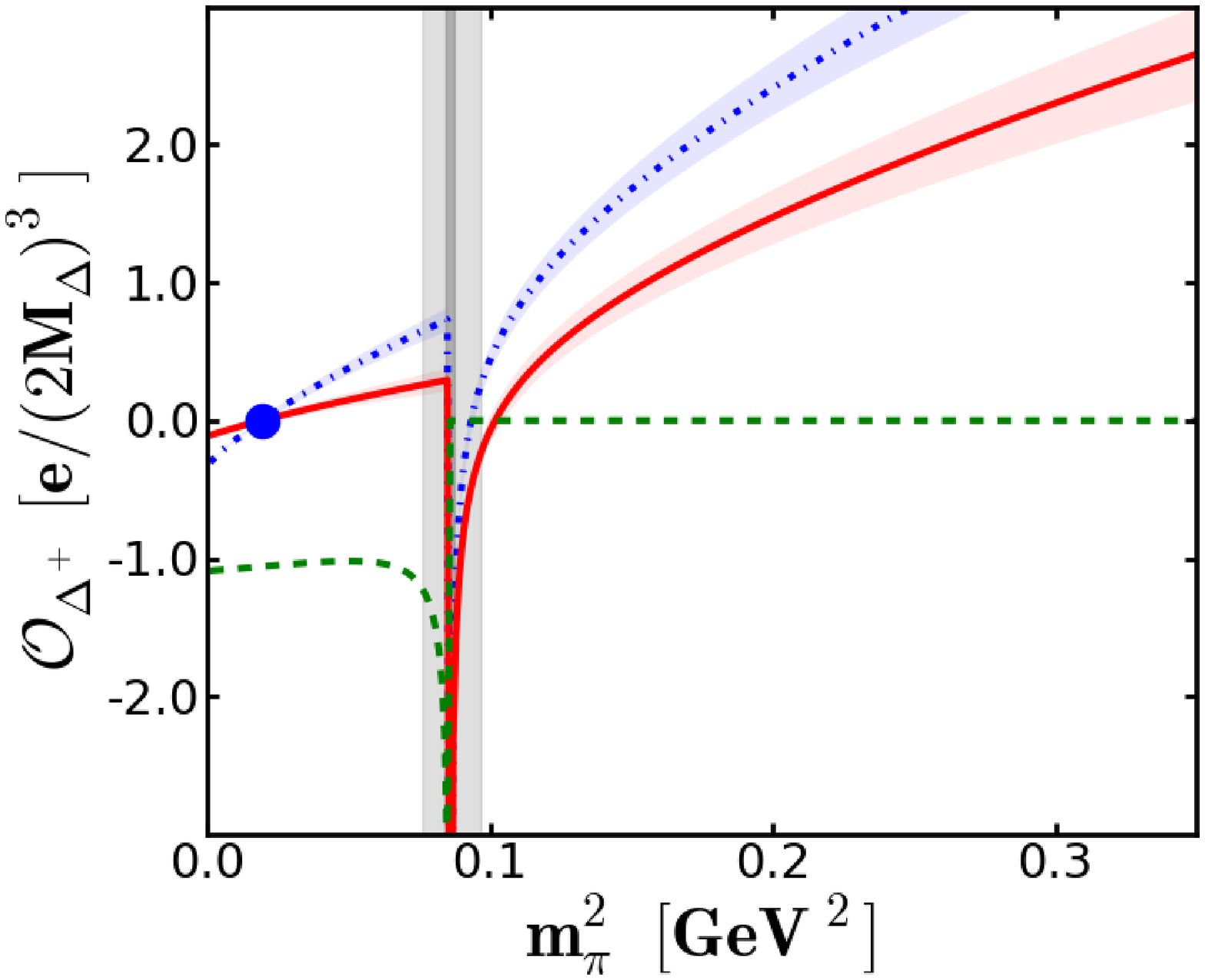}\includegraphics[scale=0.4]{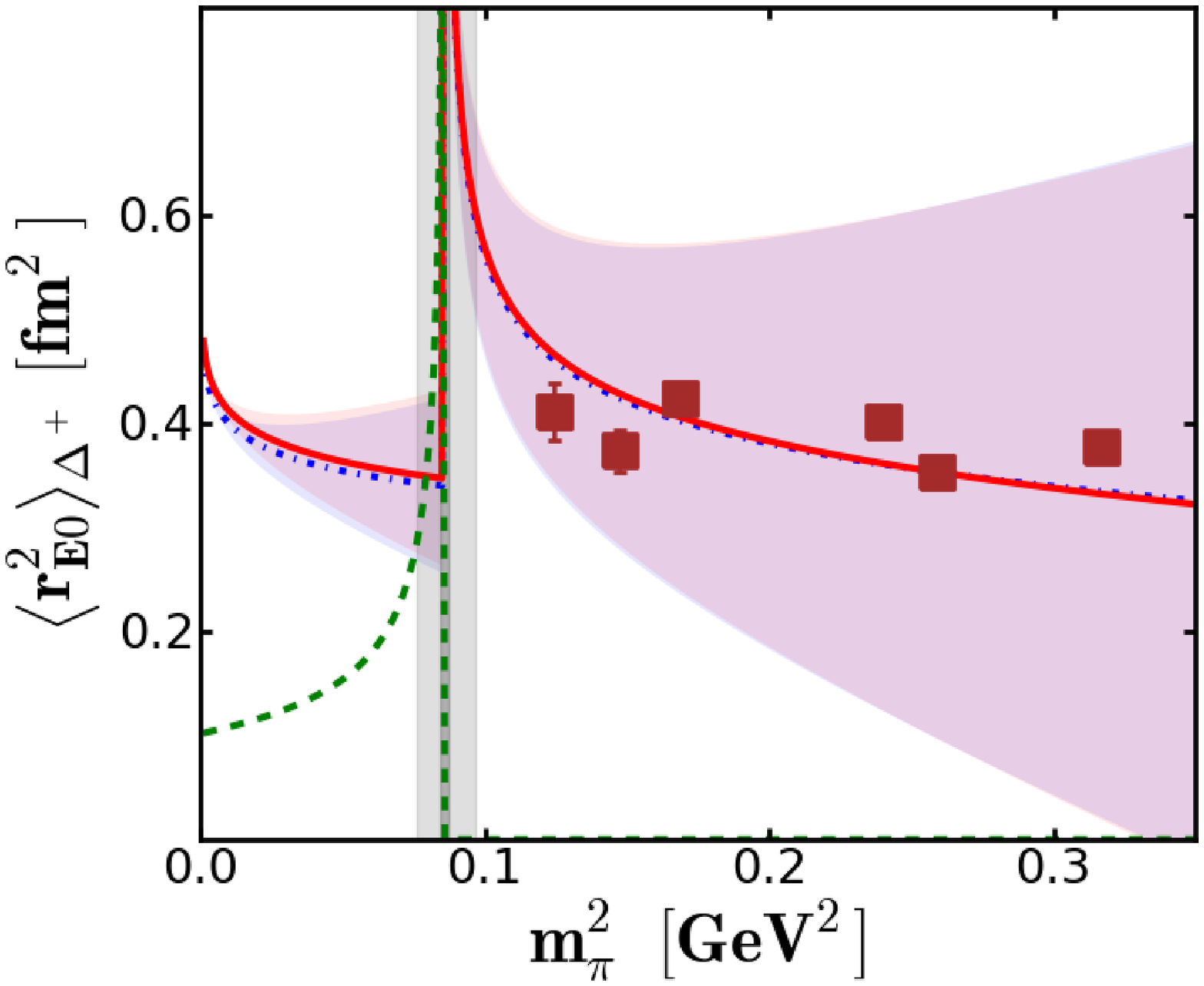}\end{center}
\end{figure}

We now proceed to the $\Delta^{+}(1232)$-isobar magnetic dipole (MDM),
$\mu_{\Delta^{+}}$, electric quadrupole (EQM), $\mathcal{Q}_{\Delta^{+}}$,
magnetic octupole (MOM), $\mathcal{O}_{\Delta^{+}}$, moments and
its charge radius (CR), $\langle r_{E0}^{2}\rangle$. The experimental
knowledge of the $\Delta^{+}(1232)$-isobar is rather scarce. For
the $\Delta^{+}(1232)$-isobar MDM a value is obtained from the radiative
pion photoproduction $\gamma N\to\pi N\gamma^{\prime}$ \cite{DeltaEXPERIMENT:MDM}:

\begin{equation}
\mu_{\Delta^{+}}=2.7_{-1.3}^{+1.0}(\mbox{stat.})\pm1.5\left(\mbox{syst.}\right)\pm3.9\left(\mbox{theor.}\right)\,\,\,.
\end{equation}
For the $\Delta^{+}(1232)$ EQM we use the following model-independent
estimation based on the large-$N_{c}$ limit. In Ref. \cite{Buchmann(2002):LargeNCDeltaEQM}
the large-$N_{c}$ relation $\mathcal{Q}_{\Delta^{+}}=\frac{2\sqrt{2}}{5}\mathcal{Q}_{p\Delta}+\mathcal{O}\left(N_{c}^{-2}\right)$
was found which, combined with the $\Delta(1232)$-nucleon electric
quadrupole moment of $\mathcal{Q}_{p\Delta}=\left(-0.0846\pm0.0033\right)\, e\,\mbox{fm}^{2}$
\cite{Tiator(2003):ExpNDeltaQuadrupole}, gives a $\Delta^{+}(1232)$-isobar
EQM estimation at the physical pion mass of:

\begin{equation}
\mathcal{Q}_{\Delta^{+}}=\left(-0.048\pm0.002\right)\,
e\,\mbox{fm}^{2}\approx-1.87\,\frac{e}{M_{\Delta}^{2}}\,\,\,.
\end{equation}
Both values for $\mu_{\Delta^{+}}$ and $\mathcal{Q}_{\Delta^{+}}$
are represented by blue circles in Fig. \ref{fig:DELTAresutls}.
There is no experimental knowledge on the $\Delta(1232)$-isobar MOM
and CR. The Fig. \ref{fig:DELTAresutls} shows our results for the
$\Delta(1232)$-isobar electromagnetic quantities compared to the
lQCD results of \cite{Aubin:2008qp,Alexandrou:2008bn}.
The red solid curves correspond to the real parts with truncated SUGRA
while the blue long-dashed curves correspond
to excluding the non-minimal couplings. The green short-dashed curve
are the imaginary parts which vanishes above
$m_{\text{\ensuremath{\pi}}}=M_{\Delta}-M_{N}$.
The lQCD studies apply
different extraction techniques for the $\Delta^{+}(1232)$-isobar
electromagnetic moments. The Ref. \cite{Aubin:2008qp}
extracted the MDM by applying the external background field technique while in
\cite{Alexandrou:2008bn} the MDM, EQM and CR are obtained through the form
factors evaluated at finite $Q^{2}$ and extrapolating to $Q^{2}=0$ by dipole and
exponential fits. 

There are two pion mass regions for the
$\Delta(1232)$-isobar. Above the threshold $m_{\pi}=M_{\Delta}-M_{N}$
the $\Delta(1232)$ is stable while below the $N\pi$ decay
channel is open. Striking features in Fig. \ref{fig:DELTAresutls}
are the cusp and singularities in the real and
imaginary parts of the moments and CR at this pion mass. They result from the fact that
resonance electromagnetic properties at and near the opening of thresholds
are not well-defined \cite{Ledwig(2010):Singularities}.  

The MDM $\mu$, taken as the example, is usually defined by
the linear energy shift of the particle in an external magnetic
field $\vec{B}$:

\begin{equation}
M(\vec{B})=M_{0}-\vec{\mu}\cdot\vec{B}+O(B^{2})\,\,.\label{eq:linearENERGYshift}
\end{equation}
However, the energy change of unstable particles depend non-analytically on
$\vec{B}$ and the above linear approximation can only be used when the following
condition is met \cite{Ledwig(2010):Singularities}:

\begin{equation}
\frac{e|\vec{B}|}{2M_{\Delta}|M_{\Delta}-M_{N}-m_{\pi}|}\ll1\,\,\,.\label{eq:RELforB}
\end{equation}
At the pion mass $m_{\pi}=M_{\Delta}-M_{N}$ this is not the case and as a consequence the conventionally used one-photon
approximation to extract a the moment is not valid. Moreover,
for a given magnetic field strength $|\vec{B}|$, there exists a pion
mass region for which the $\Delta^{+}(1232)$-isobar energy is not
accurately approximated by Eq. (\ref{eq:linearENERGYshift}),
i.e. where the MDM is not well defined. This is directly relevant for lattice
QCD investigations where the periodic boundary conditions limit the
values of $|\vec{B}|$.

To give explicit situations, we take two examples. Once a spatial
lattice of $L=32$ with spacing $a^{-1}=1$ GeV and once $L=24$ with
$a^{-1}=2$ GeV. We compare the magnetic field implementation by $eBa^{2}=2\pi/L$
as in Ref. \cite{Lee(2005):MagStrength} and by $eBa^{2}=2\pi/L^{2}$
as in \cite{Aubin:2008qp}. Further, we take Eq. (\ref{eq:RELforB})
to be unity, i.e. a completely non-fulfillment of this relation, and
solve for the region around the threshold $m_{\pi}=M_{\Delta}-M_{N}$.
Within this region higher order $\vec{B}$ contributions can not be neglected.
For the finer lattice and linear-$L$ implementation this region is
$m_{\pi}=213.4\sim372.6$ MeV and for the quadratic-$L$
case $m_{\pi}=290.5\sim295.5$ MeV. For the second setting the regions are for
the linear-$L$ $m_{\pi}=-131.8\sim717.8$ MeV and for the quadratic-$L$ $m_{\pi}=275.3\sim310.7$ MeV.
We represent the two regions for the quadratic-$L$
implementation as grey bands in Fig. \ref{fig:DELTAresutls}.

The above considerations are directly applicable to the extraction
of $\Delta(1232)$-isobar moments by the external background field
method as used, e.g., in \cite{Aubin:2008qp}. The Eq.
(\ref{eq:RELforB}) gives a relation on how to chose the parameters
in order to interpret the extracted lQCD number as a MDM. We like to notice that the work \cite{Aubin:2008qp}
uses pion masses where Eq. (\ref{eq:RELforB}) is not violated. However,
future lattices will soon allow for pion masses where this will be
the case. 

The implications of the cusp and singularities on the three-point
function method is more subtle. Form factors data points are obtained for finite $Q^{2}$
and extrapolated by dipole or exponential fits to $Q^{2}=0$. In the
present case the cusp and singularities fall on $Q^{2}=0$
for $m_{\pi}=M_{\Delta}-M_{N}$. Qualitatively, a finite $Q^{2}$
would enter as an additional energy parameter and the singularities
would shift to $Q^{2}\neq0$ related to $m_{\pi}\neq M_{\Delta}-M_{N}$.
For lattice calculations this could mean that one extrapolates across
this singularity when all data points are on the right of the singularity.

Apart from the cusp and singularities we see that the present B$\chi$PT
study seems to be consistent with the lQCD data. With respect to the 
phenomenological uncertainties of the values at $m_{\pi}=139$
MeV, we can adjust the LECs $\overset{\circ}{\mu}_\Delta$, $\overset{\circ}{\mathcal{Q}}_\Delta$,
$\overset{\circ}{\mathcal{O}}_\Delta$ and $\overset{\circ}{r_{E0}}$ such that both our results could agree with both lQCD works \cite{Aubin:2008qp,Alexandrou:2008bn}.
In the chiral limit the EQM and the CR are logarithmically divergent while the
MDM and MOM are finite. In the case of the MOM the divergent part of
$F_2(q^2)$ is canceled by an equaly divergent part in $F_4 (q^2)$.

%
%
%
%

\section{Summary and conclusion}

Thus, we have investigated the electromagnetic moments, radii, and form factors
of the nucleon and $\Delta(1232)$ baryons in the realm of chiral perturbation theory.
The principle of analyticity (microcausality) is playing an important role in these quantities,
and hence, in contrast to previous investigations, we have performed the calculations in
the EOMS scheme.
We have obtained analytical expressions for various
contributions to the nucleon isovector magnetic moments,
Dirac- and Pauli-radius as well as the $\Delta(1232)$-isobar magnetic
dipole, electric quadrupole and magnetic octupole moments and charge
radius, and compared the results to experimental data and to recent lattice 
QCD calculations where available. 

B$\chi$PT predicts the analytic structure of the form factors at small $Q^2$ and can serve the purpose of extrapolating the electron-scattering data to $Q^2=0$. 
 We calculated the value of the second derivative of the proton electric form
  factor $G_E^p(q^2)$ in B$\chi$PT. 

 We have analyzed the cusps and singularities appearing in the pion mass dependance of
  the $\Delta(1232)$ electromagnetic quantities at the point where the $\Delta
  \to N \pi$ decay channel opens. This singularities are connected to the fact
  that em properties of unstable particles at the threshold are not well
  defined in perturbation theory. The self-energy of unstable particles depend non-analytically on
  the external magnetic field. This has an impact on the
  extraction of lQCD em moments of unstable particles near their decay
  threshold. To interpret the number extracted in lQCD in the vicinity of the opening of
  decay channels as an em moment, the applied external magnetic field or the
  $Q^2$ data points of the form factor have
  to be chosen specifically. 
  
  Comparing the chiral behavior of the nucleon em quantities given by our covariant B$\chi$PT to
  recent lQCD studies, we see that our results for the iso-vector anomalous magnetic
  moment and Dirac radius are in qualitative agreement (within the B$\chi$PT
  uncertainties) with lQCD results for $m^2_\pi>0.1$ GeV$^2$. Including finite volume effects in
  our B$\chi$PT formluae is the next step and expected to resolve some of the discrepancies.
 
\begin{acknowledgments}
The work of TL was partially supported
by the Research Centre Elementarkraefte und Mathematische
Grundlagen at the Johannes Gutenberg University Mainz. JMC acknowledges the MEC contract FIS2006-03438, the EU Integrated Infrastructure Initiative Hadron Physics Project contract RII3-CT-2004-506078 and the Science and Technology Facilities Council [grant number ST/H004661/1] for support.
\end{acknowledgments}
\begin{appendix}

%
%
%
%

\section{Notations\label{App:Pre}}

%
%
%
%

\subsection{Form factors}

The e.m.\ form factors are defined through the Lorentz
decomposition of the matrix element of the vector current, $V^{\mu}=\overline{\Psi}(0)\gamma^{\mu}\Psi(0)$,
between baryon states.
In the case of the nucleon,
\begin{equation}
\langle
N(p^{\prime})|\overline{\Psi}(0)\gamma^{\mu}\Psi(0)|N(p)\rangle=\bar{u}(p^{\prime})\left[\gamma^{\mu}F^N_{1}(Q^{2})+\frac{i\sigma^{\mu\nu}q_{\nu}}{2M_{N}}F^N_{2}(Q^{2})\right]u(p)\,\,\,,\label{Eq:NucleonFF0}
\end{equation}
with $q=p^{\prime}-p$, $Q^{2}=-q^{2}$, and
$u(p)$ as the nucleon spinor with mass $M_{N}$. The invariants $F_{1}$
and $F_{2}$ are the Dirac and Pauli form factors, respectively, which at $Q^{2}=0$
yield the nucleon charge in units of $e$ and the nucleon anomalous magnetic
moment:
\begin{eqnarray}
F_{1}^N(0) & = & e_{N},\nn\\
F_{2}^N(0) & = & \kappa_{N}.
\end{eqnarray}
One distinguishes the iso-vector and iso-scalar nucleon form
factors as: 
\begin{equation}
F_{i}^{V}(Q^{2})=F_{i}^{p}(Q^{2})-F_{i}^{n}(Q^{2})\,, \qquad 
F_{i}^{S}(Q^{2})=F_{i}^{p}(Q^{2})+F_{i}^{n}(Q^{2})\,.
\end{equation}

In the case of the $\Delta(1232)$, which has spin 3/2, there are four 
independent form factors:
\begin{eqnarray}
\langle\Delta(p^{\prime})|V^{\mu}|\Delta(p)\rangle & = & -\bar{u}_{\alpha}(p^{\prime})\Big\{\,\,\,\left[F_{1}^{\De}(Q^{2})\gamma^{\mu}+\frac{i\sigma^{\mu\nu}q_{\nu}}{2M_{\Delta}}F_{2}^{\De}(Q^{2})\right]g^{\alpha\beta}\nonumber \\
 &  &
+\left[F_{3}^{\De}(Q^{2})\gamma^{\mu}+\frac{i\sigma^{\mu\nu}q_{\nu}}{2M_{\Delta}}F_{4}^{\De}(Q^{2})\right]\frac{q^{\alpha}q^{\beta}}{4M_{\Delta}^{2}}\,\,\,\Big\}u_{\beta}(p)\,\,\,,\label{Eq:DeltaFF0}
\end{eqnarray}
where $u_{\alpha}(p)$ is a Rarita-Schwinger spinor for the spin-3/2
$\Delta(1232)$-isobar state of mass $M_{\Delta}$. The multipole 
form factors $G_{E0}$, $G_{M1}$,  $G_{E2} $, $G_{M3}$ of the $\De$ are
expressed in terms of $F$'s as follows:
\begin{eqnarray}
G_{E0}(Q^{2}) & = & F_{1}^{\De}(Q^{2})-\tau F_{2}^{\De}(Q^{2})+\frac{2}{3}\tau G_{E2}(Q^{2})\,\nn,\\
G_{E2}(Q^{2}) & = & F_{1}^{\De}(Q^{2})-\tau F_{2}^{\De}(Q^{2})-\frac{1}{2}(1+\tau)\left[F_{3}^{\De}(Q^{2})-\tau F_{4}^{\De}(Q^{2})\right]\,,\nn\\
G_{M1}(Q^{2}) & = & F_{1}^{\De}(Q^{2})+F_{2}^{\De}(Q^{2})+\frac{4}{5}\tau G_{M3}(Q^{2})\,,\\
G_{M3}(Q^{2}) & = &
F_{1}^{\De}(Q^{2})+F_{2}^{\De}(Q^{2})-\frac{1}{2}(1+\tau)\left[F_{3}^{\De}(Q^{2})+F_{4}^{\De}(Q^{2})\right]\,,\nn
\end{eqnarray}
with $\tau=Q^{2}/(4M_{\Delta}^{2})$. At $Q^{2}=0$, the multipole
form factors define the static 
moments: charge $e_{\Delta}$, magnetic dipole moment
$\mu_{\Delta}$, electric quadrupole moment $\mathcal{Q}_{\Delta}$
magnetic octupole moment $\mathcal{O}_{\Delta}$, i.e.:

\begin{eqnarray}
e_{\Delta} & = & G_{E0}(0)=F_{1}^{\De}(0)\,,\nn\\
\mu_{\Delta} & = & \frac{e}{2M_{\Delta}}G_{M1}(0)=\frac{e}{2M_{\Delta}}\left[e_{\Delta}+F_{2}^{\De}(0)\right]\,,\nn\\
\mathcal{Q}_{\Delta} & = & \frac{e}{M_{\Delta}^{2}}G_{E2}(0)=\frac{e}{M_{\Delta}^{2}}\left[e_{\Delta}-\frac{1}{2}F_{3}^{\De}(0)\right]\,,\\
\mathcal{O}_{\Delta} & = &
\frac{e}{2M_{\Delta}^{3}}G_{M3}(0)=\frac{e}{2M_{\Delta}^{3}}\left[e_{\Delta}+F_{2}^{\De}(0)-\frac{1}{2}\left(F_{3}^{\De}(0)+F_{4}^{\De}(0)\right)\right]\,.\nn
\end{eqnarray}

Besides the static moments, the slopes of the
form factors are of interest as they indicate the radii of the respective e.m. distributions;
generically:
\begin{equation}
\langle r^2\rangle=\frac{6}{F(0)}\frac{dF(q^{2})}{dq^{2}}
\end{equation}

\subsection{Isospin and Lorentz structures}
The iso-spin 1/2 to 3/2 and 3/2 to 3/2 transition matrices $T^{a}$
and $\mathcal{T}^{a}$ appearing in the $N\Delta$ and $\Delta\Delta$
Lagrangians are given by:\begin{eqnarray}
T^{1}=\frac{1}{\sqrt{6}}\left(\begin{array}{cccc}
-\sqrt{3} & 0 & 1 & 0\\
0 & -1 & 0 & \sqrt{3}\end{array}\right) & \,\,\,\,\,\,\,\,\,\,\,\,\,\,\,\,\,\,\, & \mathcal{T}^{1}=\frac{2}{3}\left(\begin{array}{cccc}
0 & \frac{\sqrt{3}}{2} & 0 & 0\\
\frac{\sqrt{3}}{2} & 0 & 1 & 0\\
0 & 1 & 0 & \frac{\sqrt{3}}{2}\\
0 & 0 & \frac{\sqrt{3}}{2} & 0\end{array}\right)=\mathcal{T}^{1\dagger}\nonumber \\
T^{2}=\frac{-i}{\sqrt{6}}\left(\begin{array}{cccc}
\sqrt{3} & 0 & 1 & 0\\
0 & 1 & 0 & \sqrt{3}\end{array}\right) &  & \mathcal{T}^{2}=\frac{2i}{3}\left(\begin{array}{cccc}
0 & -\frac{\sqrt{3}}{2} & 0 & 0\\
\frac{\sqrt{3}}{2} & 0 & -1 & 0\\
0 & 1 & 0 & -\frac{\sqrt{3}}{2}\\
0 & 0 & \frac{\sqrt{3}}{2} & 0\end{array}\right)=\mathcal{T}^{2\dagger}\nonumber \\
T^{3}=\sqrt{\frac{2}{3}}\left(\begin{array}{cccc}
0 & 1 & 0 & 0\\
0 & 0 & 1 & 0\end{array}\right) &  & \mathcal{T}^{3}=\left(\begin{array}{cccc}
1 & 0 & 0 & 0\\
0 & 1/3 & 0 & 0\\
0 & 0 & -1/3 & 0\\
0 & 0 & 0 & -1\end{array}\right)=\mathcal{T}^{3\dagger}\nonumber \\
Q_{\pi}^{ab}=-i\varepsilon^{ab3} &  & Q_{N}=\frac{1}{2}(1+\tau^{3})\nonumber \\
Q_{N\Delta}=\frac{1}{2}(1+3\mathcal{T}^{3}) & \,\,\,\,\,\, & Q_{\Delta}=\frac{1}{2}(1+3\mathcal{T}^{3})\,\,\,.\end{eqnarray}
The totally antisymmetric Dirac matrices products appearing in the
$N\Delta$ and $\Delta\Delta$ Lagrangians are defined as:\begin{eqnarray}
\gamma^{\mu\nu} & = & \frac{1}{2}\left[\gamma^{\mu},\gamma^{\nu}\right]\,\,\,,\nonumber \\
\gamma^{\mu\nu\rho} & = & \frac{1}{2}\left\{ \gamma^{\mu\nu},\gamma^{\rho}\right\} =-i\varepsilon^{\mu\nu\rho\sigma}\gamma_{\sigma}\gamma_{5}\,\,\,,\nonumber \\
\gamma^{\mu\nu\rho\sigma} & = & \frac{1}{2}\left[\gamma^{\mu\nu\rho},\gamma^{\sigma}\right]=i\varepsilon^{\mu\nu\rho\sigma}\gamma_{5}\,\,\,,\end{eqnarray}
 with the convention: $\varepsilon_{0123}=-\varepsilon^{0123}=+1$.
\\

\subsection{Loop integrals}
In this work the following loop integrals appear:

\begin{eqnarray}
J_{n}(\mathcal{M})=\int\frac{d^{d}l}{(2\pi)^{d}}\frac{1}{\left[l^{2}-\mathcal{M}\right]^{n}} & = & \frac{i}{(4\pi)^{\frac{d}{2}}}(-1)^{n}\frac{\Gamma(n-\frac{d}{2})}{\Gamma(n)}\left[\mathcal{M}\right]^{\frac{d}{2}-n}\nonumber \\
\int\frac{d^{d}l}{(2\pi)^{d}}\frac{l_{\kappa}l_{\lambda}}{\left[l^{2}-\mathcal{M}\right]^{n}} & = & \frac{1}{2(n-1)}\,\, J_{n-1}(\mathcal{M})g_{\kappa\lambda}\nonumber \\
\int\frac{d^{d}l}{(2\pi)^{d}}\frac{l_{\alpha}l_{\beta}l_{\mu}l_{\nu}}{\left[l^{2}-\mathcal{M}\right]^{n}} & = & \frac{g_{\alpha\beta}g_{\mu\nu}+g_{\beta\mu}g_{\alpha\nu}+g_{\beta\nu}g_{\alpha\mu}}{4(n-1)(n-2)}\,\, J_{n-2}(\mathcal{M})\,\,\,.\end{eqnarray}
The corresponding solutions are:\begin{eqnarray}
J_{1}\left(\mathcal{M}\right) & = & \frac{-i}{\left(4\pi\right)^{2}}M_{sc}^{2}\tilde{\mathcal{M}}\left[L-1+\ln\tilde{\mathcal{M}}\right]\nonumber \\
J_{2}\left(\mathcal{M}\right) & = & \frac{-i}{\left(4\pi\right)^{2}}\left[L+\ln\tilde{\mathcal{M}}\right]\nonumber \\
J_{3}\left(\mathcal{M}\right) & = & \frac{-i}{\left(4\pi\right)^{2}}\frac{1}{2M_{sc}^{2}}\frac{1}{\tilde{\mathcal{M}}}\,\,\,,\end{eqnarray}
with $\tilde{\mathcal{M}}=\mathcal{M}/M_{sc}^{2}$, where $M_{sc}$
is the relevant mass scale, and $L=-\frac{1}{\epsilon}+\gamma_{E}+\ln\frac{M_{sc}^{2}}{4\pi\Lambda^{2}}$.
We work in $d=4-2\epsilon$ dimensions.

The $D$-dimensional spin-$3/2$ $\Delta$-isobar propagator is given by:
\begin{eqnarray}
&&S_{\Delta}^{\alpha\beta}(p)  =  \frac{\s p+M_{\Delta}}{p^{2}-M_{\Delta}^{2}+i\varepsilon}\nonumber \\
 &  & \left[-g^{\alpha\beta}+\frac{1}{D-1}\gamma^{\alpha}\gamma^{\beta}+\frac{1}{(D-1)M_{\Delta}}(\gamma^{\alpha}p^{\beta}-\gamma^{\beta}p^{\alpha})+\frac{D-2}{(D-1)M_{\Delta}^{2}}p^{\alpha}p^{\beta}\right]\,.\end{eqnarray}
We use the following parameters:
\begin{equation}
\mu=\frac{m_\pi}{M_{sc}} \;\;\;\;\;\; R=\frac{M_\Delta}{M_N}\;\;\; \;\;\;
r=\frac{M_N}{M_\Delta}\;\;\; \;\;\;\tilde{q}^2=\frac{q^2}{M_N^2}
\end{equation}
\begin{equation}
C_{NN} = \frac{g_A M_{sc}}{8f_\pi\pi}\;\;\; \;\;\; C_{ND} = \frac{h_A M_{sc}}{8f_\pi\pi}\;\;\;
\;\;\; C_{DD} = \frac{H_A M_{sc}}{8f_\pi\pi}\;\;\;.
\end{equation}

%
%
%
%

\section{Nucleon electromagnetic form factors \label{APP:NucleonFF}}
For the nucleon electromagnetic form factors we take the mass scale
$M_{sc}=M_N$ and the following functions occur:
\begin{eqnarray}
\tilde{\mathcal{M}}_1   & = & z\mu^{2}+(1-z)^2 -z^2x(1-x)\tilde{q}^2\,\,\,,\nonumber \\
\tilde{\mathcal{M}}_2   & = & (1-z)\mu^{2}+z^2 -z^2x(1-x)\tilde{q}^2\,\,\,,\nonumber \\
\tilde{\mathcal{M}}_3   & = & z\mu^{2}-z(1-z)+(1-z)R^2 -z^2x(1-x)\tilde{q}^2\,\,\,,\nonumber \\
\tilde{\mathcal{M}}_4   & = & (1-z)\mu^{2}-z(1-z) +zR^2-z^2x(1-x)\tilde{q}^2\,\,\,,\nonumber \\
\tilde{\mathcal{M}}_{78} & = & z\mu^{2}-z(1-z)+(1-z)R^2\,\,\,,
\end{eqnarray}
with $\tilde{q}^2=q^2/M_N^2$.
The iso-spin factors are given:

\begin{eqnarray}
I_{N1} =\left(\begin{array}{c}
+2\\
-2\end{array}\right)\,\,\,\,\,\, & I_{N2}=\left(\begin{array}{c}
+1\\
+2\end{array}\right) & \,\,\,\,\,\, I_{N3}=\left(\begin{array}{c}
-\frac{2}{3}\\
+\frac{2}{3}\end{array}\right)\nonumber \\
I_{N4}=\left(\begin{array}{c}
+\frac{8}{3}\\
-\frac{2}{3}\end{array}\right)\,\,\,\,\,\, & I_{N78a}=\left(\begin{array}{c}
-\frac{2}{3}\\
+\frac{2}{3}\end{array}\right) & \,\,\,\,\,\, I_{N78b}=\left(\begin{array}{c}
+\frac{8}{3}\\
-\frac{2}{3}\end{array}\right) \,\,\,\,\,\, I_{NT1}=\left(\begin{array}{c}
+1\\
-1\end{array}\right)
\end{eqnarray}
In case of the unintegrated versions the densities have to be integrated by: 
\begin{equation}
F(q^2) =  \int_{0}^{1}dz \int_{0}^{1}dx \mathcal{F}(q^2,z,x)
\end{equation}
with 
\begin{equation}
\mathcal{L}_i = \ln[\tilde{\mathcal{M}}_i] \nn
\end{equation}
The integrated iso-vector versions contain the following expression:
\begin{equation}
\mathcal{A} = \frac{1}{\Gamma} (\arctan \frac{1+R^2-\mu^2}{\Gamma} + \arctan
\frac{1-R^2+\mu^2}{\Gamma}  ) \nn
\end{equation}
\begin{equation}
 \Gamma=\sqrt{-R^4 - (-1 + \mu^2)^2 + 2 R^2 (1 + \mu^2)}
\end{equation}
In the following expressions for the nucleon quantities we renormalized
already the infinite tems proportional to $L$. However, the power-counting
breaking terms (constant in $m_\pi$ terms) are still kept at this stage.


\subsection{Contributions to $F_{1}(q^2)$ \label{AppNucleon}}
Contributions from virtual nucleons:
\begin{eqnarray}
&&\frac{\mathcal{F}_1^{N1}}{C_{NN}^2 I_{N1}} =
-6 z  + 14 z^2 - 12 x z^2 - 4 z^3 + 4 x z^3
 +   (-15 z - 3 M_1 z + 42 z^2\nn\\&& - 36 x z^2 - 17 z^3 + 
    16 x z^3)\mathcal{L}_1 + 2 \tilde{\mathcal{M}}_1 z
 + (-2 z + 10 z^2 - 4 x z^2\nn\\&& - 18 z^3 + 12 x z^3 + 
  14 z^4 - 12 x z^4 - 4 z^5 + 4 x z^5)[1/\tilde{\mathcal{M}}_1]\nn\\&& + 
 \tilde{q}^2(x z^3 - x^2 z^3)\mathcal{L}_1 + \tilde{q}^2 (
    6 x z^3 - 6 x^2 z^3 - 14 x z^4 + 26 x^2 z^4\nn\\&& - 12 x^3 z^4 + 
     4 x z^5 - 8 x^2 z^5 + 4 x^3 z^5) [1/\tilde{\mathcal{M}}_1]\\
&&\frac{\mathcal{F}_1^{N2}}{C_{NN}^2 I_{N2}}  =
-4 z - \tilde{\mathcal{M}}_2 z + 4 z^2 - 8 x z^2 + 2 z^3\nn\\&& + 
 \mathcal{L}_2 (-6 z + 6 \tilde{\mathcal{M}}_2 z + 12 z^2 - 24 x z^2 + 6 z^3) + 
 [1/\tilde{\mathcal{M}}_2] (4 z^4 - 8 x z^4 + z^5)\nn\\&& + 
 \tilde{q}^4 [1/\tilde{\mathcal{M}}_2]  (x z^3 - x^2 z^3 - x z^4 + x^2 z^4 + 
    x^2 z^5 - 2 x^3 z^5 + x^4 z^5)\nn\\&& + 
 \tilde{q}^2 (z^2 - 2 x z^3 + 2 x^2 z^3 + 
    \mathcal{L}_2 (-z + 3 z^2 - 6 x z^3 + 6 x^2 z^3)\nn\\&& + 
    [1/\tilde{\mathcal{M}}_2] (4 x z^3 - 4 x^2 z^3 + z^4 - 4 x z^4 + 
       12 x^2 z^4 - 8 x^3 z^4 - 2 x z^5 + 2 x^2 z^5))\\
&&\mathcal{F}_1^{NT1} = 
\frac{I_{NT1} M_N^2}{32 f_\pi^2 \pi^2} ( \tilde{q}^2 (-1 + x) x + \mu^2) (-1 + 
   \ln[\tilde{q}^2 (-1 + x) x + \mu^2]))
\end{eqnarray}
Contributions from virtual $\Delta(1232)$ with minimal $\gamma\Delta\Delta$ coupling:
\begin{eqnarray}
&&\frac{18R^2\,\mathcal{F}_1^{N3}}{C_{ND}^2 I_{N3}}  =
-36 R z - 36 z^2 + 80 R z^2 - 88 R x z^2 + 80 z^3 - 88 x z^3\nn\\&& + 
 \mathcal{L}_3 (18 \tilde{\mathcal{M}}_3 z - 36 R z - 36 z^2 + 96 R z^2 - 
    120 R x z^2 + 96 z^3 - 120 x z^3)\nn\\&& + 
 \tilde{q}^2 (16 \tilde{\mathcal{M}}_3 z + 20 R z + 20 z^2 - 42 R z^2 + 44 R x z^2 - 
    42 z^3 + 40 x z^3 + 4 x^2 z^3\nn\\&& + 
    \mathcal{L}_3 (-24 \tilde{\mathcal{M}}_3 z + 24 R z + 24 z^2 - 63 R z^2 + 
       60 R x z^2 - 63 z^3 + 48 x z^3 + 12 x^2 z^3)\nn\\&& + 
    [1/\tilde{\mathcal{M}}_3] (24 R x z^3 - 24 R x^2 z^3 + 24 x z^4 - 48 R x z^4 - 
       24 x^2 z^4 + 96 R x^2 z^4\nn\\&& - 48 R x^3 z^4 - 48 x z^5 + 
       96 x^2 z^5 - 48 x^3 z^5))\nn\\&& + 
 \tilde{q}^4 (-x z^3 + x^2 z^3 + \mathcal{L}_3 (6 x z^3 - 6 x^2 z^3) + 
    [1/\tilde{\mathcal{M}}_3] (-6 R x z^3 + 6 R x^2 z^3 - 6 x z^4\nn\\&& + 12 R x z^4 + 
       6 x^2 z^4 - 24 R x^2 z^4 + 12 R x^3 z^4 + 12 x z^5 - 
       24 x^2 z^5 + 12 x^3 z^5))
\end{eqnarray}
\begin{eqnarray}
&&\frac{54R^4\,\mathcal{F}_1^{N4}}{C_{ND}^2 I_{N4}}  =
-54 R^2 z + 54 \tilde{\mathcal{M}}_4 R^2 z - 108 R^3 z - 54 R^4 z + 240 R^2 z^2 + 
 240 R^3 z^2\nn\\&& - 264 R^2 x z^2 - 264 R^3 x z^2 - 186 R^2 z^3 + 
 264 R^2 x z^3\nn\\&& + 
 \mathcal{L}_4 (-54 R^2 z + 108 \tilde{\mathcal{M}}_4 R^2 z - 108 R^3 z - 54 R^4 z + 
    288 R^2 z^2 + 288 R^3 z^2\nn\\&& - 360 R^2 x z^2 - 360 R^3 x z^2 - 
    234 R^2 z^3 + 360 R^2 x z^3)\nn\\&& + 
 \tilde{q}^2 (-8 z + 30 R z + 48 R^2 z - 26 R^3 z - 36 R^4 z - 16 z^2 - 
    79 R z^2 - 138 R^2 z^2\nn\\&& - 75 R^3 z^2 - 48 R x z^2 + 36 R^2 x z^2 + 
    84 R^3 x z^2 + 56 z^3 + 100 R z^3 + 162 R^2 z^3\nn\\&& - 258 R^2 x z^3 + 
    126 R^2 x^2 z^3- 32 z^4 - 51 R z^4 + 48 R x z^4 + 
    \tilde{\mathcal{M}}_4 (30 z + 86 R z\nn\\&& + 60 R^2 z - 38 z^2 - 69 R z^2 + 
       84 R x z^2) + 
    [1/\tilde{\mathcal{M}}_4] (36 R^2 x z^3 + 72 R^3 x z^3 + 36 R^4 x z^3\nn\\&& - 
       36 R^2 x^2 z^3 - 72 R^3 x^2 z^3 - 36 R^4 x^2 z^3 - 
       144 R^2 x z^4 - 144 R^3 x z^4 + 288 R^2 x^2 z^4\nn\\&& + 
       288 R^3 x^2 z^4 - 144 R^2 x^3 z^4 - 144 R^3 x^3 z^4 + 
       108 R^2 x z^5 - 252 R^2 x^2 z^5 + 144 R^2 x^3 z^5)\nn\\&& + 
    \mathcal{L}_4 (6 z + 18 R z + 18 R^2 z + 6 R^3 z - 42 z^2 - 
       33 R z^2 - 135 R^2 z^2 - 144 R^3 z^2\nn\\&& - 36 R x z^2 + 
       108 R^2 x z^2 + 144 R^3 x z^2 + 66 z^3 + 60 R z^3 + 
       189 R^2 z^3 - 342 R^2 x z^3\nn\\&& + 162 R^2 x^2 z^3 - 30 z^4 - 
       45 R z^4 + 36 R x z^4\nn\\&& + 
       \tilde{\mathcal{M}}_4 (108 z + 120 R z + 72 R^2 z - 120 z^2 - 180 R z^2 + 
          144 R x z^2)))\\
&&\frac{R^2\,\mathcal{F}_1^{N78}}{2C_{ND}^2}  = (I_{N78a} + I_{N78b})
\mathcal{M}_{78} (R + z) \ln \tilde{\mathcal{M}}_{78}
 \end{eqnarray}
Additional contribution from non-minimal $\gamma \Delta\Delta$ coupling:
\begin{eqnarray}
&&\frac{27R^5\,\mathcal{F}_1^{nm}}{C_{ND}^2 I_{N4}}  =
\tilde{q}^2 (-24 z - 20 R z + 24 R^2 z + 12 R^3 z - 8 R^4 z + 72 z^2 + 
   116 R z^2 + 41 R^2 z^2\nn\\&& - 3 R^3 z^2 - 72 z^3 - 172 R z^3 - 
   65 R^2 z^3 + 24 z^4 + 76 R z^4\nn\\&& + 
   \tilde{\mathcal{M}}_4 (-42 z - 198 R z - 115 R^2 z + 42 z^2 + 178 R z^2)\nn\\&& + 
   \mathcal{L}_4 (-18 z - 12 R z + 27 R^2 z + 18 R^3 z - 3 R^4 z + 
      54 z^2\nn\\&& + 66 R z^2 - 15 R^2 z^2 - 27 R^3 z^2 - 54 z^3 - 
      96 R z^3 - 12 R^2 z^3 + 18 z^4 + 42 R z^4\nn\\&& + 
      \tilde{\mathcal{M}}_4 (-72 z - 198 R z - 96 R^2 z + 72 z^2 + 168 R z^2)))
\end{eqnarray}

\subsection{Contributions to $F_{2}(q^2)$}
Contributions from virtual nucleons:
\begin{eqnarray}
&&\frac{\mathcal{F}_2^{N1}}{C_{NN}^2 I_{N1}}  =
6 z - 14 z^2 + 12 x z^2 + 4 z^3 - 4 x z^3+ 
  (12 z - 42 z^2 + 36 x z^2 + 16 z^3 - 16 x z^3)\mathcal{L}_1\nn\\&& + 
  (2 z - 10 z^2 + 4 x z^2 + 18 z^3 - 12 x z^3 - 
    14 z^4 + 12 x z^4 + 4 z^5 - 4 x z^5) [1/\tilde{\mathcal{M}}_1]\nn\\&& + 
  \tilde{q}^2(-6 x z^3 + 6 x^2 z^3 + 14 x z^4 - 
    26 x^2 z^4 + 12 x^3 z^4 - 4 x z^5 + 8 x^2 z^5 - 4 x^3
    z^5) [1/\tilde{\mathcal{M}}_1]\\
&&\frac{\mathcal{F}_2^{N2}}{C_{NN}^2 I_{N2}}  = 
-8 z^2 + 8 x z^2 + \mathcal{L}_2 (8 z - 24 z^2 + 24 x z^2) + 
  [1/\tilde{\mathcal{M}}_2] (-8 z^4 + 8 x z^4)\nn\\&& + 
 \tilde{q}^2  [1/\tilde{\mathcal{M}}_2] (-8 x z^3 + 8 x^2 z^3 + 8 x z^4 - 
    16 x^2 z^4 + 8 x^3 z^4)
\end{eqnarray}
Contributions from virtual $\Delta(1232)$ with minimal $\gamma\Delta\Delta$ coupling:
\begin{eqnarray}
&&\frac{9R^2\,\mathcal{F}_2^{N3}}{C_{ND}^2 I_{N3}}  =
-28 \tilde{\mathcal{M}}_3 z + 18 R z + 18 z^2 - 44 R z^2 + 44 R x z^2 - 44 z^3 + 
 44 x z^3\nn\\&& + 
 \mathcal{L}_3 (24 \tilde{\mathcal{M}}_3 z + 18 R z + 18 z^2 - 42 R z^2 + 
    60 R x z^2 - 42 z^3 + 60 x z^3)\nn\\&& + 
 [1/\tilde{\mathcal{M}}_3] \tilde{q}^4 (3 R x z^3 - 3 R x^2 z^3 + 3 x z^4 - 
    6 R x z^4 - 3 x^2 z^4 + 12 R x^2 z^4\nn\\&& - 6 R x^3 z^4 - 6 x z^5 + 
    12 x^2 z^5 - 6 x^3 z^5)\nn\\&& + 
 \tilde{q}^2 (-10 R z - 10 z^2 + 22 R z^2 - 22 R x z^2 + 22 z^3 - 
    18 x z^3 - 4 x^2 z^3\nn\\&& + 
    \mathcal{L}_3 (-12 R z - 12 z^2 + 30 R z^2 - 30 R x z^2 + 30 z^3 - 
       36 x z^3 + 6 x^2 z^3)\nn\\&& + 
   [1/\tilde{\mathcal{M}}_3] (-12 R x z^3 + 12 R x^2 z^3 - 12 x z^4 + 24 R x z^4 + 
       12 x^2 z^4 - 48 R x^2 z^4\nn\\&& + 24 R x^3 z^4 + 24 x z^5 - 
       48 x^2 z^5 + 24 x^3 z^5))
\end{eqnarray}
\begin{eqnarray}
&&\frac{27R^4\,\mathcal{F}_2^{N4}}{2C_{ND}^2 I_{N4}} =
8 z - 30 R z - 42 R^2 z + 38 R^3 z + 42 R^4 z - 24 z^2 + 63 R z^2 + 
 18 R^2 z^2\nn\\&& - 69 R^3 z^2 + 66 R^2 x z^2 + 66 R^3 x z^2 + 24 z^3 - 
 36 R z^3 + 24 R^2 z^3 - 66 R^2 x z^3 \nn\\&&- 8 z^4 + 3 R z^4 + 
 \tilde{\mathcal{M}}_4 (50 z - 4 R z - 48 R^2 z - 50 z^2 - 15 R z^2) + 
 \mathcal{L}_4 (-6 z - 18 R z\nn\\&& + 30 R^3 z + 18 R^4 z + 18 z^2 + 45 R z^2 - 
    45 R^2 z^2 - 72 R^3 z^2 + 90 R^2 x z^2\nn\\&& + 90 R^3 x z^2 - 18 z^3 - 
    36 R z^3 + 45 R^2 z^3 - 90 R^2 x z^3 + 6 z^4 + 9 R z^4\nn\\&& + 
    \tilde{\mathcal{M}}_4 (-24 z - 78 R z - 90 R^2 z + 24 z^2 + 36 R z^2))\nn\\&& + 
 \tilde{q}^2 (2 z + 2 R z - 2 R^2 z - 2 R^3 z - 6 z^2 - 39 R z^2 - 
    12 R^2 z^2 + 21 R^3 z^2 + 12 R x z^2\nn\\&& - 9 R^2 x z^2 - 
    21 R^3 x z^2 - 2 z^3 + 17 R z^3 - 33 R^2 z^3 + 44 x z^3 + 
    24 R x z^3 + 126 R^2 x z^3\nn\\&& - 44 x^2 z^3 - 24 R x^2 z^3 - 
    93 R^2 x^2 z^3 + 6 z^4 + 12 R z^4 - 44 x z^4 - 15 R x z^4 + 
    44 x^2 z^4\nn\\&& + 3 R x^2 z^4+ 
    \tilde{\mathcal{M}}_4 (-9 z - 5 R z + 24 z^2 + 21 R z^2 - 21 R x z^2)\nn\\&& + 
    [1/\tilde{\mathcal{M}}_4] (12 x z^3 + 18 R x z^3 - 9 R^2 x z^3 - 24 R^3 x z^3 - 
      9 R^4 x z^3 - 12 x^2 z^3 - 18 R x^2 z^3\nn\\&& + 9 R^2 x^2 z^3 + 
       24 R^3 x^2 z^3 + 9 R^4 x^2 z^3 - 36 x z^4 - 36 R x z^4 + 
       36 R^2 x z^4 + 36 R^3 x z^4\nn\\&& + 36 x^2 z^4 + 36 R x^2 z^4 - 
       72 R^2 x^2 z^4 - 72 R^3 x^2 z^4 + 36 R^2 x^3 z^4 + 
       36 R^3 x^3 z^4 + 36 x z^5\nn\\&& + 18 R x z^5 - 27 R^2 x z^5 - 
       36 x^2 z^5 - 18 R x^2 z^5 + 63 R^2 x^2 z^5 - 36 R^2 x^3 z^5 - 
       12 x z^6\nn\\&& + 12 x^2 z^6)+ 
    \mathcal{L}_4 (3 z + 3 R z - 3 R^2 z - 3 R^3 z - 18 z^2 - 36 R z^2 + 
       18 R^2 z^2 + 36 R^3 z^2\nn\\&& + 9 R x z^2 - 27 R^2 x z^2 - 
       36 R^3 x z^2 + 15 z^3 + 12 R z^3 - 45 R^2 z^3 + 66 x z^3 + 
       54 R x z^3\nn\\&& + 108 R^2 x z^3 - 66 x^2 z^3 - 54 R x^2 z^3 - 
       63 R^2 x^2 z^3 + 9 R z^4 - 66 x z^4 - 18 R x z^4 + 
       66 x^2 z^4\nn\\&& + 9 R x^2 z^4 + 
       \tilde{\mathcal{M}}_4 (6 R z + 36 R z^2 - 36 R x z^2)))\\
&&\frac{9R^3\,\mathcal{F}_2^{N78}}{4 C_{ND}^2 } = -\tilde{\mathcal{M}}_{78} (R + z) (-I_{N78b} - I_{N78a} R - 
   I_{N78b} R\nn\\&& + (6 I_{N78a} R + I_{N78b} (-3 + 6 R)) \ln \mathcal{M}_{78})
\end{eqnarray}
Additional contribution from non-minimal $\gamma \Delta\Delta$ vertex:
\begin{eqnarray}
&&\frac{27R^5\,\mathcal{F}_2^{nm}}{C_{ND}^2 I_{N4}}  =
80 R z + 126 R^2 z + 12 R^3 z - 34 R^4 z - 240 R z^2 - 252 R^2 z^2 - 
 12 R^3 z^2\nn\\&& + 240 R z^3 + 126 R^2 z^3 - 80 R z^4 + 
 \tilde{\mathcal{M}}_4 (176 R z + 170 R^2 z - 176 R z^2) + 
 \mathcal{L}_4 (48 R z + 18 R^2 z\nn\\&& - 108 R^3 z - 78 R^4 z - 144 R z^2 - 
    36 R^2 z^2 + 108 R^3 z^2 + 144 R z^3 + 18 R^2 z^3 - 48 R z^4\nn\\&& + 
    \tilde{\mathcal{M}}_4 (192 R z + 174 R^2 z - 192 R z^2)) + 
 \tilde{q}^2 (24 z - 20 R z - 72 R^2 z\nn\\&& + 12 R^3 z + 40 R^4 z - 80 z^2 + 
    90 R z^2 + 164 R^2 z^2 - 6 R^3 z^2 + 88 z^3 - 52 R z^3\nn\\&& - 
    16 R^2 z^3 - 304 R x z^3 - 234 R^2 x z^3 + 304 R x^2 z^3 + 
    234 R^2 x^2 z^3 - 32 z^4 - 18 R z^4\nn\\&& + 304 R x z^4 - 
    304 R x^2 z^4 + 
    \tilde{\mathcal{M}}_4 (40 z + 84 R z + 28 R^2 z - 38 z^2 - 18 R z^2)\nn\\&& + 
    [1/\tilde{\mathcal{M}}_4] (-24 R x z^3 + 72 R^3 x z^3 + 48 R^4 x z^3 + 
       24 R x^2 z^3 - 72 R^3 x^2 z^3 - 48 R^4 x^2 z^3\nn\\&& + 72 R x z^4 - 
       72 R^3 x z^4 - 72 R x^2 z^4 + 72 R^3 x^2 z^4 - 72 R x z^5 + 
       72 R x^2 z^5 + 24 R x z^6\nn\\&& - 24 R x^2 z^6)+ 
    \mathcal{L}_4 (18 z - 12 R z - 36 R^2 z + 36 R^3 z + 42 R^4 z - 
       66 z^2 + 54 R z^2\nn\\&& + 48 R^2 z^2 - 72 R^3 z^2 + 78 z^3 - 
       24 R z^3 + 48 R^2 z^3 - 168 R x z^3 - 234 R^2 x z^3\nn\\&& + 
       168 R x^2 z^3 + 234 R^2 x^2 z^3 - 30 z^4 - 18 R z^4 + 
       168 R x z^4 - 168 R x^2 z^4\nn\\&& + 
       \tilde{\mathcal{M}}_4 (114 z + 108 R z + 24 R^2 z - 120 z^2 - 72 R z^2)))
\end{eqnarray}


\subsection{Iso-vector Dirac radius $\langle r_{1}^2 \rangle_V$}
Contributions from virtual nucleons:
\begin{eqnarray}
&&dF_1^{N}  =
-\frac{\ln \mu}{48 f_\pi^2 \pi^2} - \frac{C_{NN}^2}{3 M_N^2 (-4 + \mu^2)}(-172
+ 163 \mu^2 - 30 \mu^4 + (-80 + 372 \mu^2\nn\\&& - 208 \mu^4 + 
       30 \mu^6) \ln \mu + \mu \sqrt{-4 + \mu^2} (70 - 
       74 \mu^2 + 15 \mu^4) \ln  \frac{\mu - 
       \sqrt{-4 + \mu^2}}{\mu + \sqrt{-4 + \mu^2}} )
\end{eqnarray}
Contributions from virtual $\Delta(1232)$ with minimal $\gamma \Delta\Delta$ coupling:
\begin{eqnarray}
&&\frac{324R^4M_N^2\,dF_1^{\Delta}}{C_{ND}^2}  =
60 - 20 R + 143 R^2 + 314 R^3 - 611 R^4 + 26 R^5 + 441 R^6\nn\\&& - 44 R^7 - 
 90 R^8 - 130 \mu^2 - 80 R \mu^2 - 562 R^2 \mu^2 - 
 438 R^3 \mu^2 - 778 R^4 \mu^2\nn\\&& + 168 R^5 \mu^2 + 
 270 R^6 \mu^2 - 55 \mu^4 - 20 R \mu^4 + 287 R^2 \mu^4 - 
 204 R^3 \mu^4 - 270 R^4 \mu^4\nn\\&& + 50 \mu^6 + 80 R \mu^6 + 
 90 R^2 \mu^6 + \mathcal{A} (70 + 60 R + 32 R^2 + 124 R^3\nn\\&& - 678 R^4 - 664 R^5 + 
    1136 R^6 + 432 R^7 - 1046 R^8 + 92 R^9 + 576 R^{10}\nn\\&& - 44 R^{11} - 
    90 R^{12} - 230 \mu^2 - 280 R \mu^2 - 418 R^2 \mu^2 - 
    108 R^3 \mu^2 + 424 R^4 \mu^2\nn\\&& - 8 R^5 \mu^2 + 
    888 R^6 \mu^2 - 676 R^7 \mu^2 - 1930 R^8 \mu^2 + 
    256 R^9 \mu^2 + 450 R^{10} \mu^2 + 220 \mu^4\nn\\&& + 
    400 R \mu^4 + 972 R^2 \mu^4 + 560 R^3 \mu^4 + 
    1016 R^4 \mu^4 + 936 R^5 \mu^4 + 2284 R^6 \mu^4 - 
    584 R^7 \mu^4\nn\\&& - 900 R^8 \mu^4 + 20 \mu^6 - 120 R \mu^6 - 
    728 R^2 \mu^6 - 212 R^3 \mu^6 - 1032 R^4 \mu^6 + 
    656 R^5 \mu^6\nn\\&& + 900 R^6 \mu^6 - 130 \mu^8 - 140 R \mu^8 + 
    52 R^2 \mu^8 - 364 R^3 \mu^8 - 450 R^4 \mu^8 + 
    50 \mu^{10}\nn\\&& + 80 R \mu^{10} + 90 R^2 \mu^{10}) + (-70 - 60 R - 
    102 R^2 - 184 R^3 + 576 R^4\nn\\&& - 480 R^5 + 560 R^6 - 48 R^7 - 
    486 R^8 + 44 R^9 + 90 R^{10} + 160 \mu^2\nn\\&& + 220 R \mu^2 + 
    336 R^2 \mu^2 + 24 R^3 \mu^2 + 144 R^4 \mu^2 + 
    504 R^5 \mu^2 + 1264 R^6 \mu^2 - 212 R^7 \mu^2\nn\\&& - 
    360 R^8 \mu^2 - 60 \mu^4 - 180 R \mu^4 - 516 R^2 \mu^4 - 
    396 R^3 \mu^4 - 1020 R^4 \mu^4 + 372 R^5 \mu^4\nn\\&& + 
    540 R^6 \mu^4 - 80 \mu^6 - 60 R \mu^6 + 192 R^2 \mu^6 - 
    284 R^3 \mu^6 - 360 R^4 \mu^6 + 50 \mu^8\nn\\&& + 80 R \mu^8 + 
    90 R^2 \mu^8) \ln R + (-70 - 60 R - 102 R^2 - 184 R^3 + 576 R^4\nn\\&& + 480 R^5 - 
    560 R^6 + 48 R^7 + 486 R^8 - 44 R^9 - 90 R^{10} + 160 \mu^2\nn\\&& + 
    220 R \mu^2 + 528 R^2 \mu^2 - 24 R^3 \mu^2 - 
    144 R^4 \mu^2 - 504 R^5 \mu^2 - 1264 R^6 \mu^2 + 
    212 R^7 \mu^2\nn\\&& + 360 R^8 \mu^2 + 60 \mu^4 + 180 R \mu^4 + 
    516 R^2 \mu^4 + 396 R^3 \mu^4 + 1020 R^4 \mu^4 - 
    372 R^5 \mu^4\nn\\&& - 540 R^6 \mu^4 + 80 \mu^6 + 60 R \mu^6 - 
    192 R^2 \mu^6 + 284 R^3 \mu^6 + 360 R^4 \mu^6\nn\\&& - 
    50 \mu^8 - 80 R \mu^8 - 90 R^2 \mu^8) \ln \mu
\end{eqnarray}
Additional contribution from the non-minimal $\gamma \Delta\Delta$ coupling:
\begin{eqnarray}
&&\frac{162R^5M_N^2\,dF_1^{\Delta}}{C_{ND}^2}  =
-40 - 100 R + 260 R^3 + 45 R^4 - 185 R^5 - 430 R^6 + 30 R^7\nn\\&& + 
 100 R^8 + 130 \mu^2 + 70 R \mu^2 - 30 R^2 \mu^2 + 
 230 R^3 \mu^2 + 240 R^4 \mu^2 - 190 R^5 \mu^2 \nn\\&&- 
 300 R^6 \mu^2 - 105 \mu^4 + 155 R \mu^4 + 160 R^2 \mu^4 + 
 290 R^3 \mu^4 + 300 R^4 \mu^4 + 30 \mu^6\nn\\&& - 130 R \mu^6 - 
 100 R^2 \mu^6 + 
 \mathcal{A} (-30 - 170 R - 120 R^2 + 650 R^3 + 820 R^4\nn\\&& - 960 R^5 - 
    1540 R^6 + 680 R^7 + 1350 R^8 - 230 R^9 - 580 R^10 + 30 R^11\nn\\&& + 
    100 R^12 + 150 \mu^2 + 550 R \mu^2 + 410 R^2 \mu^2 - 
    520 R^3 \mu^2 - 630 R^4 \mu^2 - 620 R^5 \mu^2\nn\\&& - 
    830 R^6 \mu^2 + 840 R^7 \mu^2 + 1400 R^8 \mu^2 - 
    250 R^9 \mu^2 - 500 R^10 \mu^2 - 300 \mu^4 - 
    500 R \mu^4\nn\\&& - 410 R^2 \mu^4 - 360 R^3 \mu^4 - 
    340 R^4 \mu^4 - 640 R^5 \mu^4 - 750 R^6 \mu^4 + 
    700 R^7 \mu^4 + 1000 R^8 \mu^4\nn\\&& + 300 \mu^6 - 
    100 R \mu^6 - 30 R^2 \mu^6 - 320 R^3 \mu^6 - 
    350 R^4 \mu^6 - 900 R^5 \mu^6 - 1000 R^6 \mu^6\nn\\&& - 
    150 \mu^8 + 350 R \mu^8 + 250 R^2 \mu^8 + 550 R^3 \mu^8 + 
    500 R^4 \mu^8 + 30 \mu^{10} - 130 R \mu^{10}\nn\\&& - 
    100 R^2 \mu^10) + (30 + 170 R + 150 R^2 - 480 R^3 - 670 R^4 - 
    480 R^5\nn\\&& - 870 R^6 + 200 R^7 + 480 R^8 - 30 R^9 - 100 R^10 - 
    120 \mu^2 - 380 R \mu^2\nn\\&& - 320 R^2 \mu^2 - 480 R^5 \mu^2 - 
    720 R^6 \mu^2 + 220 R^7 \mu^2 + 400 R^8 \mu^2 + 
    180 \mu^4 + 120 R \mu^4\nn\\&& + 90 R^2 \mu^4 + 60 R^3 \mu^4 + 
    30 R^4 \mu^4 - 480 R^5 \mu^4 - 600 R^6 \mu^4 - 
    120 \mu^6 + 220 R \mu^6\nn\\&& + 180 R^2 \mu^6 + 420 R^3 \mu^6 + 
    400 R^4 \mu^6 + 30 \mu^8 - 130 R \mu^8 - 
    100 R^2 \mu^8) \ln R\nn\\&& + (30 + 170 R + 150 R^2 - 480 R^3 - 670 R^4 + 480 R^5 + 
    870 R^6\nn\\&& - 200 R^7 - 480 R^8 + 30 R^9 + 100 R^10 - 120 \mu^2 - 
    380 R \mu^2 - 320 R^2 \mu^2\nn\\&& + 480 R^5 \mu^2 + 
    720 R^6 \mu^2 - 220 R^7 \mu^2 - 400 R^8 \mu^2 - 
    180 \mu^4 - 120 R \mu^4 - 90 R^2 \mu^4\nn\\&& - 60 R^3 \mu^4 - 
    30 R^4 \mu^4 + 480 R^5 \mu^4 + 600 R^6 \mu^4 + 
    120 \mu^6 - 220 R \mu^6 - 180 R^2 \mu^6\nn\\&& - 420 R^3 \mu^6 - 
    400 R^4 \mu^6 - 30 \mu^8 + 130 R \mu^8 + 
    100 R^2 \mu^8) \ln \mu
\end{eqnarray}

\subsection{Iso-vector anomalous magnetic moment $\kappa_V$}
Contributions from virtual nucleons:
\begin{eqnarray}
&&\frac{F_2^{N}}{4 C_{NN}^2}  =
\frac{1}{\sqrt{-4 + \mu^2}} ((5 - 6 \mu^2) \sqrt{-4 + \mu^2} + 
   2 \mu^2 \sqrt{-4 + \mu^2} (-7 + 
      3 \mu^2) \ln \mu\nn\\&& + \mu (8 - 13 \mu^2 + 
      3 \mu^4) \ln \frac{\mu - \sqrt{-4 + \mu^2}}{\mu + 
      \sqrt{-4 + \mu^2}})
\end{eqnarray}
Contributions from virtual $\Delta(1232)$ with minimal $\gamma \Delta\Delta$ coupling:
\begin{eqnarray}
&&\frac{81R^4\,F_2^\Delta}{ C_{N\Delta}^2} = 
-40 - 100 R - 153 R^2 + 220 R^3 + 697 R^4 + 88 R^5 - 191 R^6\nn\\&& + 
 40 R^7 + 54 R^8 + 150 \mu^2 + 140 R \mu^2 + 522 R^2 \mu^2 + 
 356 R^3 \mu^2 + 230 R^4 \mu^2\nn\\&& - 120 R^5 \mu^2 - 
 162 R^6 \mu^2 - 35 \mu^4 + 60 R \mu^4 - 49 R^2 \mu^4 + 
 120 R^3 \mu^4 + 162 R^4 \mu^4\nn\\&& + 10 \mu^6 - 40 R \mu^6 - 
 54 R^2 \mu^6 + 
 \mathcal{A} (-10 - 40 R - 64 R^2 + 188 R^3 + 434 R^4\nn\\&& - 292 R^5 - 
    800 R^6 + 220 R^7 + 658 R^8 - 116 R^9 - 272 R^{10} + 40 R^{11} + 
    54 R^{12}\nn\\&& + 50 \mu^2 + 120 R \mu^2 + 342 R^2 \mu^2 - 
    4 R^3 \mu^2 - 552 R^4 \mu^2 - 648 R^5 \mu^2 - 
    632 R^6 \mu^2\nn\\&& + 444 R^7 \mu^2 + 774 R^8 \mu^2 - 
    200 R^9 \mu^2 - 270 R^{10} \mu^2 - 100 \mu^4 - 80 R \mu^4 - 
    588 R^2 \mu^4\nn\\&& - 356 R^3 \mu^4 - 328 R^4 \mu^4 - 
    420 R^5 \mu^4 - 700 R^6 \mu^4 + 400 R^7 \mu^4 + 
    540 R^8 \mu^4 + 100 \mu^6\nn\\&& - 80 R \mu^6 + 352 R^2 \mu^6 - 
    28 R^3 \mu^6 + 176 R^4 \mu^6 - 400 R^5 \mu^6 - 
    540 R^6 \mu^6 - 50 \mu^8\nn\\&& + 120 R \mu^8 + 12 R^2 \mu^8 + 
    200 R^3 \mu^8 + 270 R^4 \mu^8 + 10 \mu^{10} - 40 R \mu^{10} - 
    54 R^2 \mu^{10})\nn\\&& + (10 + 40 R + 74 R^2 - 148 R^3 - 360 R^4 - 
    576 R^5 - 440 R^6\nn\\&& + 76 R^7 + 218 R^8 - 40 R^9 - 54 R^{10} - 
    40 \mu^2 - 80 R \mu^2 - 288 R^2 \mu^2\nn\\&& - 144 R^3 \mu^2 + 
    72 R^4 \mu^2 - 288 R^5 \mu^2 - 448 R^6 \mu^2 + 
    160 R^7 \mu^2 + 216 R^8 \mu^2 + 60 \mu^4\nn\\&& + 
    300 R^2 \mu^4 + 132 R^3 \mu^4 + 252 R^4 \mu^4 - 
    240 R^5 \mu^4 - 324 R^6 \mu^4 - 40 \mu^6 + 80 R \mu^6\nn\\&& - 
    32 R^2 \mu^6 + 160 R^3 \mu^6 + 216 R^4 \mu^6 + 10 \mu^8 - 
    40 R \mu^8 - 54 R^2 \mu^8) \ln R \nn\\&&+ (10 + 40 R + 74 R^2 - 148 R^3 - 360 R^4 + 144 R^5 + 440 R^6\nn\\&& - 
    76 R^7 - 218 R^8 + 40 R^9 + 54 R^{10} - 40 \mu^2 - 80 R \mu^2 - 
    432 R^2 \mu^2\nn\\&& - 288 R^3 \mu^2 - 72 R^4 \mu^2 + 
    288 R^5 \mu^2 + 448 R^6 \mu^2 - 160 R^7 \mu^2 - 
    216 R^8 \mu^2 - 60 \mu^4\nn\\&& - 300 R^2 \mu^4 - 
    132 R^3 \mu^4 - 252 R^4 \mu^4 + 240 R^5 \mu^4 + 
    324 R^6 \mu^4 + 40 \mu^6 - 80 R \mu^6\nn\\&& + 32 R^2 \mu^6 - 
    160 R^3 \mu^6 - 216 R^4 \mu^6 - 10 \mu^8 + 40 R \mu^8 + 
    54 R^2 \mu^8) \ln \mu
\end{eqnarray}
Additional contribution from the non-minimal $\gamma \Delta\Delta$ coupling:
\begin{eqnarray}
&&\frac{81R^4\,F_2^\Delta}{ C_{N\Delta}^2} = 
40 + 180 R + 280 R^2 + 340 R^3 + 140 R^4 + 80 R^5 - 160 R^6\nn\\&& - 
 100 R^7 - 120 \mu^2 - 100 R \mu^2 + 120 R^2 \mu^2 + 
 40 R^3 \mu^2 + 280 R^4 \mu^2 + 300 R^5 \mu^2\nn\\&& + 140 \mu^4 - 
 120 R \mu^4 - 80 R^2 \mu^4 - 300 R^3 \mu^4 - 40 \mu^6 + 
 100 R \mu^6\nn\\&& + 
 \mathcal{A} (40 + 130 R - 20 R^2 - 320 R^3 - 20 R^4 + 280 R^5 - 
    220 R^6\nn\\&& - 220 R^7 + 380 R^8 + 230 R^9 - 160 R^{10} - 100 R^{11} - 
    200 \mu^2 - 420 R \mu^2\nn\\&& + 60 R^2 \mu^2 + 220 R^3 \mu^2 - 
    120 R^4 \mu^2 + 120 R^5 \mu^2 - 340 R^6 \mu^2 - 
    420 R^7 \mu^2 + 600 R^8 \mu^2\nn\\&& + 500 R^9 \mu^2 + 
    400 \mu^4 + 380 R \mu^4 - 60 R^2 \mu^4 + 20 R^3 \mu^4 - 
    260 R^4 \mu^4 - 120 R^5 \mu^4\nn\\&& - 800 R^6 \mu^4 - 
    1000 R^7 \mu^4 - 400 \mu^6 + 80 R \mu^6 + 20 R^2 \mu^6 + 
    580 R^3 \mu^6 + 400 R^4 \mu^6\nn\\&& + 1000 R^5 \mu^6 + 
    200 \mu^8 - 270 R \mu^8 - 500 R^3 \mu^8 - 40 \mu^{10} + 
    100 R \mu^{10})\nn\\&& + (-40 - 130 R - 20 R^2 + 190 R^3 + 90 R^5 - 
    220 R^6 - 130 R^7\nn\\&& + 160 R^8 + 100 R^9 + 160 \mu^2 + 
    290 R \mu^2 + 90 R^5 \mu^2 - 440 R^6 \mu^2 - 
    400 R^7 \mu^2\nn\\&& - 240 \mu^4 - 90 R \mu^4 + 60 R^2 \mu^4 + 
    210 R^3 \mu^4 + 360 R^4 \mu^4 + 600 R^5 \mu^4 + 
    160 \mu^6\nn\\&& - 170 R \mu^6 - 40 R^2 \mu^6 - 400 R^3 \mu^6 - 
    40 \mu^8 + 100 R \mu^8) \ln R + (-40 - 130 R\nn\\&& - 20 R^2 + 190 R^3 - 90 R^5 + 220 R^6 + 
    130 R^7 - 160 R^8 - 100 R^9\nn\\&& + 160 \mu^2 + 290 R \mu^2 - 
    90 R^5 \mu^2 + 440 R^6 \mu^2 + 400 R^7 \mu^2 + 
    240 \mu^4 + 90 R \mu^4\nn\\&& - 60 R^2 \mu^4 - 210 R^3 \mu^4 - 
    360 R^4 \mu^4 - 600 R^5 \mu^4 - 160 \mu^6 + 170 R \mu^6 + 
    40 R^2 \mu^6\nn\\&& + 400 R^3 \mu^6 + 40 \mu^8 - 
    100 R \mu^8) \ln \mu
\end{eqnarray}

\subsection{Iso-vector Pauli radius $\langle r_{2}^2 \rangle_V$}
Contributions from virtual nucleons:
\begin{eqnarray}
&&dF_2^{N}  =
\frac{2 C_{NN}^2}{3 M_N^2 \mu (-4 + \mu^2)^{3/2}} (\mu \sqrt{-4 + \mu^2} (-124 + 105 \mu^2 - 
      18 \mu^4)\nn\\&& + 
   6 \mu \sqrt{-4 + \mu^2} (-16 + 44 \mu^2 - 22 \mu^4 + 
      3 \mu^6) \ln \mu \nn\\&& + (16 - 216 \mu^2 + 246 \mu^4 - 
      84 \mu^6 + 9 \mu^8) \ln \frac{\mu - 
      \sqrt{-4 + \mu^2}}{\mu + 
      \sqrt{-4 + \mu^2}}  )
\end{eqnarray}
Contributions from virtual $\Delta(1232)$ with minimal $\gamma \Delta\Delta$ coupling::
\begin{eqnarray}
&&\frac{486R^4M_N^2\,dF_2^{\Delta}}{C_{ND}^2}  = 
-90 + 280 R - 171 R^2 - 1106 R^3 + 108 R^4 + 186 R^5 - 675 R^6\nn\\&& + 
 132 R^7 + 162 R^8 - 340 R \mu^2 + 198 R^2 \mu^2 + 
 162 R^3 \mu^2 + 1116 R^4 \mu^2 - 504 R^5 \mu^2\nn\\&& - 
 486 R^6 \mu^2 + 270 \mu^4 + 300 R \mu^4 - 261 R^2 \mu^4 + 
 612 R^3 \mu^4 + 486 R^4 \mu^4 - 180 \mu^6\nn\\&& - 240 R \mu^6 - 
 162 R^2 \mu^6 + 
 \mathcal{A} (300 R - 126 R^2 - 1500 R^3 + 378 R^4 + 2712 R^5\nn\\&& - 
    540 R^6 - 1488 R^7 + 1188 R^8 - 12 R^9 - 918 R^{10} + 132 R^{11} + 
    162 R^{12}\nn\\&& - 600 R \mu^2 + 342 R^2 \mu^2 + 948 R^3 \mu^2 + 
    576 R^4 \mu^2 + 576 R^5 \mu^2 - 1044 R^6 \mu^2 + 
    348 R^7 \mu^2\nn\\&& + 2952 R^8 \mu^2 - 768 R^9 \mu^2 - 
    810 R^{10} \mu^2 + 180 \mu^4 + 480 R \mu^4 - 
    648 R^2 \mu^4 + 216 R^3 \mu^4\nn\\&& - 936 R^4 \mu^4 - 
    3168 R^6 \mu^4 + 1752 R^7 \mu^4 + 1620 R^8 \mu^4 - 
    540 \mu^6 - 600 R \mu^6 + 252 R^2 \mu^6\nn\\&& - 996 R^3 \mu^6 + 
    972 R^4 \mu^6 - 1968 R^5 \mu^6 - 1620 R^6 \mu^6 + 
    540 \mu^8 + 660 R \mu^8 + 342 R^2 \mu^8\nn\\&& + 
    1092 R^3 \mu^8 + 810 R^4 \mu^8 - 180 \mu^10 - 
    240 R \mu^{10} - 162 R^2 \mu^{10}) + (-300 R + 126 R^2\nn\\&& + 
    1440 R^3 + 108 R^4 + 1368 R^5 - 432 R^6 - 120 R^7 + 756 R^8 - 
    132 R^9\nn\\&& - 162 R^{10} + 300 R \mu^2 - 216 R^2 \mu^2 + 
    192 R^3 \mu^2 - 72 R^4 \mu^2 - 96 R^5 \mu^2 - 
    1872 R^6 \mu^2\nn\\&& + 636 R^7 \mu^2 + 648 R^8 \mu^2 - 
    180 \mu^4 - 180 R \mu^4 + 252 R^2 \mu^4 - 204 R^3 \mu^4 + 
    1296 R^4 \mu^4\nn\\&& - 1116 R^5 \mu^4 - 972 R^6 \mu^4 + 
    360 \mu^6 + 420 R \mu^6 + 852 R^3 \mu^6 + 648 R^4 \mu^6 - 
    180 \mu^8 \nn\\&&- 240 R \mu^8 - 162 R^2 \mu^8) \ln R + (-300 R + 270 R^2 + 1344 R^3 - 108 R^4 - 1368 R^5\nn\\&& + 432 R^6 + 
    120 R^7 - 756 R^8 + 132 R^9 + 162 R^{10} + 300 R \mu^2 + 
    216 R^2 \mu^2\nn\\&& - 192 R^3 \mu^2 + 72 R^4 \mu^2 + 
    96 R^5 \mu^2 + 1872 R^6 \mu^2 - 636 R^7 \mu^2 - 
    648 R^8 \mu^2 + 180 \mu^4\nn\\&& + 180 R \mu^4 - 252 R^2 \mu^4 + 
    204 R^3 \mu^4 - 1296 R^4 \mu^4 + 1116 R^5 \mu^4 + 
    972 R^6 \mu^4 - 360 \mu^6\nn\\&& - 420 R \mu^6 - 852 R^3 \mu^6 - 
    648 R^4 \mu^6 + 180 \mu^8 + 240 R \mu^8 + 
    162 R^2 \mu^8) \ln \mu
\end{eqnarray}
Additional contribution from the non-minimal $\gamma \Delta\Delta$ coupling:
\begin{eqnarray}
&&\frac{243R^5M_N^2\,dF_2^{\Delta}}{C_{ND}^2}  = 
120 + 60 R + 25 R^2 + 390 R^3 - 125 R^4 + 420 R^6 - 120 R^8\nn\\&&  - 
 270 \mu^2 + 60 R \mu^2 + 65 R^2 \mu^2 - 720 R^3 \mu^2 - 
 270 R^4 \mu^2 + 360 R^5 \mu^2 + 360 R^6 \mu^2\nn\\&&  + 90 \mu^4 - 
 540 R \mu^4 - 150 R^2 \mu^4 - 720 R^3 \mu^4 - 
 360 R^4 \mu^4 + 360 R \mu^6 + 120 R^2 \mu^6\nn\\&&  + 
 \mathcal{A} (90 + 180 R - 60 R^2 - 540 R^3 - 600 R^4 + 540 R^5 + 
    1380 R^6\nn\\&&  - 180 R^7 - 1290 R^8 + 600 R^10 - 120 R^12 - 
    360 \mu^2 - 540 R \mu^2\nn\\&&  - 60 R^2 \mu^2 + 360 R^3 \mu^2 + 
    750 R^4 \mu^2 + 1080 R^5 \mu^2 + 540 R^6 \mu^2 - 
    1260 R^7 \mu^2 - 1470 R^8 \mu^2\nn\\&&  + 360 R^9 \mu^2 + 
    600 R^10 \mu^2 + 540 \mu^4 + 180 R \mu^4 + 
    210 R^2 \mu^4 + 720 R^3 \mu^4 + 540 R^4 \mu^4\nn\\&&  + 
    1440 R^5 \mu^4 + 810 R^6 \mu^4 - 1440 R^7 \mu^4 - 
    1200 R^8 \mu^4 - 360 \mu^6 + 900 R \mu^6 + 
    120 R^2 \mu^6\nn\\&&  + 900 R^3 \mu^6 + 390 R^4 \mu^6 + 
    2160 R^5 \mu^6 + 1200 R^6 \mu^6 + 90 \mu^8 - 
    1080 R \mu^8 - 330 R^2 \mu^8\nn\\&&  - 1440 R^3 \mu^8 - 
    600 R^4 \mu^8 + 360 R \mu^{10} + 120 R^2 \mu^{10}) + (-90 - 
    180 R - 30 R^2\nn\\&&  + 360 R^3 + 1170 R^4 + 180 R^5 + 810 R^6 - 
    480 R^8 + 120 R^10 + 270 \mu^2\nn\\&&  + 360 R \mu^2 + 
    120 R^2 \mu^2 + 300 R^4 \mu^2 + 900 R^5 \mu^2 + 
    750 R^6 \mu^2 - 360 R^7 \mu^2 - 480 R^8 \mu^2\nn\\&&  - 
    270 \mu^4 + 180 R \mu^4 - 180 R^3 \mu^4 - 60 R^4 \mu^4 + 
    1080 R^5 \mu^4 + 720 R^6 \mu^4 + 90 \mu^6\nn\\&&  - 720 R \mu^6 - 
    210 R^2 \mu^6 - 1080 R^3 \mu^6 - 480 R^4 \mu^6 + 
    360 R \mu^8 + 120 R^2 \mu^8) \ln R\nn\\&&  + (-90 - 180 R - 30 R^2 + 360 R^3 + 570 R^4 - 180 R^5 - 
    810 R^6\nn\\&&  + 480 R^8 - 120 R^10 + 270 \mu^2 + 360 R \mu^2 + 
    120 R^2 \mu^2 - 300 R^4 \mu^2 - 900 R^5 \mu^2\nn\\&&  - 
    750 R^6 \mu^2 + 360 R^7 \mu^2 + 480 R^8 \mu^2 + 
    270 \mu^4 - 180 R \mu^4 + 180 R^3 \mu^4 + 60 R^4 \mu^4\nn\\&&  - 
    1080 R^5 \mu^4 - 720 R^6 \mu^4 - 90 \mu^6 + 720 R \mu^6 + 
    210 R^2 \mu^6 + 1080 R^3 \mu^6\nn\\&&  + 480 R^4 \mu^6 - 
    360 R \mu^8 - 120 R^2 \mu^8) \ln \mu
\end{eqnarray}

\subsection{Renormalized constants}
The constants are with $\tilde{\delta}=\delta^2R^2(R+1)^2$:

\begin{eqnarray}
c_{r} \cdot(f_\pi8\pi)^2 &=&  -\frac{43 g_A^2}{3} +\frac{h_A^2}{324R^5} (60 R - 20 R^2 + 143 R^3 + 314 R^4- 611 R^5 + 26 R^6\nn\\&&  + 
   441 R^7 - 44 R^8 -    90 R^9 + (12 R^3 + 24 R^4 - 96 R^5 - 960 R^6\nn\\&&  + 1120 R^7 - 96 R^8 - 
      972 R^9 + 88 R^{10} + 180 R^{11}) \ln R \nn\\&&  + (1 + R)^2 (-35 R + 40 R^2 - 96 R^3 + 60 R^4 + 264 R^5 - 
      348 R^6 + 152 R^7\nn\\&&  + 68 R^8 - 
      45 R^9) \ln \tilde{\delta} + \frac{h_A^2\kappa_{nm}}{324R^5} (-80 - 200 R + 520 R^3\nn\\&&  + 
      90 R^4 - 370 R^5 - 860 R^6 + 60 R^7 + 
      200 R^8 + (-1920 R^5 - 3480 R^6\nn\\&& + 800 R^7 + 1920 R^8 - 
         120 R^9 - 400 R^{10}) \ln R + (1 + R)^2 (30 + 110 R\nn\\&& - 100 R^2 - 390 R^3 + 210 R^4 + 
         450 R^5 - 240 R^6\nn\\&& - 170 R^7 + 100 R^8) \ln \tilde{\delta} \,\,\,,
\end{eqnarray}

\begin{eqnarray}
c_{\kappa} \cdot\frac{(f_\pi8\pi)^2}{M_N^2} &=& 20 g_A^2 + \frac{h_A^2}{81R^4} (-40 - 100 R - 153 R^2 + 220 R^3\nn\\&& + 697 R^4 + 88 R^5 - 191 R^6 + 
   40 R^7 + 
   54 R^8\nn\\&& + (-720 R^5 - 880 R^6 + 152 R^7 + 436 R^8 - 80 R^9 - 
      108 R^{10}) \ln R\nn\\&& + \delta R (1 + R)^3 (-5  - 10 R  - 
      17  R^2  + 98 R^3  -       41  R^4 - 34  R^5\nn\\&&  + 
      27  R^6) \ln\tilde{\delta} + \frac{h_A^2\kappa_{nm}}{81R^4} (40 + 
      180 R + 280 R^2 + 340 R^3 + 140 R^4\nn\\&& + 80 R^5 - 160 R^6 - 
      100 R^7 + (180 R^5 - 440 R^6 - 260 R^7 + 320 R^8 \nn\\&& + 200 R^9) \ln
      R + \delta R(1 + R)^3 (20  + 25 R  -  40 R^2  + 25  R^3 \nn\\&& +  20
      R^4 - 50 R^5) \ln \tilde{\delta} )\,\,\,.
\end{eqnarray}
%
%
%
%

\section{$\Delta(1232)$ electromagnetic form factors \label{APP:DeltaFF}}

For the nucleon electromagnetic form factors we take the mass scale
$M_{sc}=M_{\Delta}$.

The following functions occur in the Feynman-graphs of Fig. \ref{fig:diagramsDelta}:\begin{eqnarray}
\tilde{\mathcal{M}}_{1} & = & z\mu^{2}+\left(1-z\right)^{2}\nonumber\,\,\,, \\
\tilde{\mathcal{M}}_{2} & = & \left(1-z\right)\mu^{2}+z^{2}\nonumber\,\,\,, \\
\tilde{\mathcal{M}}_{3} & = & z\mu^{2}+\left(1-z\right)r^{2}-z\left(1-z\right)\nonumber\,\,\,, \\
\tilde{\mathcal{M}}_{4} & = & \left(1-z\right)\mu^{2}+zr^{2}-z\left(1-z\right)\,\,\,,\nn\\
\tilde{\mathcal{M}}_{56} & = &z\mu^{2}+\left(1-z\right)^{2} \,\,\,.
\end{eqnarray}

For better reading we introduce the functions:

\begin{equation}
\tilde{J}_{1}\left(\tilde{\mathcal{M}}\right)  =  \tilde{\mathcal{M}}\left[L-1+\ln\tilde{\mathcal{M}}\right] \,\,\,,\,\,\,
\tilde{J}_{2}\left(\tilde{\mathcal{M}}\right)  =  \left[L+\ln\tilde{\mathcal{M}}\right] \,\,\,,\,\,\,
\tilde{J}_{3}\left(\tilde{\mathcal{M}}\right)  =   \frac{1}{\tilde{\mathcal{M}}}\,\,\,.
\end{equation}

The iso-spin factors are given by:

\begin{eqnarray}
I_{1}=\left(\begin{array}{c}
\frac{2}{3}\\
\frac{2}{9}\\
-\frac{2}{9}\\
-\frac{2}{3}\end{array}\right)\,\,\,\,\,\,\,\,\, & I_{2}=\left(\begin{array}{c}
\frac{8}{3}\\
\frac{13}{9}\\
\frac{2}{9}\\
-1\end{array}\right) & \,\,\,\,\,\,\,\,\, I_{3}=\left(\begin{array}{c}
1\\
\frac{1}{3}\\
-\frac{1}{3}\\
-1\end{array}\right)\nonumber \\
I_{4}=\left(\begin{array}{c}
1\\
\frac{2}{3}\\
\frac{1}{3}\\
0\end{array}\right)\,\,\,\,\,\,\,\,\, & I_{56}^{a}=\left(\begin{array}{c}
\frac{2}{3}\\
\frac{2}{9}\\
-\frac{2}{9}\\
-\frac{2}{3}\end{array}\right) & \,\,\,\,\,\,\,\,\, I_{56}^{b}=\left(\begin{array}{c}
\frac{8}{3}\\
\frac{13}{9}\\
\frac{2}{9}\\
-1\end{array}\right)\,\,\,\,\,\,\,\,\, I_{T}=\left(\begin{array}{c}
\frac{2}{3}\\
\frac{2}{9}\\
-\frac{2}{9}\\
-\frac{2}{3}\end{array}\right)\end{eqnarray}
In order to project on the individual Lorentz-structures we use the
following identieis: \begin{eqnarray}
\overline{u}_{\alpha}(p^{\prime})\gamma^{\mu}u^{\alpha}(p) & = & \frac{1}{2M_{\Delta}}\overline{u}_{\alpha}(p^{\prime})\left[n^{\mu}-\gamma^{\mu\nu}q_{\nu}\right]u^{\alpha}(p)\\
\overline{u}_{\alpha}(p^{\prime})\left[q^{\alpha}g^{\mu\beta}-q^{\beta}g^{\mu\alpha}\right]u_{\beta}(p) & = & \overline{u}_{\alpha}(p^{\prime})\left[2M_{\Delta}\left(1+\tau\right)g^{\alpha\beta}\gamma^{\mu}-g^{\alpha\beta}n^{\mu}+\frac{1}{M_{\Delta}}q^{\alpha}q^{\beta}\gamma^{\mu}\right]u_{\beta}(p)\,\,\,,\nonumber \end{eqnarray}
with $n=p^{\prime}+p$. The first one is the Gordon-identity for $\Delta$
spinors and the second a $\Delta$ spinor identity given, e.g. in \cite{Nozawa(1990):DeltaMATELEM}.

In the following subsections we give the individual contributions
from the Feynman-diagrams of Fig. \ref{fig:diagramsDelta} to the
$\Delta(1232)$-isobar form factors. We only list non-vanishing
contributions. These are the non-renormalized expressions.

\subsection{Contributions to $F_{2}^{\Delta}(0)$}

\begin{eqnarray}
F_{2}^{\Delta 1}(0) & = & C_{\Delta\Delta}\,\, I_{1}\,\,\int_{0}^{1}dz\,\,2z\,\left[\frac{5}{6}-\frac{1}{3}z-\frac{7}{6}z^{2}-\epsilon_{d}\left(\frac{13}{9}-\frac{4}{9}z-\frac{17}{9}z^{2}\right)\right]\,\,\tilde{J}_{2}\left(\tilde{\mathcal{M}}_{1}\right)\nonumber \\
F_{2}^{\Delta 2}(0) & = & C_{\Delta\Delta}\,\, I_{2}\,\,\int_{0}^{1}dz\,\,2z\,\,\Big\{-\left[-\frac{1}{9}+\frac{5}{9}z+\epsilon_{d}\left(\frac{11}{27}-\frac{199}{108}z\right)\right]\tilde{J}_{1}\left(\tilde{\mathcal{M}}_{2}\right)\nonumber \\
 &  & -\left[-\frac{4}{9}+3z-\frac{5}{3}z^{2}+\frac{5}{36}z^{3}+\epsilon_{d}\left(\frac{32}{27}-\frac{50}{9}z+\frac{10}{3}z^{2}-\frac{23}{54}z^{3}\right)\right]\tilde{J}_{2}\left(\tilde{\mathcal{M}}_{2}\right)\Big\}\nonumber \\
F_{2}^{\Delta 3}(0) & = & C_{N\Delta}\,\, I_{3}\,\,\int_{0}^{1}dz\,\,2z\,\,\Big\{\,\,-\left[-\frac{1}{2}z+z^{2}-\frac{1}{2}r+zr\right]\,\,\,\tilde{J}_{2}\left(\tilde{\mathcal{M}}_{3}\right)\,\,\Big\}\nonumber \\
F_{2}^{\Delta 4}(0) & = & C_{N\Delta}\,\, I_{4}\,\,\int_{0}^{1}dz\,\,2z\,\,\Big\{-\left[z-z^{2}+rz\right]\,\,\,\tilde{J}_{2}\left(\tilde{\mathcal{M}}_{4}\right)\,\,\Big\}\end{eqnarray}
Non-minimal contribution to $D2$ with $\kappa_{1}=\kappa_{2}=\kappa_{nm}$
(without the $\gamma^{\beta\delta\gamma}$ part) :\begin{eqnarray}
F_{2}^{\Delta 2}(0) & = & \kappa_{nm}C_{\Delta\text{\ensuremath{\Delta}}}\, I_{2}\,\,\int_{0}^{1}dz\,\,2z\,\,\Big\{-\left[\frac{47}{18}-\frac{8}{9}z+\epsilon_{d}\left(-\frac{271}{54}+\frac{70}{27}z\right)\right]\tilde{J}_{1}\left(\tilde{\mathcal{M}}_{2}\right)\\
 &  & -\left[\frac{22}{9}-\frac{10}{3}z+\frac{3}{2}z^{2}-\frac{2}{9}z^{3}+\epsilon_{d}\left(-\frac{188}{27}+\frac{28}{3}z-\frac{37}{9}z^{2}+\frac{16}{27}z^{3}\right)\right]\tilde{J}_{2}\left(\tilde{\mathcal{M}}_{2}\right)\Big\}\nonumber \end{eqnarray}

\subsection{Contributions to $F_{3}^{\Delta}(0)$}

\begin{eqnarray}
F_{3}^{\Delta 1}(0) & = & C_{\Delta\Delta}\,\, I_{1}\,\,\int_{0}^{1}dz\,\,2z\,\Big\{\left[3-2\epsilon_{d}\right]\,\,\tilde{J}_{1}\left(\tilde{\mathcal{M}_{1}}\right)\nonumber \\
 &  & -\left[4+\frac{2}{3}z-\frac{70}{9}z^{2}+\epsilon_{d}\left(-\frac{44}{9}z+\frac{76}{27}z^{2}\right)\right]\,\,\tilde{J}_{2}\left(\tilde{\mathcal{M}_{1}}\right)\nonumber \\
 &  & -\frac{8}{9}z^{2}\left[1-z^{2}\right]\,\,\tilde{J}_{3}\left(\tilde{\mathcal{M}_{1}}\right)\Big\}\nonumber \\
F_{3}^{\Delta 2}(0) & = & C_{\Delta\Delta}\,\, I_{2}\,\,\int_{0}^{1}dz\,\,2z\Big\{-\left[-\frac{5}{3}+\frac{2}{9}z+\epsilon_{d}\left(-\frac{10}{3}-\frac{59}{54}z\right)\right]\,\,\,\tilde{J}_{1}\left(\tilde{\mathcal{M}_{2}}\right)\nonumber \\
 &  & -\left[\frac{16}{3}-\frac{104}{9}z+5z^{2}+\frac{1}{18}z^{3}+\epsilon_{d}\left(\frac{32}{9}+\frac{88}{27}z-\frac{174}{27}z^{2}-\frac{7}{27}z^{3}\right)\right]\tilde{J}_{2}\left(\tilde{\mathcal{M}_{2}}\right)\nonumber \\
 &  & -\left[\frac{16}{9}z^{2}-\frac{16}{9}z^{3}+\frac{4}{9}z^{4}\right]\,\,\,\tilde{J}_{3}\left(\tilde{\mathcal{M}_{1}}\right)\,\,\Big\}\nonumber \\
F_{3}^{\Delta 3}(0) & = & C_{N\Delta}\,\, I_{3}\,\,\int_{0}^{1}dz\,\,2z\,\,\Big\{\,\,-\left[\frac{4}{3}z^{2}+rz\right]\,\,\,\tilde{J}_{2}\left(\tilde{\mathcal{M}_{3}}\right)\nonumber \\
 &  & +\left[\frac{2}{3}z^{3}-\frac{2}{3}z^{4}+\frac{2}{3}rz^{2}-\frac{2}{3}rz^{3}\right]\,\,\,\tilde{J}_{3}\left(\tilde{\mathcal{M}_{3}}\right)\,\,\Big\}\nonumber \\
F_{3}^{\Delta 4}(0) & = & C_{N\Delta}\,\, I_{4}\,\,\int_{0}^{1}dz\,\,2z\,\,\Big\{-\left[z+rz-\epsilon_{d}\frac{1}{3}z^{2}\right]\,\,\,\tilde{J}_{2}\left(\tilde{\mathcal{M}_{3}}\right)\,\,\nonumber \\
 &  & -\left[-\frac{1}{3}z^{2}+\frac{2}{3}z^{3}-\frac{1}{3}z^{4}-\frac{2}{3}rz^{2}-\frac{1}{3}r^{2}z^{2}+\frac{2}{3}rz^{3}\right]\,\,\,\tilde{J}_{3}\left(\tilde{\mathcal{M}_{4}}\right)\,\,\,\Big\}\nonumber \\
F_{3}^{\Delta 56}(0) & = & C_{\Delta\Delta}\,\, I_{56}^{a}\,\,\int_{0}^{1}dx\,\,\left[4+6z-\epsilon_{d}4z\right]\,\,\tilde{J}_{1}\left(\tilde{\mathcal{M}}_{56}\right)\nonumber \\
 &  & +C_{\Delta\Delta}\,\, I_{56}^{b}\,\,\int_{0}^{1}dx\,\,\frac{1}{9}\left[12+48z+\epsilon_{d}\left(44-28z\right)\right]\,\,\tilde{J}_{1}\left(\tilde{\mathcal{M}}_{56}\right)\end{eqnarray}
Non-minimal contribution to $D2$ with $\kappa_{1}=\kappa_{2}=\kappa_{nm}$
(without the $\gamma^{\beta\delta\gamma}$ part) :

\begin{eqnarray}
F_{3}^{\Delta 2}(0) & = & \kappa_{nm}C_{\Delta\Delta}I_{2}\int_{0}^{1}dz\,\,2z\Big\{\left[4-\frac{28}{9}z-\epsilon_{d}\left(-\frac{118}{9}-\frac{269}{27}z\right)\right]\tilde{J}_{1}\left(\tilde{\mathcal{M}_{2}}\right) \\
 &  &
-\left[\frac{8}{3}+\frac{8}{9}z-\frac{8}{3}z^{2}+\frac{7}{9}z^{3}+\epsilon_{d}\left(-\frac{272}{9}+\frac{704}{27}z-\frac{8}{9}z^{2}-\frac{62}{27}z^{3}\right)\right]\tilde{J}_{2}\left(\tilde{\mathcal{M}_{2}}\right)\Big\}\nn
\end{eqnarray}

\subsection{Contributions to $F_{4}^{\Delta}(0)$}

\begin{eqnarray}
F_{4}^{\Delta 1}(0) & = & C_{\Delta\Delta}\,\, I_{1}\,\,\int_{0}^{1}dz\,\,2z\,\,\Big\{\left[4-8z^{2}-\epsilon_{d}4z\left(1-z\right)\right]\,\,\,\tilde{J}_{2}\left(\tilde{\mathcal{M}}_{1}\right)\nonumber \\
 &  & +\frac{8}{9}z^{2}\left[1-z^{2}\right]\,\,\,\,\tilde{J}_{3}\left(\tilde{\mathcal{M}}_{1}\right)\,\,\,\Big\}\nonumber \\
F_{4}^{\Delta 2}(0) & = & C_{\Delta\Delta}I_{2}\int_{0}^{1}dz\,\,2z\Big\{-\left[\frac{128}{9}z-8z^{2}-\frac{10}{27}z^{3}+\epsilon_{d}\left(-\frac{64}{27}z+\frac{104}{27}z^{2}+\frac{86}{81}z^{3}\right)\right]\tilde{J}_{2}\left(\tilde{\mathcal{M}}_{2}\right)\nonumber \\
 &  & -\left[-\frac{32}{27}z^{2}+\frac{16}{9}z^{3}-\frac{4}{9}z^{4}-\frac{2}{27}z^{5}\right]\,\,\tilde{J}_{3}\left(\tilde{\mathcal{M}}_{2}\right)\,\,\,\Big\}\nonumber \\
F_{4}^{\Delta 3}(0) & = & C_{N\Delta}\,\, I_{3}\,\,\int_{0}^{1}dz\,\,2z\,\,\Big\{\,\,-\frac{2}{3}\left[z^{3}-z^{4}+rz^{2}-rz^{3}\right]\,\,\,\tilde{J}_{3}\left(\tilde{\mathcal{M}}_{3}\right)\Big\}\nonumber \\
F_{4}^{\Delta 4}(0) & = & C_{N\Delta}\,\, I_{4}\,\,\int_{0}^{1}dz\,\,2z\,\,\Big\{-\frac{2}{3}\left[-z^{3}+z^{4}-rz^{3}\right]\,\,\,\tilde{J}_{3}\left(\tilde{\mathcal{M}}_{4}\right)\,\,\Big\}\end{eqnarray}
Non-minimal contribution to $D2$ with $\kappa_{1}=\kappa_{2}=\kappa_{nm}$
(without the $\gamma^{\beta\delta\gamma}$ part) :\begin{eqnarray}
\frac{1}{\text{\ensuremath{\kappa}}}F_{4}^{\Delta 2}(0) & = & C_{\Delta\Delta}I_{2}\int_{0}^{1}dz\,\,2z\Big\{-\left[16-8z+\epsilon_{d}\left(-\frac{16}{3}+\frac{38}{3}z\right)\right]\tilde{J}_{1}\left(\tilde{\mathcal{M}}_{2}\right)\nonumber \\
 &  & -\left[-\frac{152}{9}z+\frac{112}{9}z^{2}-\frac{74}{27}z^{3}+\epsilon_{d}\left(\frac{544}{27}z-\frac{136}{9}z^{2}+\frac{148}{81}z^{3}\right)\right]\tilde{J}_{2}\left(\tilde{\mathcal{M}}_{2}\right)\nonumber \\
 &  & -\left[\frac{32}{27}z^{2}-\frac{16}{9}z^{3}+\frac{8}{9}z^{4}-\frac{4}{27}z^{5}\right]\tilde{J}_{3}\left(\tilde{\mathcal{M}}_{2}\right)\Big\}\end{eqnarray}

\subsection{Contributions to the charge radius $\langle r_{E0}^2 \rangle$}

\begin{eqnarray}
\left.\frac{d}{dq^{2}}\right|_{q^{2}=0}F_{1}^{\Delta 1} & = & -\frac{C_{\Delta\Delta}}{M_{\Delta}^{2}}\, I_{1}\,\int_{0}^{1}dz\,\,2z\,\,\frac{1}{216}\,\Big\{-132z+110z^{2}-27\tilde{\mathcal{M}_{1}}\nonumber \\
 &  & +\left[-66z^{2}+66z^{4}\right]\frac{1}{\tilde{\mathcal{M}}_{1}}+\left[-108-18z+231z^{2}\right]\ln\tilde{\mathcal{M}}_{1}+81\tilde{\mathcal{M}_{1}}\ln\tilde{\mathcal{M}_{1}}\Big\}\nonumber \\
\left.\frac{d}{dq^{2}}\right|_{q^{2}=0}F_{1}^{\Delta 2} & = & -\frac{C_{\Delta\Delta}}{M_{\Delta}^{2}}\,\, I_{2}\,\int_{0}^{1}dz\,\,2z\,\,\frac{1}{216}\,\,\Big\{-\left[-128+192z-113z^{2}+23z^{3}\right]\nonumber \\
 &  & -\left[128+\frac{139}{2}z\right]\tilde{\mathcal{M}_{2}}+\left[-132z^{2}+132z^{3}-33z^{4}\right]\frac{1}{\tilde{\mathcal{M}_{2}}}\nonumber \\
 &  & +\left[-132+216z-15z^{2}-\frac{15}{2}z^{3}\right]\ln\tilde{\mathcal{M}_{2}}+\left[51-30z\right]\tilde{\mathcal{M}_{2}}\ln\tilde{\mathcal{M}_{2}}\,\,\,\Big\}\nonumber \\
\left.\frac{d}{dq^{2}}\right|_{q^{2}=0}F_{1}^{\Delta 3} & = & -\frac{C_{N\Delta}}{M_{\Delta}^{2}}\,\, I_{3}\,\int_{0}^{1}dz\,\,2z\,\frac{1}{24}\,\Big\{-z^{2}2\left[z-z^{2}+r-rz\right]\frac{1}{\tilde{\mathcal{M}}_{3}}\nonumber \\
 &  & +\left[-2z^{2}-3rz\right]\ln\tilde{\mathcal{M}}_{3}\,\,\,\Big\}\nonumber \\
\left.\frac{d}{dq^{2}}\right|_{q^{2}=0}F_{1}^{\Delta 4} & = & -\frac{C_{N\Delta}}{M_{\Delta}^{2}}\,\, I_{4}\,\int_{0}^{1}dz\,\,2z\,\frac{1}{24}\,\Big\{+z^{2}-z^{2}\left[1-2z+z^{2}+2r-2rz+r^{2}\right]\frac{1}{\tilde{\mathcal{M}_{4}}}\nonumber \\
 &  & +\left[-3z+6z^{2}-3rz\right]\ln\tilde{\mathcal{M}_{4}}\,\,\,\,\,\,\Big\}\nonumber \\
\left.\frac{d}{dq^{2}}\right|_{q^{2}=0}F_{1}^{\Delta T12} & = & -\frac{2}{3}\frac{C_{\Delta\Delta}}{H_{A}^{2}M_{\Delta}^{2}}\, I_{T}\,\ln\left(\mu^{2}\right)\end{eqnarray}
Non-minimal contribution to $D2$ with $\kappa_{1}=\kappa_{2}=\kappa_{nm}$
(without the $\gamma^{\beta\delta\gamma}$ part) :

\begin{eqnarray}
\left.\frac{d}{dq^{2}}\right|_{q^{2}=0}F_{1}^{\Delta 2} & = & -\kappa_{nm}\frac{C_{\Delta\Delta}}{M_{\Delta}^{2}}I_{2}\int_{0}^{1}dz\,\frac{2z}{108}\Big\{-472+676z-236z^{2}+8z^{3}-\left[225-23z\right]\tilde{\mathcal{M}_{2}}\nonumber \\
 &  & +\left[-60+42z-12z^{2}+3z^{3}\right]\ln\tilde{\mathcal{M}_{2}}+\left[54+12z\right]\tilde{\mathcal{M}_{2}}\ln\tilde{\mathcal{M}_{2}}\,\,\,\,\Big\}\end{eqnarray}

\subsection{Renormalized constants}
The constants are with $\overline{\delta}=\delta^2(1+r)^2$:

\begin{eqnarray}
c_\mu        &=& r\frac{C_{\Delta\Delta}}{972} (-965 - 1781 \kappa_{nm})
 -r\frac{C_{N\Delta}}{36\delta}    (-12 - 28 r - 18 \mu^2 r + 28 r^2 + 54 r^3 - 
      15 r^4\nn\\&& - 51 r^5 + 6 r^6 + 18 r^7 - 
      12 r^3 (-8 + 4 r + 13 r^2 - 3 r^3\nn\\&& - 10 r^4 + r^5 + 3 r^6) \ln r + (3 + 6 r  - 
      9 r^2  - 24 r^3 \nn\\&& +  12 r^4 + 39 r^5  - 9 r^6  - 30 r^7  +  3 r^8  + 9 r^9) \ln \overline{\delta} ) \,\,\,,\\
c_\mathcal{Q} &=&\frac{C_{\Delta\Delta}}{243} (-604 - 4121 \kappa_{nm}) + 
 \frac{C_{N\Delta}}{18}  (4 r^2 (-4 - 2 r + 4 r^2 - 2 r^3 - 4 r^4 + 
       3 r^5 + 3 r^6) \ln r \nn\\&& +\frac{1}{\delta^2} (
    11 - 13 r - 13 r^2 + 22 r^3 - 3 r^4 - 15 r^5 + 11 r^6 + 6 r^7 - 
     6 r^8\nn\\&& - \overline{\delta} (3 - 3 r - r^2 + 3 r^3 - r^4 - 3 r^5 + 
        3 r^6) \ln \overline{\delta})\,\,\,,
\end{eqnarray} 
\begin{eqnarray}
c_\mathcal{O} &=&
\frac{C_{\Delta\Delta}}{2916} (-6247 + 20293 \kappa_{nm}) 
- \frac{C_{N\Delta}}{36\delta(1+r)} (-8 - 10 r + 2 r^3 + 7 r^4 + 
    20 r^5\nn\\&& + 7 r^6 - 12 r^7 - 6 r^8 + 
    4 r (-8 + 12 r^2 + 4 r^3 + 3 r^4\nn\\&& - 2 r^5 - 13 r^6 - 5 r^7 + 
       6 r^8 + 3 r^9) \ln r \nn\\&& - \overline{\delta} (-3 - 5 r - 3 r^2 - r^3 + r^4 + 6 r^5 + 
       3 r^6) \ln \overline{\delta})\,\,\,,\\
c_{r0}       &=& \frac{C_{\Delta\Delta}}{1944M_\Delta^2\overline{\delta}}(3850 - 7700 r + 3850
r^2 + (-5395 + 10790 r - 5395 r^2) \kappa_{nm}) 
\nn\\&& + \frac{C_{N\Delta}}{1944M_\Delta^2}(( -4320 r^2 - 2160 r^3 + 4320 r^4 - 1188 r^5 - 367vv2 r^6 + 3240 r^7\nn\\&& + 
   3240 r^8) \ln r+\frac{1}{\delta^2}(2268 - 3078 r - 2430 r^2 + 4806 r^3 - 324 r^4\nn\\&& - 3888 r^5 + 2646 r^6 + 
 1620 r^7 -  1620 r^8 + (-648 + 729 r + 1566 r^2\nn\\&& - 2187 r^3 - 1080 r^4 + 
    2997 r^5 - 756 r^6 - 2349 r^7 + 1728 r^8\nn\\&& + 810 r^9 - 
    810 r^{10}) \ln \overline{\delta}))\,\,\,.
\end{eqnarray} 

\end{appendix}

\end{document}